\documentclass[twocolumn,preprintnumbers,floats,prd,amssymb,floatfix,nofootinbib,balancelastpage,superscriptaddress,amsmath]{revtex4-1}

\pdfoutput=1
\usepackage[utf8]{inputenc}
\usepackage{amssymb}
\usepackage{amsmath}
\usepackage{amsfonts}
\usepackage{graphicx}
\usepackage{color}
\usepackage{xspace}
\usepackage{comment}
\usepackage{hyperref}
\usepackage[normalem]{ulem}
\usepackage[section]{placeins}
\usepackage{afterpage}
\usepackage{slashed}
\usepackage{enumerate}
\usepackage{enumitem}
\usepackage{caption}
\usepackage{subfigure}
\usepackage{float}
\usepackage{ulem}
\usepackage{verbatim}
\usepackage{multirow,rotating}
\usepackage[dvipsnames]{xcolor}

\definecolor{dcolour}{rgb}{.5, .5, .5}
\def\gsim{\raise0.3ex\hbox{$\;>$\kern-0.75em\raise-1.1ex\hbox{$\sim\;$}}}
\def\lsim{\raise0.3ex\hbox{$\;<$\kern-0.75em\raise-1.1ex\hbox{$\sim\;$}}}
\def\gsim{\raise0.3ex\hbox{$\;>$\kern-0.75em\raise-1.1ex\hbox{$\sim\;$}}}
\def\lsim{\raise0.3ex\hbox{$\;<$\kern-0.75em\raise-1.1ex\hbox{$\sim\;$}}}

\newcommand{\ba}[1]{\begin{eqnarray} \label{(#1)}}
	\newcommand{\ea}{\end{eqnarray}}

\newcommand{\iab}{\rm ab^{-1}}

\newcommand{\mltp}{{\mkern-2mu\times\mkern-2mu}}

\newcommand{\met}{\slashed{E}_T}

\begin{document}

\title{Probe the Mixing Parameter $|V_{\tau N}|^2$ for Heavy Neutrinos}% Force line breaks with \\
	
\author{Lingxiao Bai}
\email{305858@whut.edu.cn}
\affiliation{Department of Physics, School of Science, \\Wuhan University of Technology, 430070 Wuhan, Hubei, China }
	
\author{Ying-nan Mao}
\email{ynmao@whut.edu.cn}
\affiliation{Department of Physics, School of Science, \\Wuhan University of Technology, 430070 Wuhan, Hubei, China }
	
\author{Kechen Wang}
\email{kechen.wang@whut.edu.cn (Corresponding author)}
\affiliation{Department of Physics, School of Science, \\Wuhan University of Technology, 430070 Wuhan, Hubei, China }
	
	
\begin{abstract}
Because of the difficulty in detecting final state taus, the mixing parameter $|V_{\tau N}|^2$ for heavy neutrino $N$ is not well studied at current experiments, compared with other mixing parameters $|V_{e N}|^2$ and $|V_{\mu N}|^2$. In this paper, we focus on a challenging scenario where $N$ mixes with active neutrino of tau flavour only,  i.e. $ |V_{\tau N}|^2 \neq 0 $ and $|V_{e N}|^2 = |V_{\mu N}|^2 = 0$. We derive current constraints on $|V_{\tau N}|^2$ from the rare $Z$-boson decay and electroweak precision data (EWPD). To forecast the future limits,  we also investigate the signal $p p \to \tau^{\pm} \tau^{\pm} j j $ via  a Majorana heavy neutrino at future proton-proton colliders. To suppress the background, both taus are required to decay leptonically into muons, leading to the final state containing two same sign muons, at least two jets plus moderate missing energy. The signal and relevant background processes are simulated at the HL-LHC and SppC/FCC-hh with center-of-mass energy of 14 TeV and 100 TeV. The preselection and multivariate analyses based on machine-learning are performed to reduce background. Limits on $|V_{\tau N}|^2$ are shown for heavy neutrino mass in the range 10-1000 GeV based on measurements from the rare $Z$-boson decay and EWPD, and searches at the HL-LHC and SppC/FCC-hh with integrated luminosities of 3 and 20 ab$^{-1}$.
\end{abstract}
%%%%%%%%%%%%%%%%%%%%%%%%%%%%%%%%%%%%%%%%%%%%%%%%%%%%%%%%%%%%%%%%%%%%%%
	
%\arxivnumber{}
%\pacs{} f
	
\vskip10mm
	
\maketitle
\flushbottom
	
%%%%%%%%%%%%%%%%%%%%%%%%%%%%%%%%%%%%%%%%%%%%%%%%%%%%%%%%%%%%%%%%%%%%%%
%%%%%%%%%%%%%%%%%%%%%%%%%%%%%%%%%%%%%%%%%%%%%%%%%%%%%%%%%%%%%%%%%%%%%%
%\noindent\rule[0pt]{\linewidth}{0.6pt}
%\vspace{-0.9cm}
%\tableofcontents
%\vspace{0.4cm}
%\noindent\rule[0pt]{\linewidth}{0.6pt}
%%%%%%%%%%%%%%%%%%%%%%%%%%%%%%%%%%%%%%%%%%%%%%%%%%%%%%%%%%%%%%%%%%%%%%

%%%%%%%%%%%%%%%%%%%%%%%%%%%%%%%%%%%%%%%%%%%%%%%
%%%%%%%%%%%%%%%%%%%%%%%%%%%%%%%%%%%%%%%%%%%%%%%
%.   SECTION 
%%%%%%%%%%%%%%%%%%%%%%%%%%%%%%%%%%%%%%%%%%%%%%%
%%%%%%%%%%%%%%%%%%%%%%%%%%%%%%%%%%%%%%%%%%%%%%%
\section{Introduction}
\label{sec:intro}
	
The neutrino oscillation experiments~\cite{Super-Kamiokande:1998kpq,MINOS:2006foh,MINOS:2011amj,PhysRevLett.108.131801,Ling:2013fta,Kim:2013sza} have proved that at least two of three active neutrinos in standard model (SM) are massive. This problem needs to be solved by extending the standard model.
The seesaw mechanism~\cite{FRITZSCH1975256,Minkowski:1977sc,Yanagida:1979as,Sawada:1979dis,Mohapatra:1979ia,Glashow:1979nm,GellMann:1980vs,Keung:1983uu,Foot:1988aq,Mohapatra:1986aw,MAGG198061} 
is one of the simplest solutions, and usually predicts the existence of heavy neutrinos.
Therefore, heavy neutrinos are important candidates beyond the standard model, and searching for them is a vital way to test the seesaw mechanism and probe the new physics.

Heavy neutrinos can mix with standard model active neutrinos of all flavours ($e, \mu, \tau$), in principle.
At colliders, heavy neutrinos are usually also called the heavy neutral leptons, and have been extensively explored.
They can be probed via the effects of heavy neutrino oscillations~\cite{Cvetic:2015ura, Cvetic:2018elt, Cvetic:2019rms, Tapia:2019coy, Cvetic:2020lyh, Tapia:2021gne, Cvetic:2021itw}, or searched from the decays of mesons~\cite{Cvetic:2013eza, Cvetic:2014nla, Cvetic:2015naa, Moreno:2016cfz, Zamora-Saa:2016ito}, Higgs bosons~\cite{Gao:2021one, Gao:2019tio}, $W$-bosons~\cite{Antusch:2018bgr, Dib:2017iva, Dib:2017vux, Dib:2016wge} and $Z$-bosons~\cite{Wang:2019xvx}.
Summaries of collider searches of heavy neutrinos can be found in 
Refs.~\cite{Atre:2009rg,Deppisch:2015qwa,Das:2015toa,Das:2016hof,Cai:2017mow,Das:2017rsu,Bolton:2019pcu,Ding:2019tqq,Shen:2022ffi} 
and references therein.
Recent experimental studies on heavy neutrino searches can be found in Refs.~\cite{CMS:2018iaf, CMS:2018jxx, CMS:2021lzm, ATLAS:2019kpx, LHCb:2020wxx, NA62:2020mcv, Belle:2013ytx, T2K:2019jwa}, and are reviewed in Ref.~\cite{Gu:2022muc}.

The mixing parameter $|V_{\ell N}|^2$ is related to the matrix element describing the mixing of the heavy neutrino $N$ with the SM neutrino of flavor $\ell = e, \mu, \tau$.
Compared with plentiful studies focusing on the mixing parameters $|V_{e N}|^2$ and $|V_{\mu N}|^2$, because of the challenges in detecting the final state taus, the mixing parameter $|V_{\tau N}|^2$ is more difficult to be probed, making it not well studied at current experiments.
However, there do exist some phenomenological studies to probe the mixing parameter $|V_{\tau N}|^2$ in different heavy neutrino mass ranges.
Among them, most studies assume the heavy neutrino mixes with active neutrinos of not only tau flavor but also muon and/or electron flavors.
For example, 
Ref.~\cite{Pascoli:2018rsg} 
consider mixing parameters $|V_{\tau N}|^2 = |V_{e N}|^2 \neq 0$ and $ |V_{\mu N}|^2 = 0$, and searches for heavy Dirac neutrinos in the mass range between 150 and 1000 GeV at the HL-LHC, while Refs.~\cite{Bondarenko:2018ptm,Abada:2018sfh} consider $|V_{\tau N}|^2 = |V_{e N}|^2 = |V_{\mu N}|^2$, and investigate heavy neutrinos with mass below 20 GeV.
In Ref.~\cite{Gu:2022nlj}, some of our authors consider the scenario where $|V_{\tau N}|^2, |V_{e N}|^2 \neq 0$ and $ |V_{\mu N}|^2 = 0$, and utilize the lepton number violation signal process $p\, e^- \to \tau^+ jjj$ to search for heavy Majorana neutrinos with mass between 10 and 1000 GeV at future proton-electron colliders.

So far, only a few phenomenological studies investigate the challenging scenario where $N$ mixes with active neutrino of tau flavour only,  i.e. $|V_{\tau N}|^2 \neq 0 $ and $|V_{e N}|^2 = |V_{\mu N}|^2 = 0$.
For instance, 
Refs.~\cite{Dib:2019tuj,Zhou:2021ylt} utilize large $e^- e^+ \to \tau^- \tau^+$ samples collected at $B$-factories to search for long-lived heavy neutrinos produced via the tau decay, and constrain $|V_{\tau N}|^2$ for $m_N$ between 0.1 and 1.7 GeV.
Ref.~\cite{Cvetic:2019shl} searches for heavy neutrinos from $B$-meson decays at the LHCb, and sets limits on $|V_{\tau N}|^2$ for $m_N$ between 2 and 4.5 GeV.
Ref.~\cite{Cottin:2018nms} investigates long-lived heavy neutrinos decaying inside the inner trackers of the detectors, and probe $|V_{\tau N}|^2$ for $m_N$ between 5 and 20 GeV at the LHC.
Ref.~\cite{Cheung:2020buy} considers the tri-lepton final state  ($\tau_h \tau_h + e/\mu$)  containing two hadronic decaying tau-jets plus one electron or muon at the LHC, and constrains $|V_{\tau N}|^2$ for $m_N$ between 25 and 150 GeV.
It is worth noting that although it is assumed that only $|V_{\tau N}|^2 \neq 0$ and $|V_{e N}|^2 = |V_{\mu N}|^2 = 0$ in this study,  the scenario with $|V_{\tau N}|^2 \neq 0 $ and $|V_{e N}|^2, |V_{\mu N}|^2 \neq 0$ can also induce the same tri-lepton final state  ($\tau_h \tau_h + e/\mu$), and hence this final state cannot confirm the scenario with only $|V_{\tau N}|^2 \neq 0$.
Ref.~\cite{Florez:2017xhf} probes $|V_{\tau N}|^2$ for $m_N > 150$ GeV via the final state ($\tau_h \tau_h + 4 j$) including two hadronic decaying tau-jets plus four regular jets at the LHC.
Moreover, Ref.~\cite{Pascoli:2018heg} considers heavy neutrinos produced at $pp$ colliders and decaying into multiple charged leptons. Assuming only  $|V_{\tau N}|^2 \neq 0$, two charged lepton flavor-conserving signal processes $p p \to \tau_h \ell^+_i \ell^-_j $ and $p p \to \tau_h^+ \tau_h^- \ell_X + \tau_h^{\pm} \tau_h^{\pm} l_i$ are investigated, where $\ell_X \in \{e, \mu, \tau_h \}$ and $\ell_i, \ell_j \in \{e, \mu\}$. Sensitivities on $|V_{\tau N}|^2$ are reported for $m_N$ between 150 and 2900 GeV at different center-of-mass energies of 14, 27 and 100 TeV.

In this study, we focus on the challenging scenario where $ |V_{\tau N}|^2 \neq 0 $ and $|V_{e N}|^2 = |V_{\mu N}|^2 = 0$.
In this scenario, for heavy neutrinos with mass above 10 GeV, we derive current constraints on the mixing parameter $|V_{\tau N}|^2$ from the measurements of rare $Z$-boson decay by the DELPHI collaboration~\cite{DELPHI:1996qcc} and the electroweak precision data (EWPD)~\cite{delAguila:2008pw,Akhmedov:2013hec,Basso:2013jka,deBlas:2013gla,Antusch:2015mia,Chrzaszcz:2019inj,Cheung:2020buy}.

To forecast future limits on $|V_{\tau N}|^2$, we also investigate the relevant signal at future proton-proton colliders. 
Assuming $N$ mixes with active neutrino of tau flavour only,  a heavy Majorana neutrino $N$ can lead to the lepton number violating signal process of $p\, p \to \tau^\pm  \tau^\pm jj$.
To suppress the background, both taus are required to decay leptonically into muons, leading to the final state containing two same sign muons, at least two jets plus moderate missing energy.
We develop the search strategy and simulate the signal and background event samples at the high-luminosity Large Hadron collider (HL-LHC) running with center-of-mass energy $\sqrt{s} =$ 14 TeV, and at the Super proton-proton Collider (SppC)~\cite{CEPC-SPPCStudyGroup:2015esa,Gao:2017ssn,CEPCStudyGroup:2018rmc,Gao:2021bam,Tang:2022fzs} or the proton-proton collision mode of Future Circular Collider (FCC-hh)~\cite{Benedikt:2018csr,FCC:2018vvp,Benedikt:2022kan} running with $\sqrt{s} =$ 100 TeV. 
The preselection and multivariate analyses based on machine-learning are performed to reduce background.

Limits on $|V_{\tau N}|^2$ are shown for heavy neutrino mass in the range 10-1000 GeV based on measurements from the rare $Z$-boson decay and EWPD, and searches at the HL-LHC and SppC/FCC-hh with integrated luminosities of 3 and 20 ab$^{-1}$.
We emphasize that the signal process $p\, p \to \tau^\pm  \tau^\pm jj$ depends on the mixing parameter $|V_{\tau N}|^2$ only and exists once $|V_{\tau N}|^2 \neq 0$, so it provides an opportunity to probe $|V_{\tau N}|^2$ independent of other mixing parameters.
Therefore, although this signal is challenging to be investigated at colliders because of the difficulty in detecting final state taus and large background, it is still meaningful and deserves careful studies.

The article is organised as follows. 
In Sec.~\ref{sec:EWPD}, we derive current constraints on $|V_{\tau N}|^2$ from measurements of rare $Z$-boson decay and EWPD, and show details of derivation process.
In Sec.~\ref{sec:ppCollider}, we forecast future limits on  $|V_{\tau N}|^2$ by searching for heavy neutrinos at future $pp$ colliders, including the introduction of signal and background processes, search strategy and data analysis.
The combined constraints on $|V_{\tau N}|^2$ are presented in Sec.~\ref{sec:result}.
We summarize and discuss in Sec.~\ref{sec:sum}.
More details of this study are listed in 
Appendices~\ref{appendix:obs} - \ref{appendix:Effmixings}.

\section{Constraints from rare $Z$-boson decay and EWPD}
\label{sec:EWPD}

In this section, we show details of deriving the constraints on  $|V_{\tau N}|^2$ from rare $Z$-boson decay and EWPD, assuming $|V_{eN}|^2 = |V_{\mu N}|^2 =0$ and only $|V_{\tau N}|^2 \neq 0$. 

When $m_N<m_Z$, the ratio of partial decay widths for $Z\rightarrow N\nu_{\tau},N\bar{\nu}_{\tau}$ rare decays is \cite{DELPHI:1996qcc,Antusch:2015mia}
\begin{equation}
\frac{\Gamma_{Z\rightarrow N\nu_{\tau}}}{\Gamma_{Z\rightarrow\nu_{\tau}\bar{\nu}_{\tau}}}=
\frac{\Gamma_{Z\rightarrow N\bar{\nu}_{\tau}}}{\Gamma_{Z\rightarrow\nu_{\tau}\bar{\nu}_{\tau}}}=
|V_{\tau N}|^2f\left(\frac{m_N}{m_Z}\right) \, ,
\end{equation}
to the leading order of $|V_{\tau N}|^2$, where the function
\begin{equation}
f(x)\equiv\left\{\begin{array}{cc}\left(1-x^2\right)\left(1+x^2/2\right),&(x\leq1);\\0&(x>1).\end{array}\right.
\end{equation}
Note that if $m_N<m_Z/2$, two-body $Z\rightarrow NN$ decay mode also exists, but its partial decay width $\Gamma_{Z\rightarrow NN}\propto |V_{\tau N}|^4\ll\Gamma_{Z\rightarrow N\nu_{\tau},\, N\bar{\nu}_{\tau}}$, so one can always ignore this exotic decay channel. 
We then have the branching ratio for exotic $Z$ decay is 
\begin{equation}
\textrm{Br}_{\textrm{exo}} \simeq 0.13\, |V_{\tau N}|^2\, f\left(\frac{m_N}{m_Z}\right).
\end{equation}

Comparing with the experimental result $\textrm{Br}_{\textrm{exo}}\lesssim1.3\times10^{-6}$ at $95\%$ C.L. \cite{DELPHI:1996qcc}, we finally obtain the constraint on $|V_{\tau N}|^2$ in the $m_N<m_Z$ region as
\begin{equation}
|V_{\tau N}|^2\lesssim\frac{10^{-5}}{f(m_N/m_Z)}.
\end{equation}
The corresponding limits on$|V_{\tau N}|^2$ are presented  
in Sec.~\ref{sec:result}.
%Fig. \ref{fig:sensitivity}.

When $m_N>m_Z$, the EWPD constraint becomes important. Since $|V_{eN}|^2=|V_{\mu N}|^2=0$, $G_F$ does not modify from its value in the SM, and only the width of $Z\rightarrow\nu_{\tau}\bar{\nu}_{\tau}$ is modified as
\begin{equation}
\frac{\Gamma_{Z\rightarrow\nu_{\tau}\bar{\nu}_{\tau}}}{\Gamma_{Z\rightarrow\nu_{\tau}\bar{\nu}_{\tau},\, \textrm{SM}}}=
\left(1-|V_{\tau N}|^2\right)^2\, \simeq1-2\, |V_{\tau N}|^2 \, ,
\end{equation}
to the leading order of $|V_{\tau N}|^2$. 
The total decay width of $Z-$boson is modified as
\begin{equation}
\label{eq:GaZ}
\frac{\Gamma_Z}{\Gamma_{Z,\, \textrm{SM}}}=1+0.133\, |V_{\tau N}|^2\left(f\left(\frac{m_N}{m_Z}\right)-1\right).
\end{equation}

Another important observable is the cross section of the process $e^+e^-\rightarrow Z\rightarrow\textrm{had}$, denoted as $\sigma_H$, where ``had" means all kinds of hadrons. Since 
\begin{equation}
\sigma_H=\frac{12\pi\, \Gamma_{Z\rightarrow e^+e^-}\, \Gamma_{Z\rightarrow\textrm{had}}}{m^2_Z\, \Gamma^2_Z}\, ,
\end{equation}
its modification comparing with the SM value is 
\begin{equation}
\label{eq:sigH}
\frac{\sigma_H}{\sigma_{H,\, \textrm{SM}}}=\left(\frac{\Gamma_Z}{\Gamma_{Z,\textrm{SM}}}\right)^{-2}\simeq
1 - 0.267\, |V_{\tau N}|^2\left(f\left(\frac{m_N}{m_Z}\right)-1\right).
\end{equation}

When $m_N>m_Z$, Eqs.~(\ref{eq:GaZ}) and (\ref{eq:sigH}) become 
\begin{eqnarray}
\frac{\Gamma_Z}{\Gamma_{Z,\, \textrm{SM}}}&=&1-0.133\, |V_{\tau N}|^2\, ;\\
\frac{\sigma_H}{\sigma_{H,\, \textrm{SM}}}&=&1+0.267\, |V_{\tau N}|^2\, .
\end{eqnarray}
We perform the global-fit 
\footnote{
The constraints come from the modifications in $\Gamma_Z$ and $\sigma_H$. 
Different from the case with both $|V_{\tau N}|^2$ and $|V_{e N}|^2$ nonzero in \cite{Gu:2022nlj}, in this study we have assumed only $|V_{\tau N}|^2 \neq 0$ and $|V_{e N}|^2 =0$, and thus do not have the modifications of $R_{\ell}\equiv\Gamma_{Z\rightarrow
\textrm{had}}/\Gamma_{Z\rightarrow\ell^+\ell^-}$ and $R_q\equiv\Gamma_{Z\rightarrow q\bar{q}}/\Gamma_{Z\rightarrow\textrm{had}}$ (where $\ell=e,\mu,\tau$ denotes a lepton and $q=c,b$ denotes a quark), comparing with the SM.
} 
and present corresponding limits on $|V_{\tau N}|^2$ at $95\%$ confidence level in 
in Sec.~\ref{sec:result}.
In the $m_N>m_Z$ region, $|V_{\tau N}|^2$ is constrained to be less than $\sim 5\times10^{-3}$, which is consistent with the results in Refs. \cite{Antusch:2015mia,Cheung:2020buy}.

\section{Constraints from $pp$ collider searches}
\label{sec:ppCollider}

In this section, we develop the search strategy to search for a heavy Majorana neutrino in the tau final state, and forecast the limits on $ |V_{\tau N}|^2$ at $pp$ colliders.
To simplify the analyses, we consider the simplified Type-I seesaw model and the scenario that only one generation of heavy neutrinos $N$ is light and can be produced at colliders. 
The same as the above section, the $N$ is assumed to mix with active neutrinos of tau flavour only, i.e. $ |V_{\tau N}|^2 \neq 0 $ and $|V_{e N}|^2 = |V_{\mu N}|^2 = 0 $.

\subsection{The signal process}
\label{subsec:sig}

\begin{figure}[h]
\centering
\includegraphics[width=5cm,height=3.5cm]{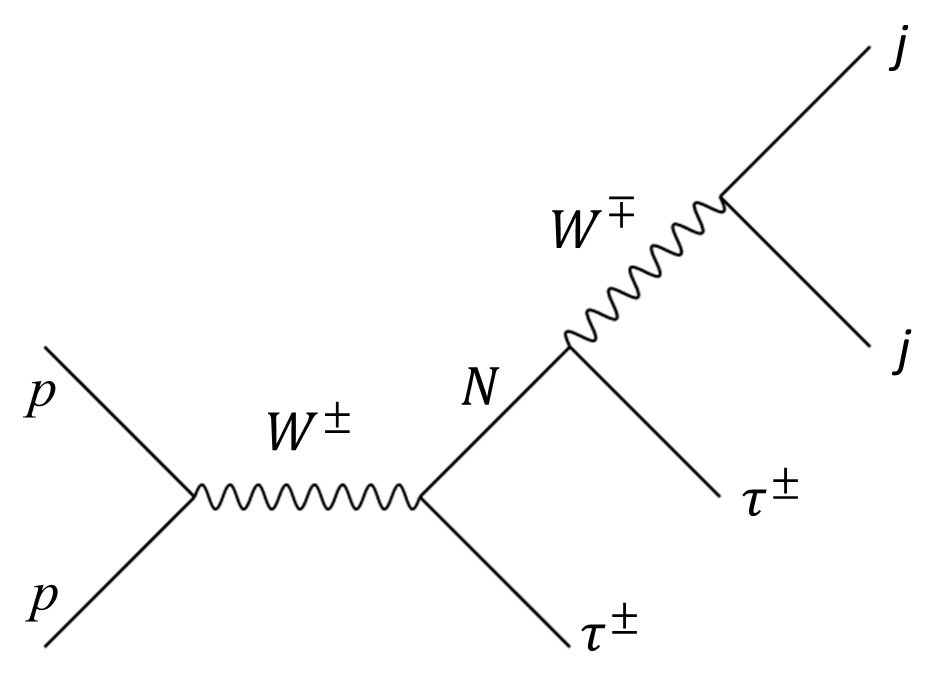}  
\caption{
The production process of the LNV signal via a Majorana heavy neutrino $N$ at $pp$ colliders.
}
\label{fig:Feynman}
\end{figure}

As shown in Fig.~\ref{fig:Feynman}, in this scenario, a heavy Majorana neutrino $N$ can be produced associated with a tau
at $pp$ colliders, and decay into one tau plus two jets, leading to the same sign di-tau plus di-jet final state.
The lepton number of this process changes from $0$ to $\pm 2$, so it violates the conservation of lepton number.

For the data simulation of signal and background processes in this study, we utilize the MadGraph5 program~\cite{Alwall:2014hca} to calculate the production cross sections and generate the collision events.
The Pythia8 program~\cite{Sjostrand:2006za} is used to decay taus and perform the parton showering and hadronization, while the configuration card files for the HL-LHC and FCC-hh detectors are implemented to the Delphes program~\cite{deFavereau:2013fsa} to complete the detector simulation.

\begin{figure}[h]
\centering
\includegraphics[width=8cm,height=6cm]{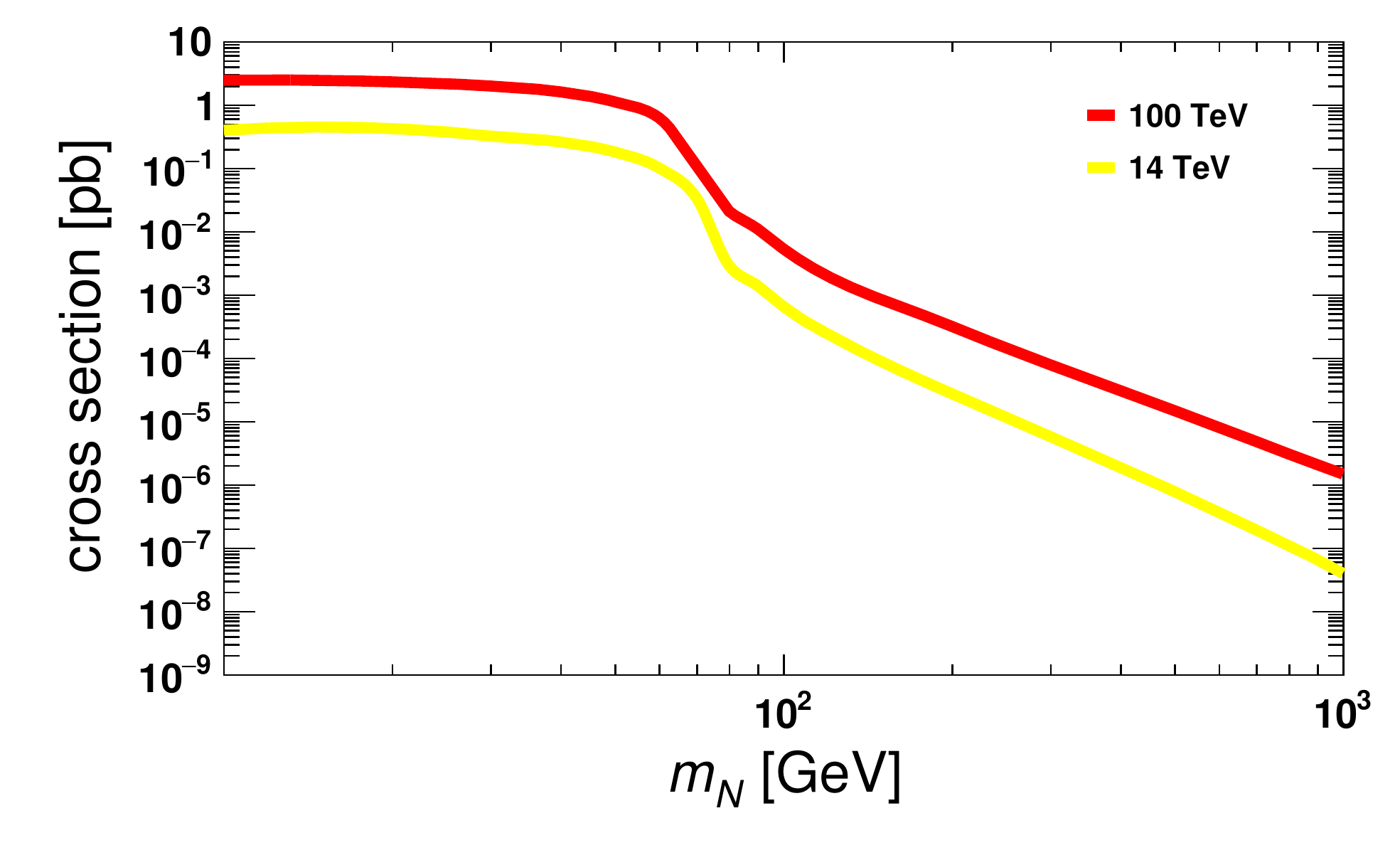}
\caption{
The production cross sections of the signal process $p\, p \to \tau^\pm  \tau^\pm jj $ via a heavy Majorana neutrino $N$ when varying $m_N$ between 10 and 1000 GeV at the HL-LHC and SppC/FCC-hh. Here, the mixing parameter $|V_{\tau N}|^2$ is fixed to be $10^{-4}$.
}
\label{fig:crs}
\end{figure}

For the signal process,
we implement the Universal FeynRules Output model file~\cite{Alva:2014gxa,Degrande:2016aje} which extends the SM with additional heavy neutrinos interacting with active neutrinos, into the MadGraph5 program to calculate the production cross sections and generate the collision events for the signal process $p\, p \to \tau^\pm  \tau^\pm jj$.
The energies of both proton beams are set to be 7 and 50 TeV for HL-LHC and SppC/FCC-hh, respectively.
Representative heavy neutrino masses between 10 GeV and 1000 are selected and the corresponding cross sections for the signal process $p\, p \to \tau^\pm  \tau^\pm jj$ are plotted in Fig.~\ref{fig:crs},
where the yellow and red curves represent the results at HL-LHC and FCC-hh, respectively.
When calculating the production cross sections and generating the collision events, loose cuts are applied at the parton level to maximally accept events and make the simulation more realistic and closer to experiments:
the minimal $p_T$ for jets and leptons is 0.5 GeV;
the maximal absolute value of rapidity is 10 for the jets, and 5 for the leptons and photons;
the minimum angular distance $\Delta R$ between two objects is 0.1.

The curves in Fig.~\ref{fig:crs} show that the cross sections at 14 TeV and 100 TeV have similar trends with the change of heavy neutrino mass.
When $m_N \lesssim$ 50 GeV, the signal cross section does not change significantly with $m_N$.
For $m_N$ between 60 GeV and 90 GeV, the signal cross section decreases rapidly with increasing $m_N$.
For larger $m_N$, the signal cross section decreases steadily as $m_N$ increases.
	
Since taus are unstable, they decay either leptonically into muons and electrons, or hadronically into mesons, leading to final state leptons or tau-jets at colliders.
The leptonic and hadronic final states have different kinematics and background, and thus need individual analyses. 

For hadronic tau decay case, the final state has two same sign tau-jets, at least two regular jets, and missing energy.
Because of the difficulty in reconstructing the tau-jet and identifying its charge, the realistic detector effects must be considered.
To estimate the misidentification rate of recognizing a regular jet ($j$) as a tau-jet ($\tau_h$), we simulate the $Z \to j j$ events at the HL-LHC, and implement the CMS detector configuration file to perform the detector simulation.
By selecting final state tau-jets in this sample, we find that the misidentification rate of $j \to \tau_h$ is around $1\%$ at the HL-LHC, which is consistent with the performance of the ATLAS detector as shown in studies~\cite{ATLAS:2021srw, ATLAS:2022aip}.
Similarly, to estimate the misidentification rate for the charge of $\tau_h$, we simulate the $Z \to \tau^+ \tau^-$ events.
By selecting final state tau-jets with same charges in this sample, we find that the misidentification rate for the charge of $\tau_h$ is more than $1\%$ at the HL-LHC.

Therefore, considering these misidentification rates and huge jet production at the $pp$ colliders, the hadronically decaying tau final state will suffer from too large background.
In order to suppress the background, we select the final state where both taus decaying leptonically into muons (i.e. $\tau^- \to \mu^- \bar{\nu}_\mu \nu_\tau$).
The final state has two same sign muons $\mu^\pm \mu^\pm$, two jets and moderate missing energy.

%\newpage
\subsection{Background processes}
\label{subsec:bkg}
	
Since the signal process $p\, p \to \tau^\pm \tau^\pm jj$ violates the lepton number conservation explicitly, it has little SM background in theory.
However, considering the leptonic decay of taus, there are still some SM processes which can have similar final state containing two same sign muons $\mu^\pm \mu^\pm$, two jets and moderate missing energy.
We consider six SM background processes and label them as ``B1-B6" in this article.
They include the single boson process $W^{\pm} j j$ and $Z j j$, the di-boson process $W^{\pm} Z j j$, $W^+ W^- j j$, $Z Z j j$, and di-top process $t \bar{t}$.
The production cross sections of signal process with benchmark $m_N$ = 50 GeV and $|V_{\tau N}|^2 = 10^{-4}$, and of dominant background processes at the HL-LHC and SppC/FCC-hh are listed 
in Table~\ref{tab:crsc1}.
	
\begin{table}[h]
\centering
\begin{tabular}{cccc}
\hline 
\hline
& $\sigma$ [pb] & HL-LHC  & SppC/FCC-hh  \\ 	
\hline 
S & $\tau^\pm \tau^\pm j j  $ &  $ 0.180$  &  $1.12$  \\ 
B1 & $ W^\pm j j$&$1.21\mltp 10^{5}$  & $1.31\mltp 10^{6}$ \\
B2 & $ Z j j$& $1.10\mltp 10^{4}$  & $1.25\mltp 10^{5}$  \\ 
B3 & $ W^\pm Z j j$& $3.44\mltp 10^{2}$  & $5.67\mltp 10^{3}$ \\  
B4 & $ W^+ W^- j j$& $1.68\mltp 10^{2}$  & $2.32\mltp 10^{3}$ \\ 
B5 & $ Z Z j j$& $1.03\mltp 10^{2}$  & $1.56\mltp 10^{3}$ \\ 
B6 & $t \bar{t}$& $5.97\mltp 10^{2}$  & $2.47\mltp 10^{4}$ \\
\hline 
\hline
\end{tabular} 
\caption{
Production cross sections of signal process with $m_N$ = 50 GeV and $|V_{\tau N}|^2 = 10^{-4}$, and dominant background processes at the HL-LHC and SppC/FCC-hh.
}
\label{tab:crsc1}
\end{table}

We note that due to limited computing resources, the following conditions during the generating procedure are required to guarantee that large number of events can be produced and survive after applying all selection cuts, and thus the statistic uncertainty can be small:
for B1 and B2, $W$- and $Z$-bosons decay into muons.
Besides, we check and find that because of small production cross section for the $W^\pm W^\pm j j$ process, its contribution to total background after all selection cuts is negligible compared with the  $ W^+ W^- j j$ process.
Therefore, for B4, only $W^+ W^- j j$ events are generated.

\subsection{Search strategy and data analysis}
\label{subsec:analysis}

We apply the following preselection cuts to select the signal and reject the background events at the first stage.
(i) Exactly two muons with the same charge, i.e. same sign di-muon pair $\mu^{\pm} \mu^{\pm}$; 
two muons are sorted by their  transverse momenta $p_T$, and
the first (second) leading muon has $p_T$ bigger than 10 (5) GeV, i.e. $p_T(\mu_1) >$ 10 GeV and $p_T(\mu_2) >$ 5 GeV.
(ii) At least two regular jets, i.e. $N(j) \geq 2$;
the jets are sorted by $p_T$, and $p_T(j_1) \geq 20$  GeV, $p_T(j_2) \geq 10$ GeV.
(iii) Events that has final state b-jets or leptons other than muons are rejected, i.e. $N(b$-${\rm jet}) = 0$ and $N(e) = 0$.
	
For signal data simulation, we choose representative heavy neutrino masses $m_N$ from 10 to 1000 GeV, and generate around one million events at the HL-LHC and SppC/FCC-hh for each $m_N$ case.
The chosen masses are $m_N = $ 10, 20, 30, 40, 50, 60, 70, 80, 90, 100, 110, 120, 300, 500 GeV for HL-LHC, 
and  $m_N = $ 10, 20, 30, 40, 50, 60, 80, 90, 100, 200, 300, 500, 700, 800, 900, 1000 GeV for SppC/FCC-hh, respectively.
Due to limited computational resources, we are not able to generate huge number of events for every background processes.
One million to ten million of events for different background processes are generated, and
the number of simulated events for each background process is determined according to its importance in reducing the statistical uncertainty on final limits.
The efficiencies of preselection cuts for signal with different $m_N$ assumptions and six background processes are summarized 
in Table~\ref{tab:CutEffiHLLHC} 
for HL-LHC and 
Table~\ref{tab:CutEffi100}
 for SppC/FCC-hh.

Moreover, to help researchers understand the kinematics of both signal and background processes,
In Appendix~\ref{appendix:obs}, we present distributions of representative kinematic observables for the signal with benchmark $m_N =$ 50 GeV and dominant background processes at the HL-LHC and SppC/FCC-hh.

To further reject the background, we input the following twenty-four observables into the TMVA package~\cite{Hocker:2007ht} to perform the multivariate analysis (MVA).
\begin{enumerate}[label*=\Alph*.]
\item The four-momenta of final state muons:
$E(\mu_1)$, $p_{x}(\mu_1)$, $p_{y}(\mu_1)$, $p_{z}(\mu_1)$;
$E(\mu_2)$, $p_{x}(\mu_2)$, $p_{y}(\mu_2)$, $p_{z}(\mu_2)$.
\item The number of regular jets $N(j)$ and four-momenta of the first two leading jets:
$E(j_{1})$, $p_{x}(j_{1})$, $p_{y}(j_{1})$, $p_{z}(j_{1})$;
$E(j_{2})$, $p_{x}(j_{2})$, $p_{y}(j_{2})$, $p_{z}(j_{2})$.
\item The magnitude and azimuthal angle of the missing transverse momentum:
$\met$, $\phi(\met)$.
\item Some additional observables of muons and jets which are also useful to reject background:
$p_{T,\, {\rm max}}^{\rm iso} (\mu)$; $N_{\rm track}(j_1)$, $N_{\rm track}(j_2)$; $R_{E}(j_1)$, $R_{E}(j_2)$.
\end{enumerate}

Details of description of the last set of observables can be found in Ref.~\cite{lhcoFormat}.
Among them,
$p_{T}^{\rm iso} (\mu)$ is the summed $p_T$ of other objects excluding the muon in a $R=0.4$ cone around the muon, which measures the behaviors around a muon and indicates its  isolation quality, 
and $p_{T,\, {\rm max}}^{\rm iso} (\mu)$ denotes the $p_{T}^{\rm iso} (\mu)$ with bigger value for two final state muons.
$N_{\rm track}(j_i)$ is the number of tracks associated with the jet.
$R_{E}(j_i)$ is the ratio of the hadronic versus electromagnetic energy deposited in the calorimeter cells associated with the jet, which is typically larger than one for a real jets and can be used to reject fake jets.

\begin{figure}[h]
\centering
\includegraphics[width=4.2cm,height=3cm]{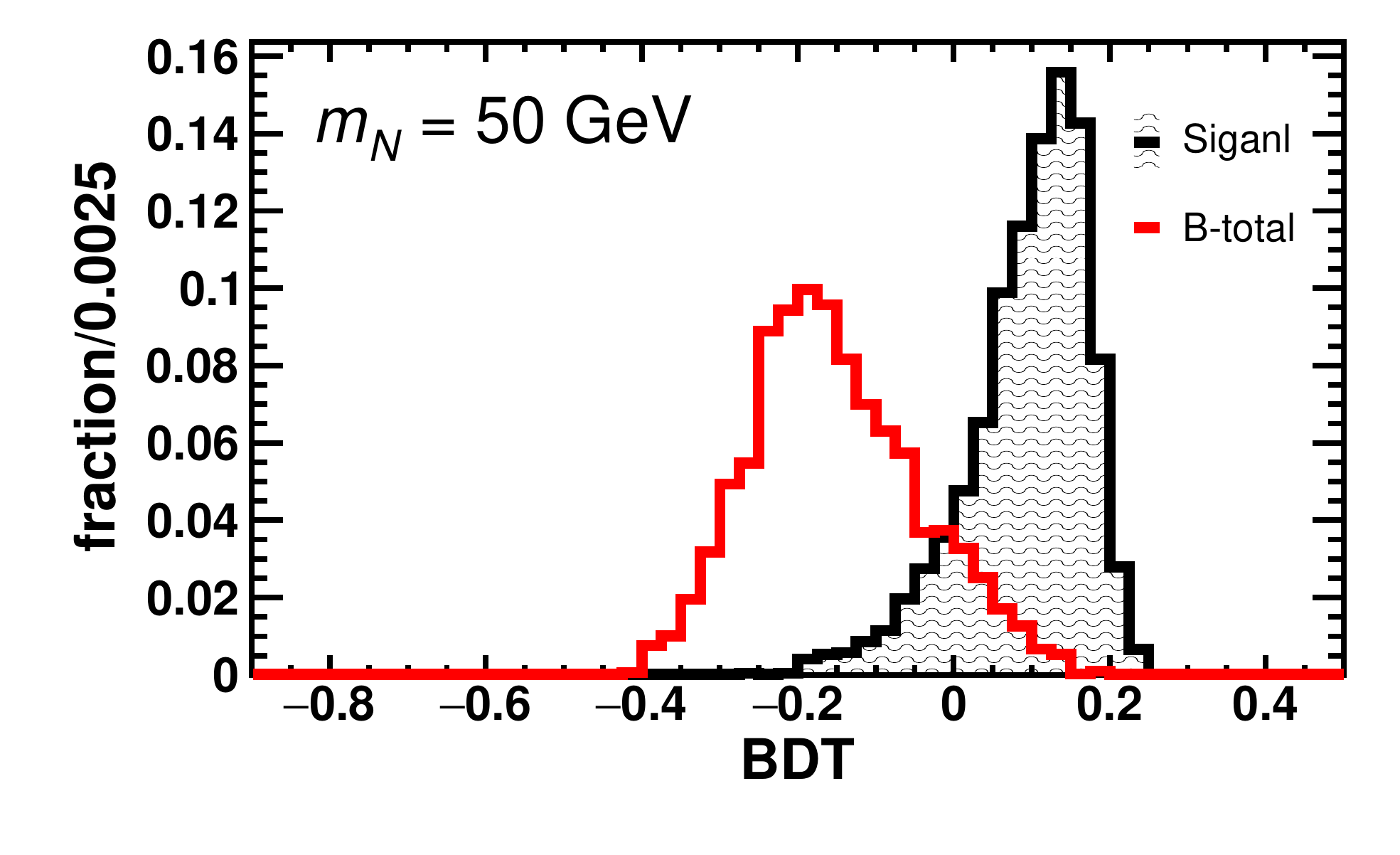}
\includegraphics[width=4.2cm,height=3cm]{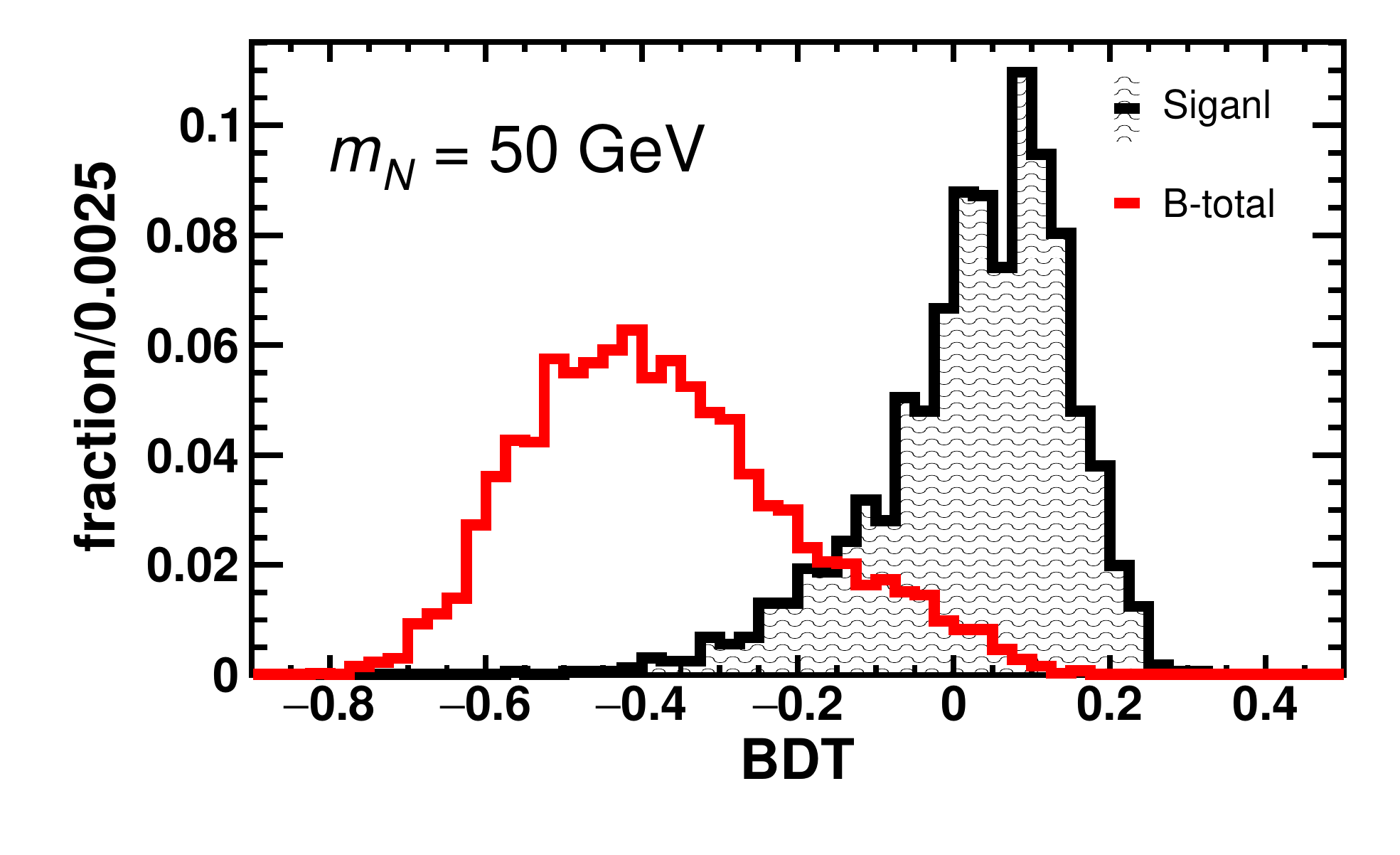}
\caption{
Distributions of BDT responses for the signal with benchmark $m_N$ = 50 GeV (black) and total background (red) at the HL-LHC (left) and SppC/FCC-hh (right).
}
\label{fig:BDT50GeV}
\end{figure}
	
The Boosted Decision Tree (BDT) algorithm in the TMVA package is adopted to perform the MVA and separate the background from the signal.
Fig.~\ref{fig:BDT50GeV} shows the distributions of BDT responses for the total background and the signal with benchmark $m_N$ = 50 GeV at the HL-LHC and SppC/FCC-hh.
Compare the left and right plots, the trends of distributions at the HL-LHC and SppC/FCC-hh are similar.
The BDT distributions of signal and background are well separated, which means that BDT cuts can be effectively applied to  reject background at both colliders.
	
Since the signal kinematics varies with $m_N$,  for each representative $m_N$ case, signal data are generated and analyzed individually to optimize the selection cuts and obtain the corresponding best limits.
In Appendix~\ref{appendix:BDTdistribution},
we show distributions of BDT responses for the signal and dominant SM background processes at the HL-LHC in the scenarios with different $m_N$ assumptions, 
while the BDT distributions at the SppC/FCC-hh are presented 
in the same appendix.

After the preselection, the BDT cut is optimized according to the signal statistical significance calculated by Eq.~(\ref{eqn:statSgf}) for each mass case.
\begin{equation}
\sigma_{\rm stat} = \sqrt{2 \left[ \left(N_s+N_b\right)\, {\rm ln}\left(1+\frac{N_s}{N_b}\right) - N_s \right] } \,\, ,
\label{eqn:statSgf}
\end{equation}
where $N_s$($N_b$) is the number of signal (total background) events after all selection (including the preselection and BDT selection) cuts.

Table~\ref{tab:CutEffiHLLHC}
shows the selection efficiencies of preselection and BDT cuts for both signal and background processes at the HL-LHC in the scenarios with different $m_N$ assumptions.
The selection efficiencies at the  SppC/FCC-hh are presented 
in Table~\ref{tab:CutEffi100}.
The total selection efficiency is the product of preselection and BDT cut efficiencies.
The number of signal and background events after all cuts can be calculated as the product of collider luminosity, their respective production cross section and total selection efficiency.
	
\section{Results}
\label{sec:result}
	
In this section, based on our analyses, we show the upper limits on the mixing parameter $|V_{\tau N }|^2$ for the heavy neutrino mass $m_N$ in the range of 10 to 1000 GeV.
	
\begin{figure}[h]
\centering
\includegraphics[width=8cm,height=6cm]{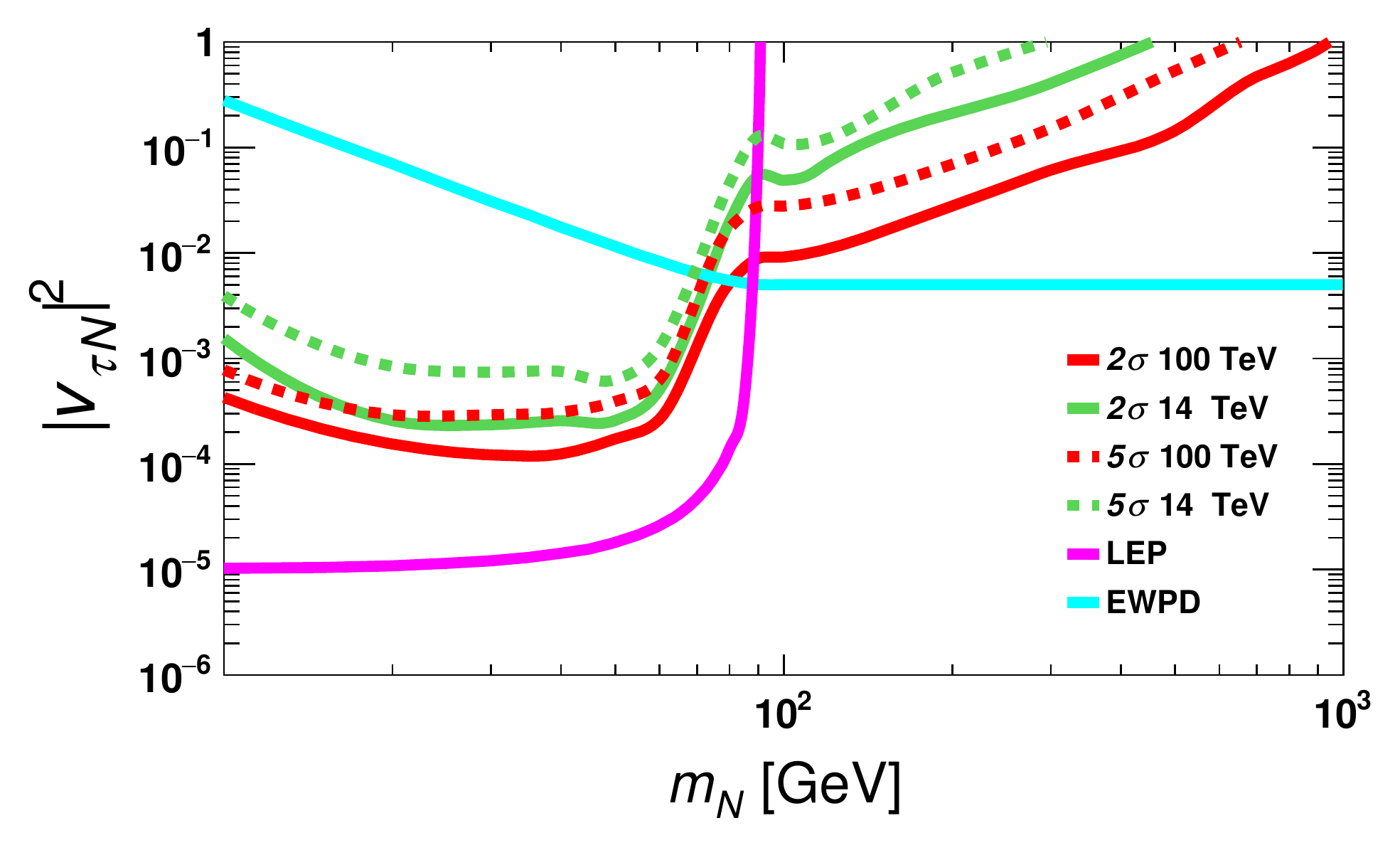}   
\caption{
Upper limits on mixing parameter $|V_{\tau N }|^2$ for the heavy neutrino mass in the range of 10 to 1000 GeV, based on our derivation from measurements of rare $Z$-boson decay at the LEP~\cite{DELPHI:1996qcc} and the electroweak precision data (EWPD)~\cite{delAguila:2008pw,Akhmedov:2013hec,Basso:2013jka,deBlas:2013gla,Antusch:2015mia,Cheung:2020buy}, and analyses of the $p p \to \tau^{\pm} \tau^{\pm} j j $ signal at the HL-LHC and SppC/FCC-hh.
}
\label{fig:sensitivity}
\end{figure}
	
In Fig.~\ref{fig:sensitivity}, we show upper limits at 95\% CL. on mixing parameter $|V_{\tau N }|^2$ for the heavy neutrino mass in the range of 10 to 1000 GeV, based on our derivation from measurements of rare $Z$-boson decay by the DELPHI collaboration at the LEP~\cite{DELPHI:1996qcc} and the electroweak precision data (EWPD)~\cite{delAguila:2008pw,Akhmedov:2013hec,Basso:2013jka,deBlas:2013gla,Antusch:2015mia,Cheung:2020buy}.
Limits at 2- and 5-$\sigma$ significances from analyses of the $p p \to \tau^{\pm} \tau^{\pm} j j $ signal at the HL-LHC and SppC/FCC-hh are also displayed in the same figure.
We emphasize again that for all limit curves, the heavy neutrino is assumed to mix with active neutrinos of tau flavour only, i.e. $ |V_{\tau N}|^2 \neq 0 $ and $|V_{e N}|^2 = |V_{\mu N}|^2 = 0 $.
	
At both colliders, the 5-$\sigma$ limits are slightly weaker than those for 2-$\sigma$.
At the HL-LHC, the 2-$\sigma$ upper limits on $|V_{\tau N }|^2$ decrease from $1.56 \times 10^{-3}$ to $2.56 \times 10^{-4}$ when $m_N$ changes from 10 GeV to 20 GeV; limits are relatively flat for $m_N$ in the middle range between 20 GeV and 50 GeV, and increase rapidly to $5.29 \times 10^{-2}$ at around 100 GeV; limits then increased steadily as $m_N$ changes afterwards. 

The limits at the SppC/FCC-hh is lower than those at the HL-LHC, but results are just slightly stronger, particularly for heavy neutrino mass below 100 GeV.
This is counterintuitive because usually one expects the limits can be improved greatly when the center-of-mass energy $\sqrt{s}$ is increased from 14 TeV to 100 TeV at $pp$ colliders.
We check and find that the reason is that at $pp$ colliders, huge number of mesons, for example $D$-, $\pi$-, $B$-, $\Omega$-meson, are produced.
These mesons can decay into muons and contribute to the background.
The number of mesons is much larger at $\sqrt{s} = 100$ TeV than that at 14 TeV, leading to much larger background at 100 TeV.
Our analyses show that, for example, when $m_N = 50$ GeV, the number of total background (signal) events after preselection is a factor of 521 (31) larger at the SppC/FCC-hh than that at the HL-LHC.
The increment of signal renders better limits at 100 TeV, but the improvement is restricted by the increment of background.
Therefore, in order to obtain stronger sensitivities at future experiments, we suggest that sophisticated methods need to be adopted to reject the meson background more efficiently.

\section{Summary and discussion}
\label{sec:sum}

Heavy neutrinos predicted by the seesaw mechanism is one important candidate beyond the standard model.
They can mix with active neutrinos of all flavours, in principle.
Compared with plentiful studies focusing on the mixing parameters $|V_{e N}|^2$ and $|V_{\mu N}|^2$, because of the challenges in detecting the final state taus, the mixing parameter $|V_{\tau N}|^2$ is more difficult to be probed, making it not well studied at current experiments.

In this article, we concentrate on the challenging scenario where $N$ mixes with active neutrino of tau flavour only,  i.e. $ |V_{\tau N}|^2 \neq 0 $ and $|V_{e N}|^2 = |V_{\mu N}|^2 = 0$.
We derive current constraints on the mixing parameter $|V_{\tau N}|^2$ for heavy neutrinos mass in the range between 10 and 1000 GeV, from the measurements of rare $Z$-boson decay at the LEP and the EWPD.

To forecast future limits on $|V_{\tau N}|^2$, we also search for a heavy Majorana neutrinos via the lepton number violating signal process of $p\, p \to \tau^\pm  \tau^\pm jj$ at future $pp$ colliders.
We consider the HL-LHC (SppC/FCC-hh) running with center-of-mass energy of 14 (100) TeV and an integrated luminosity of 3 (20) $\iab$.
To simplify the analyses, we consider the simplified Type-I model and assume that only one generation of heavy neutrinos $N$ is light and can be produced at colliders. 
Because the hadronically decaying tau final state suffers from too large background, in order to suppress the background, we select the final state where both taus decaying leptonically into muons.
We develop the search strategy, and simulate the collision events for the signal and six SM background processes, which include the single boson process $W^{\pm} j j$ and $Z j j$, the di-boson process $W^{\pm} Z j j$, $W^+ W^- j j$, $Z Z j j$, and di-top process $t \bar{t}$.
Multivariate analyses are performed individually to optimize the selection cuts and obtain the corresponding best limits for each representative $m_N$ case.

In Fig.~\ref{fig:sensitivity}, we show upper limits on $|V_{\tau N }|^2$ for $m_N$ in the range between 10 and 1000 GeV, based on our derivation from measurements of rare $Z$-boson decay at the LEP and EWPD, and analyses of the $p p \to \tau^{\pm} \tau^{\pm} j j $ signal at the HL-LHC and SppC/FCC-hh.
We find that much more mesons are produced when $\sqrt{s}$ is increased from 14 TeV to 100 TeV at $pp$ colliders.
They can contribute to the background, which lead to just slightly stronger limits at the SppC/FCC-hh than those at the HL-LHC.

We assume that $N$ decays promptly in this study. However, when the lifetime of heavy neutrinos are long enough, they can have probability to travel through the detector and will not contribute to our signal.  Therefore, in such case, the number of signal events needs to be multiplied by the average probability $\bar{P}$ of heavy neutrinos decaying inside the detector's fiducial volume.
The long-lived effects have been studied by both the ATLAS and CMS collaborations at the LHC (see e.g. Refs.~\cite{CMS:2018iaf, ATLAS:2019kpx, CMS:2021lzm}).
The results show that when $m_N \lesssim 20$ GeV, the long decay path begin to cause large signal efficiency losses, and searches from the displaced vertex signature become more sensitive to $|V_{\ell N}|^2$ than the prompt signature when $m_N \lesssim 15$ GeV.
Therefore, careful studies focusing on the displaced vertex signature are needed for lower mass region, especially when $m_N \lesssim 20$ GeV, which can improve the limits from this study in this mass region.

In this study, we focus on the scenario where $|V_{\tau N}|^2 \neq 0 $ and $|V_{e N}|^2 = |V_{\mu N}|^2 = 0$, and select the final state with both taus decaying leptonically into muons.
The scenario with $|V_{\mu N}|^2 \neq 0$ could also lead to final state muons.
However, for the tau decay, the energy and momenta of tau are split among the final state muon and neutrinos, leading to softer muons and moderate missing energy.
Therefore, in principle, kinematics are different for various mixing parameter assumptions.
We investigate the effects of varying assumptions of mixing parameters, and show the results in Appendix~\ref{appendix:Effmixings}.

We note that although our results show that the limits at $pp$ colliders are weaker than those from the measurements at LEP and EWPD, the LNV signal process considered in this study depends on the mixing parameter $|V_{\tau N}|^2$ only and exists once $|V_{\tau N}|^2 \neq 0$, so it provides an opportunity to search for heavy Majorana neutrinos and probe $|V_{\tau N}|^2$ independent of other mixing parameters.
Therefore, although this signal are challenging to be detected at colliders because of the final state are taus, this signal is still meaningful and deserves careful study.

Our analyses indicate that limits at $pp$ colliders are restricted by the meson background.
In the literature (e.g. Ref~\cite{CMS:2018jxx}), it is usually categorized into the ``misidentified-lepton'' background, and is rejected mainly
by exploiting the fact that these misidentified leptons are generally less isolated and tend to have larger impact parameters.
In this study, the isolation and other kinematical information related to muons has been adopted in the multivariate analysis.
However, since the impact parameter observable of the muon cannot be easily exported by our simulation programs, it has not been exploited in this study. 
Therefore, the impact parameter observable can be used to reduce the background further and enhance the limits for this study.
Moreover, we suggest that, in order to obtain stronger sensitivities at future experiments, more sophisticated methods need to be developed to reject the meson background more efficiently.

%\newpage
\appendix

\section{Distributions of kinematic observables}
\label{appendix:obs}

Here, $\met$ is the magnitude of the missing transverse momentum.
$p_T$ is the transverse momentum of the final state object.
$m(\mu+j_1+j_2)$ is the invariant mass of the system consisting of the first two leading jets and the closest muon $\mu$ which has the smallest solid angle distance $\Delta R$ to the di-jet $(j_1+j_2)$.
$p_{T,\, {\rm max}}^{\rm iso} (\mu)$ is explained in Sec.~\ref{subsec:analysis}.

\begin{figure}[H] 
	\centering
	\addtocounter{figure}{-1}
	\subfigure{
		\includegraphics[width=4.5cm,height=3cm]{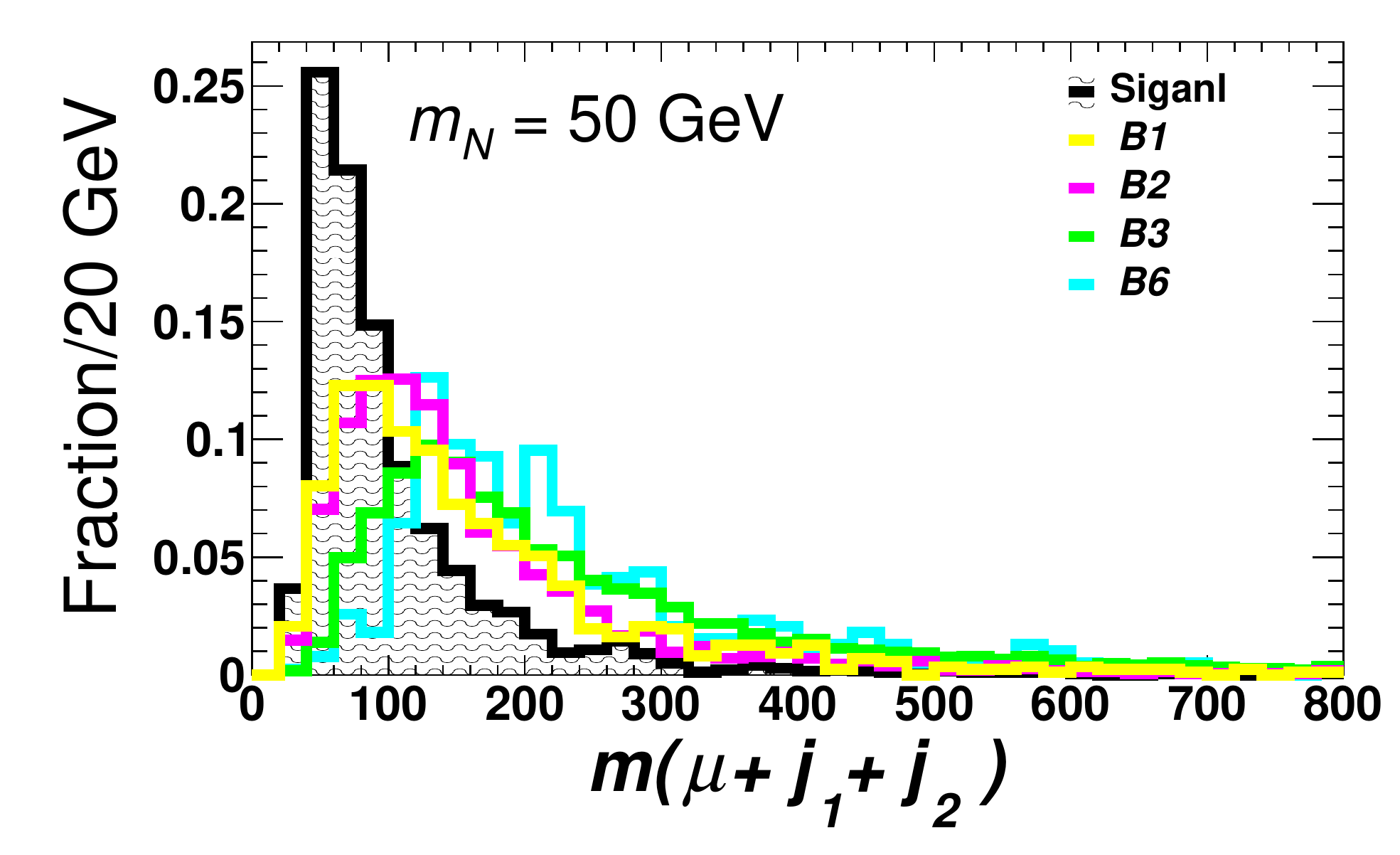}
		\includegraphics[width=4.5cm,height=3cm]{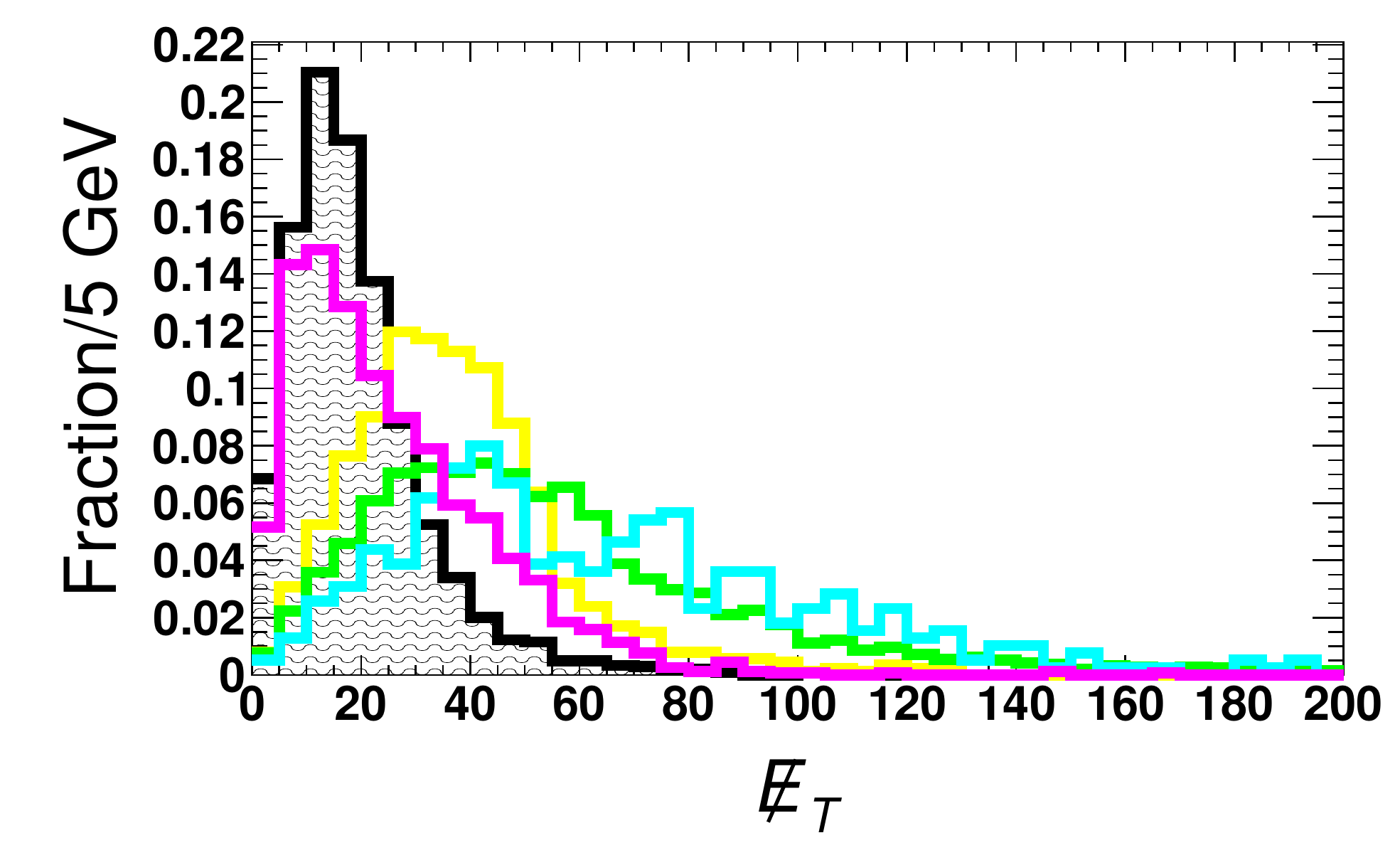}
	}
\end{figure}
\vspace{-1.0cm}
\begin{figure}[H] 
	\centering
	\addtocounter{figure}{1}
	\subfigure{
		\includegraphics[width=4.5cm,height=3cm]{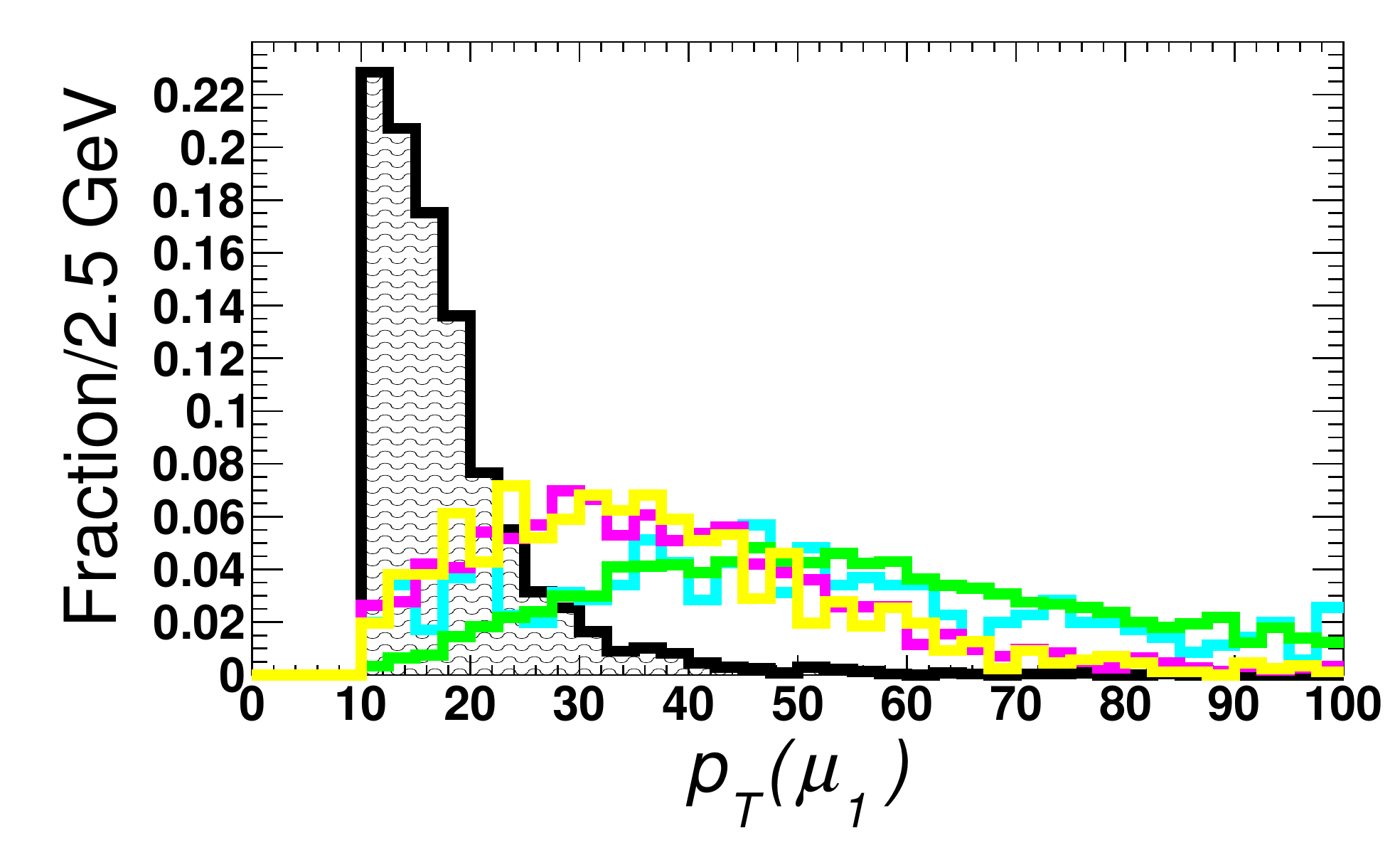}
		\includegraphics[width=4.5cm,height=3cm]{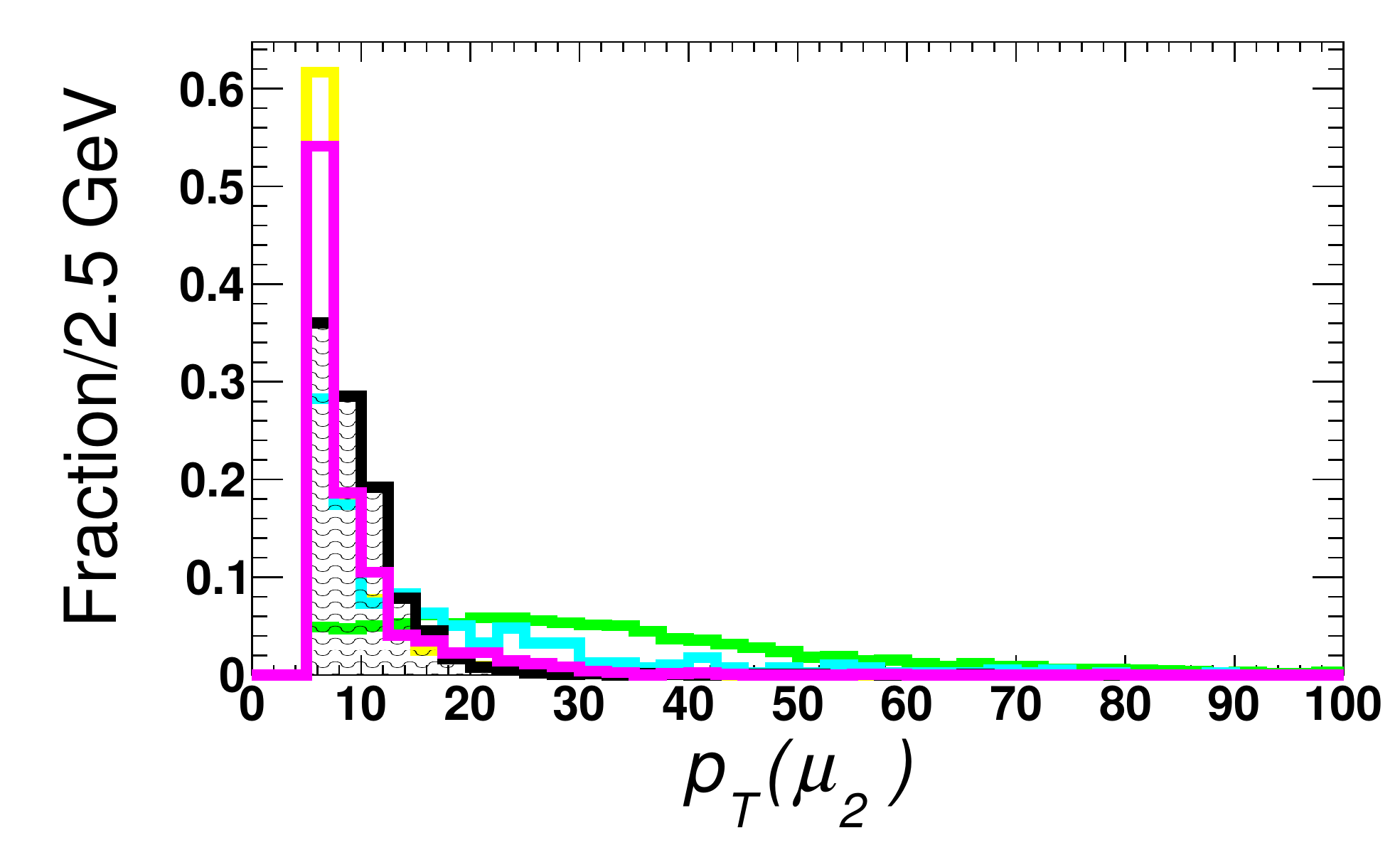}
	}
\end{figure}
\vspace{-1.0cm}
\begin{figure}[H] 
	\centering
%	\addtocounter{figure}{}
	\subfigure{
		\includegraphics[width=4.5cm,height=3cm]{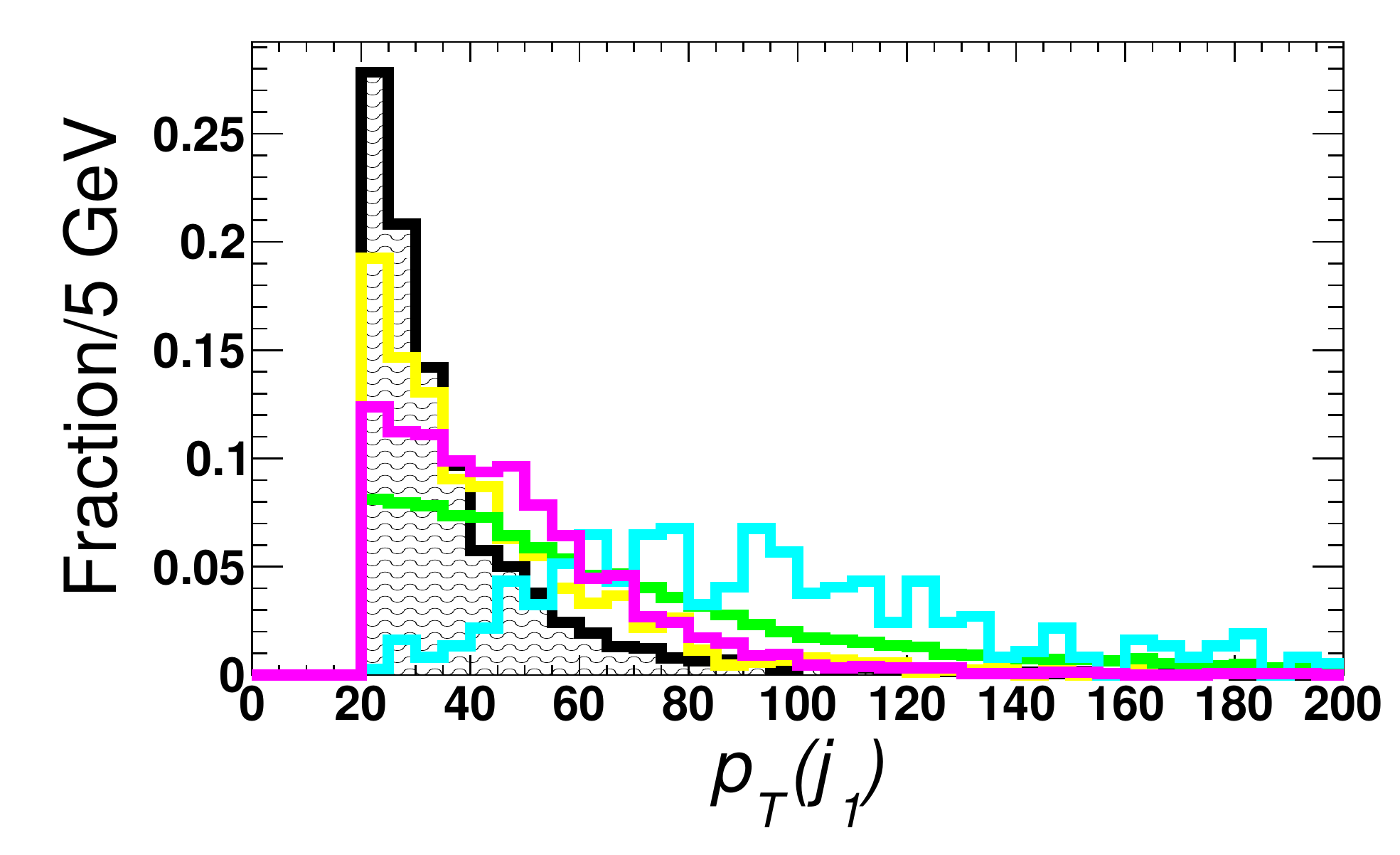}
		\includegraphics[width=4.5cm,height=3cm]{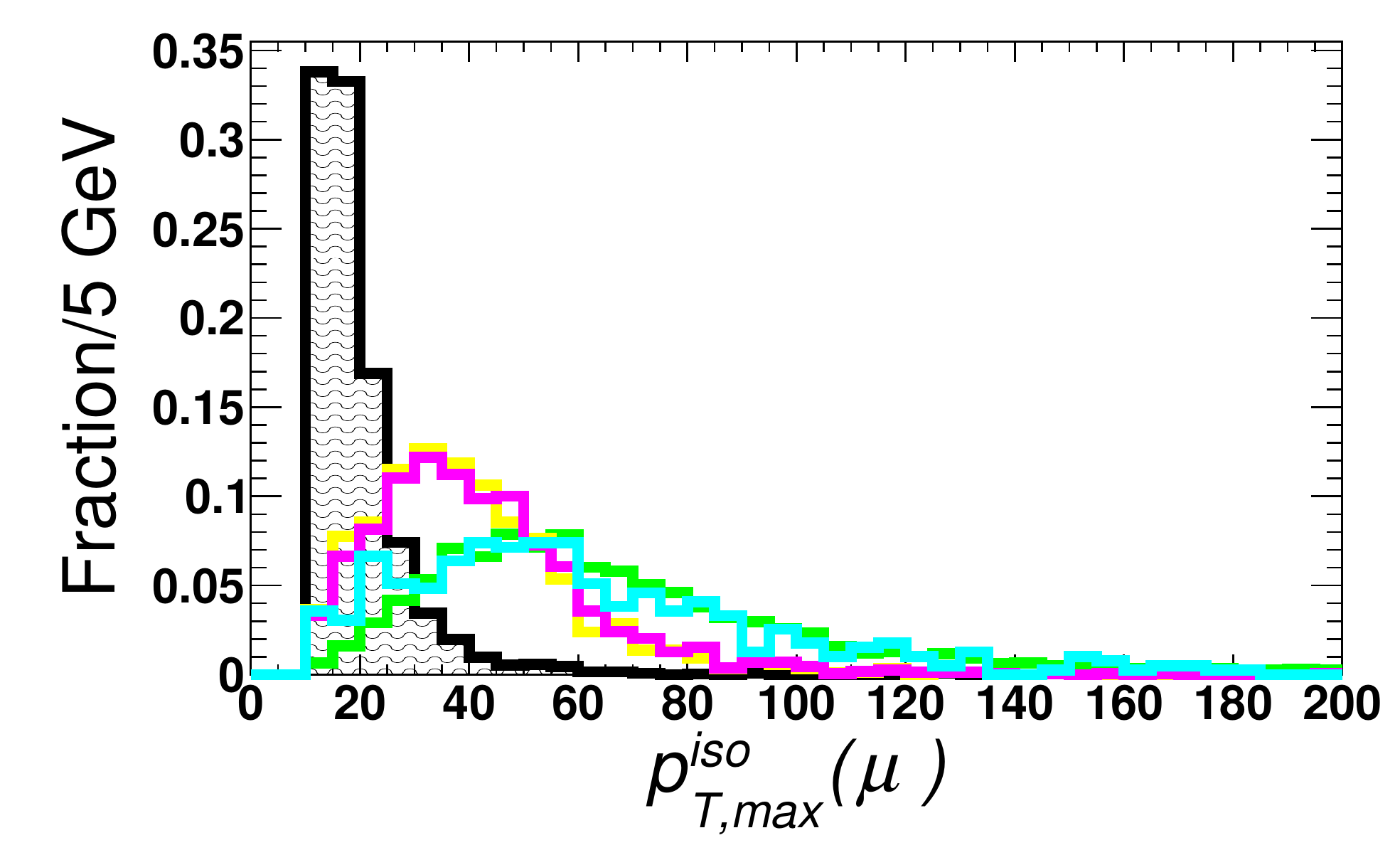}
	}
\caption{Distributions of kinematical observables at the HL-LHC for signal with $m_N$ = 50 GeV and dominant background processes.}
\label{fig:obs14TeV}
\end{figure}

\begin{figure}[H] 
	\centering
	\addtocounter{figure}{-1}
	\subfigure{
	    \includegraphics[width=4.5cm,height=3cm]{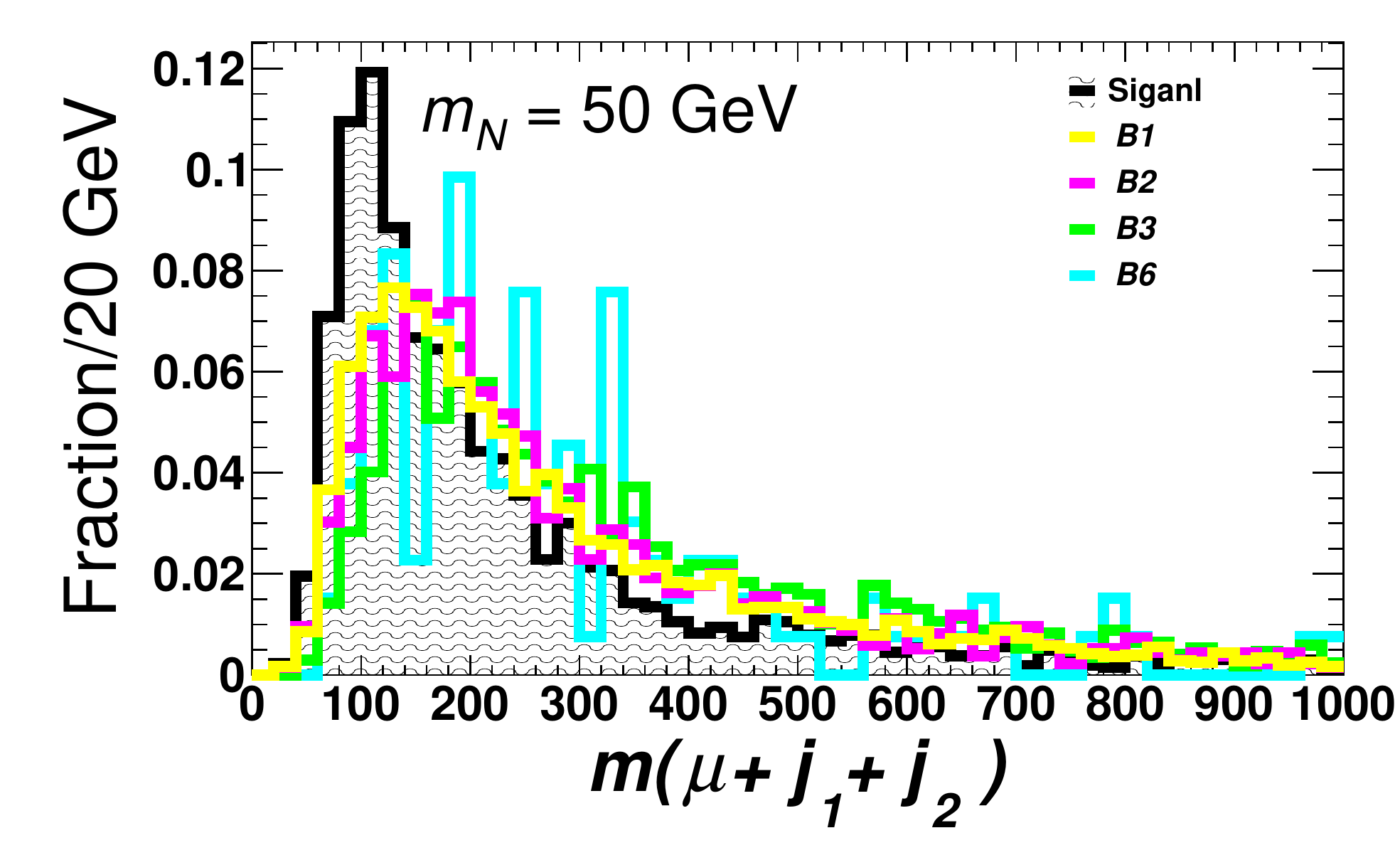}
		\includegraphics[width=4.5cm,height=3cm]{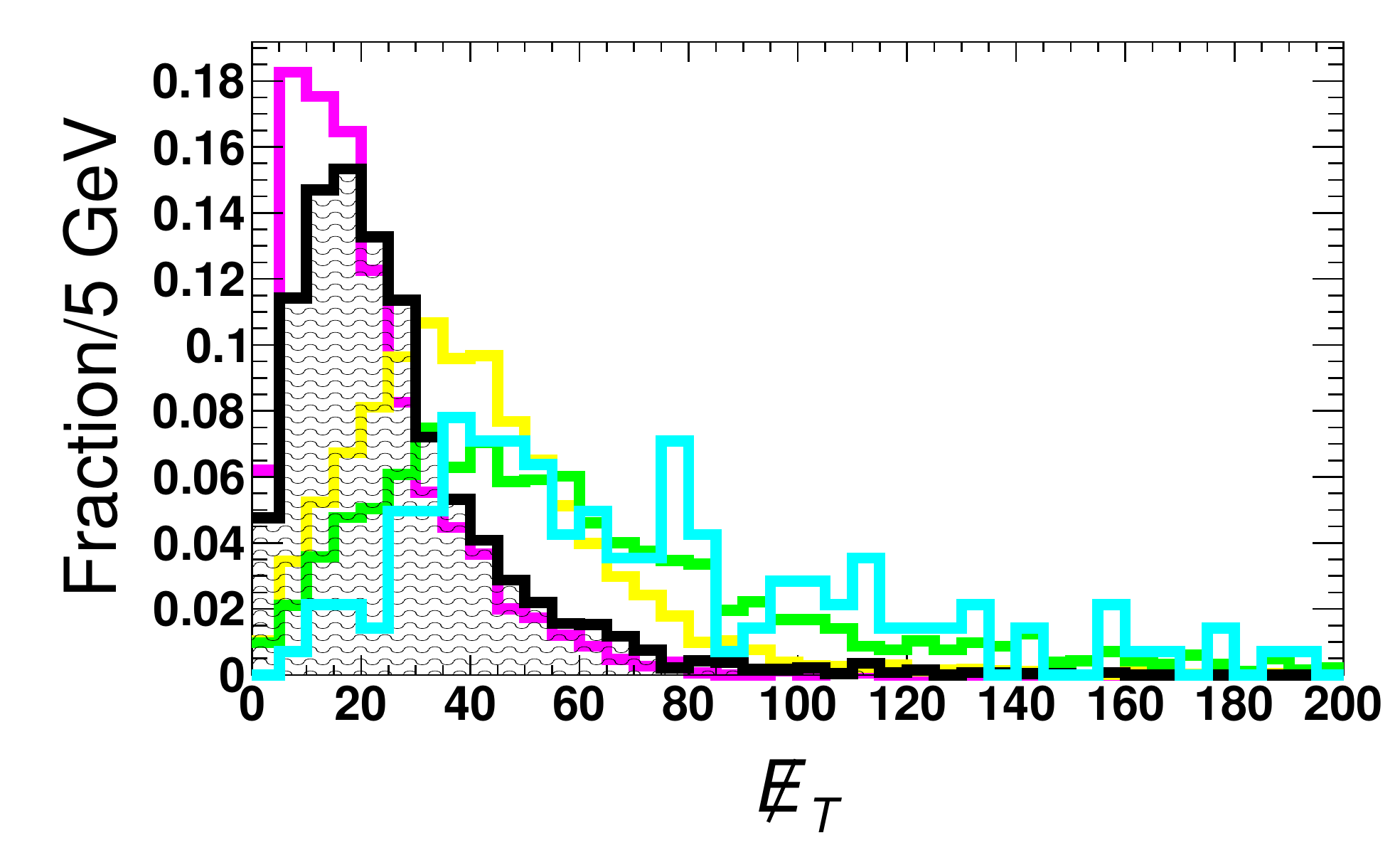}
	}
\end{figure}
\vspace{-1.0cm}
\begin{figure}[H] 
	\centering
	\addtocounter{figure}{1}
	\subfigure{
		\includegraphics[width=4.5cm,height=3cm]{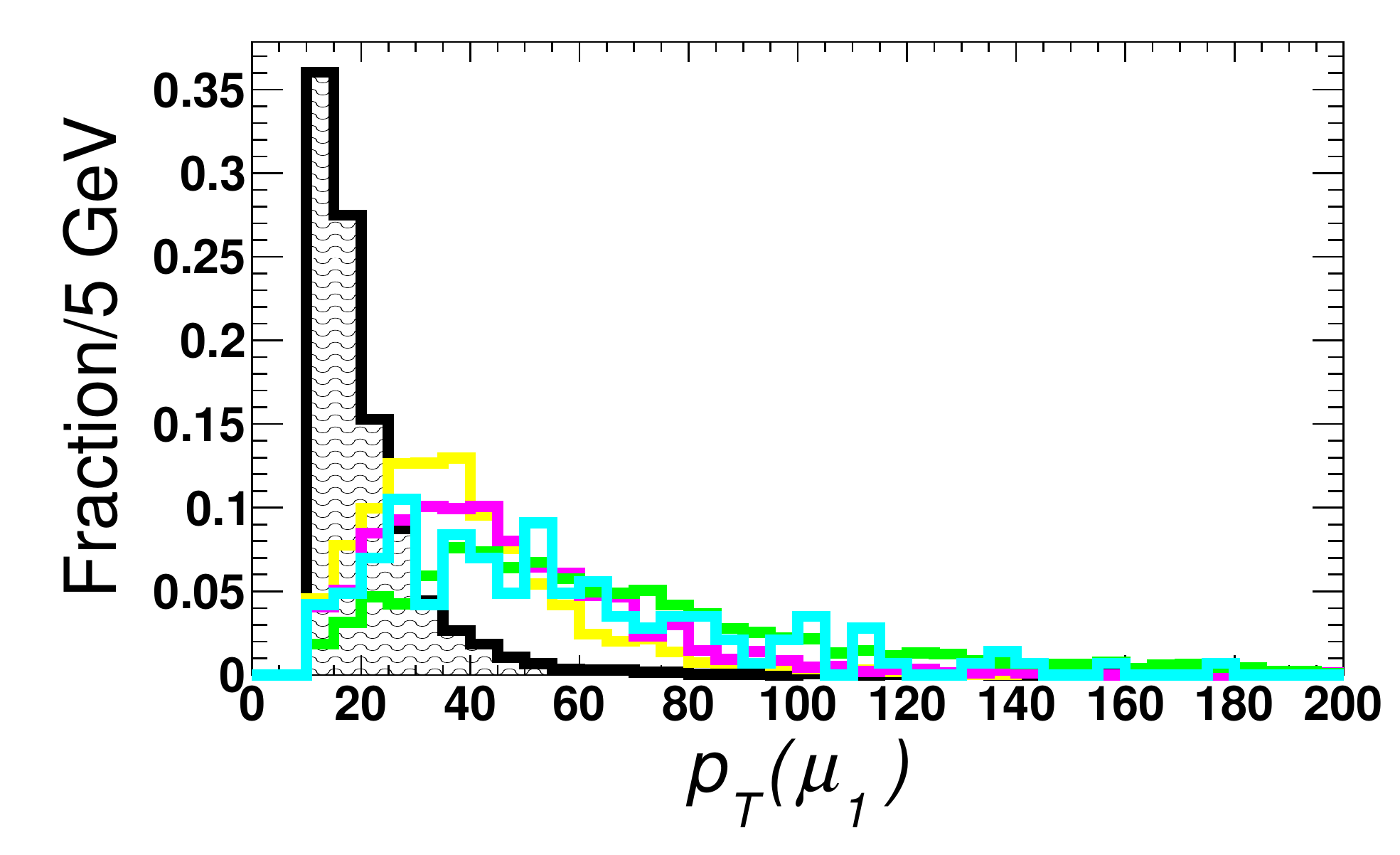}
		\includegraphics[width=4.5cm,height=3cm]{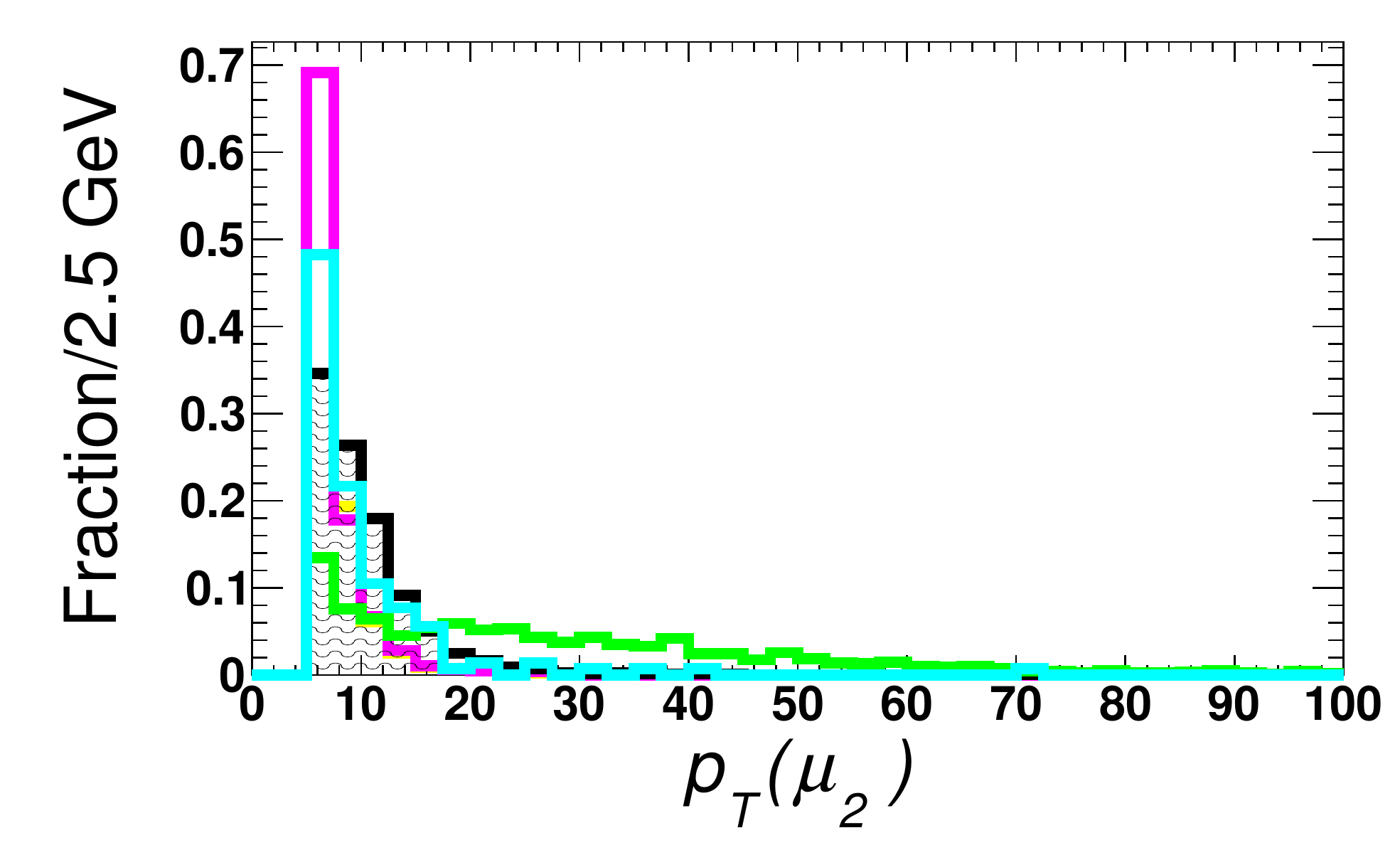}
	}
\end{figure}
\vspace{-1.0cm}
\begin{figure}[H] 
	\centering
%	\addtocounter{figure}{}
	\subfigure{
		\includegraphics[width=4.5cm,height=3cm]{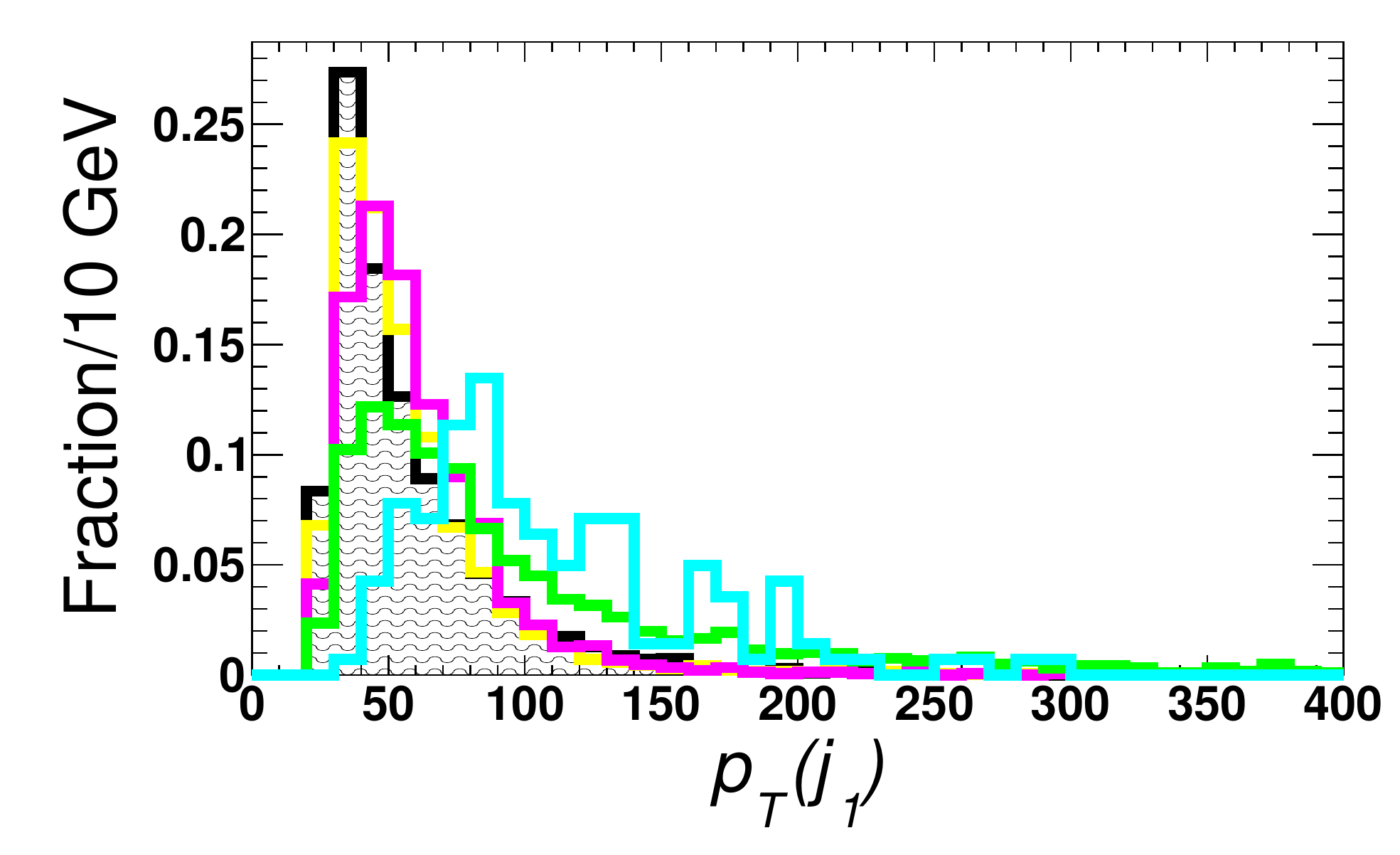}
		\includegraphics[width=4.5cm,height=3cm]{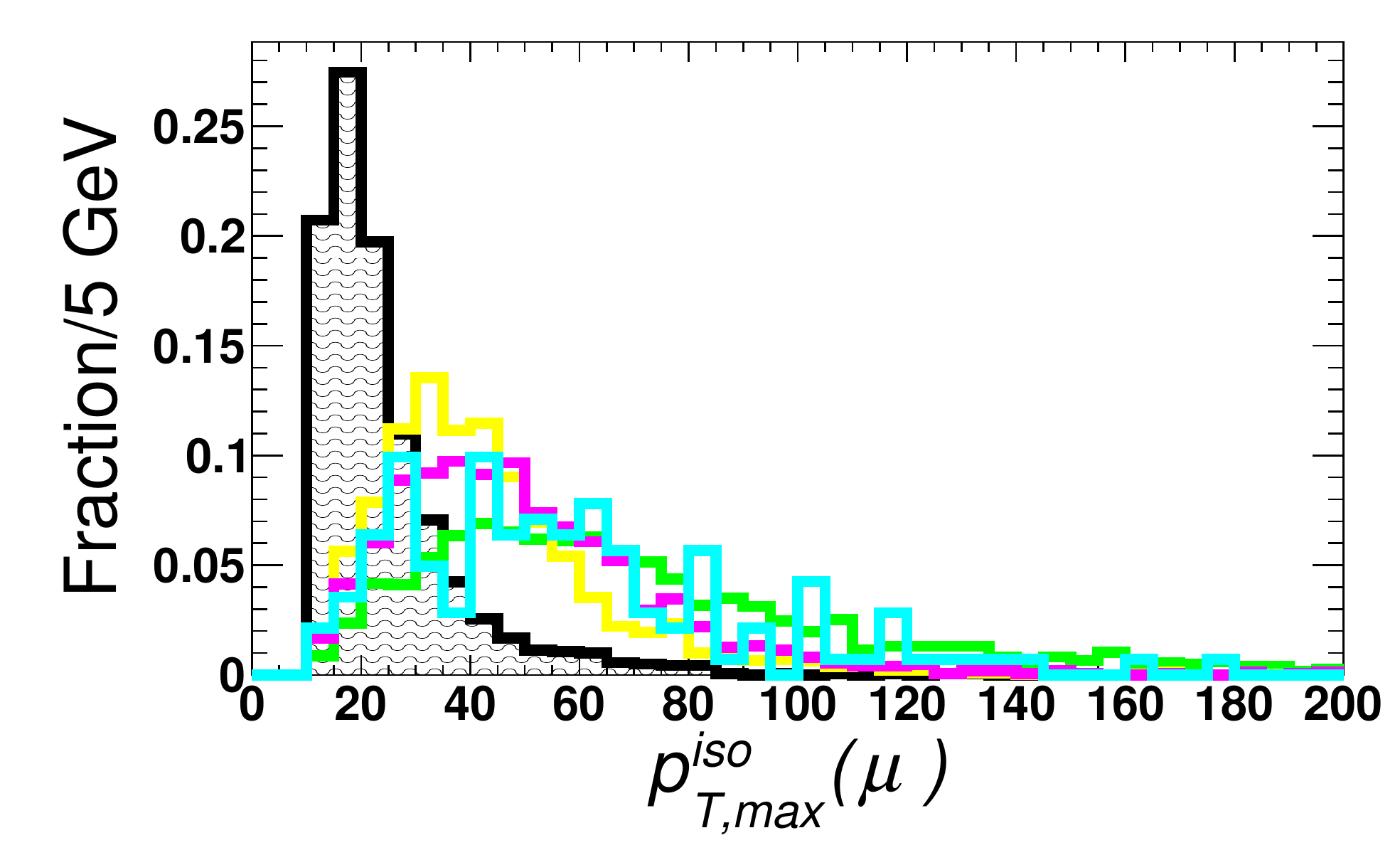}
	}
\caption{Distributions of kinematical observables at the SppC/FCC-hh for signal with $m_N$ = 50 GeV and dominant background processes.}
\label{fig:obs100TeV}
\end{figure}

\section{Distributions of BDT responses}
\label{appendix:BDTdistribution}

\begin{figure}[H]
	\centering
	\subfigure{
		\includegraphics[width=4.5cm,height=3cm]{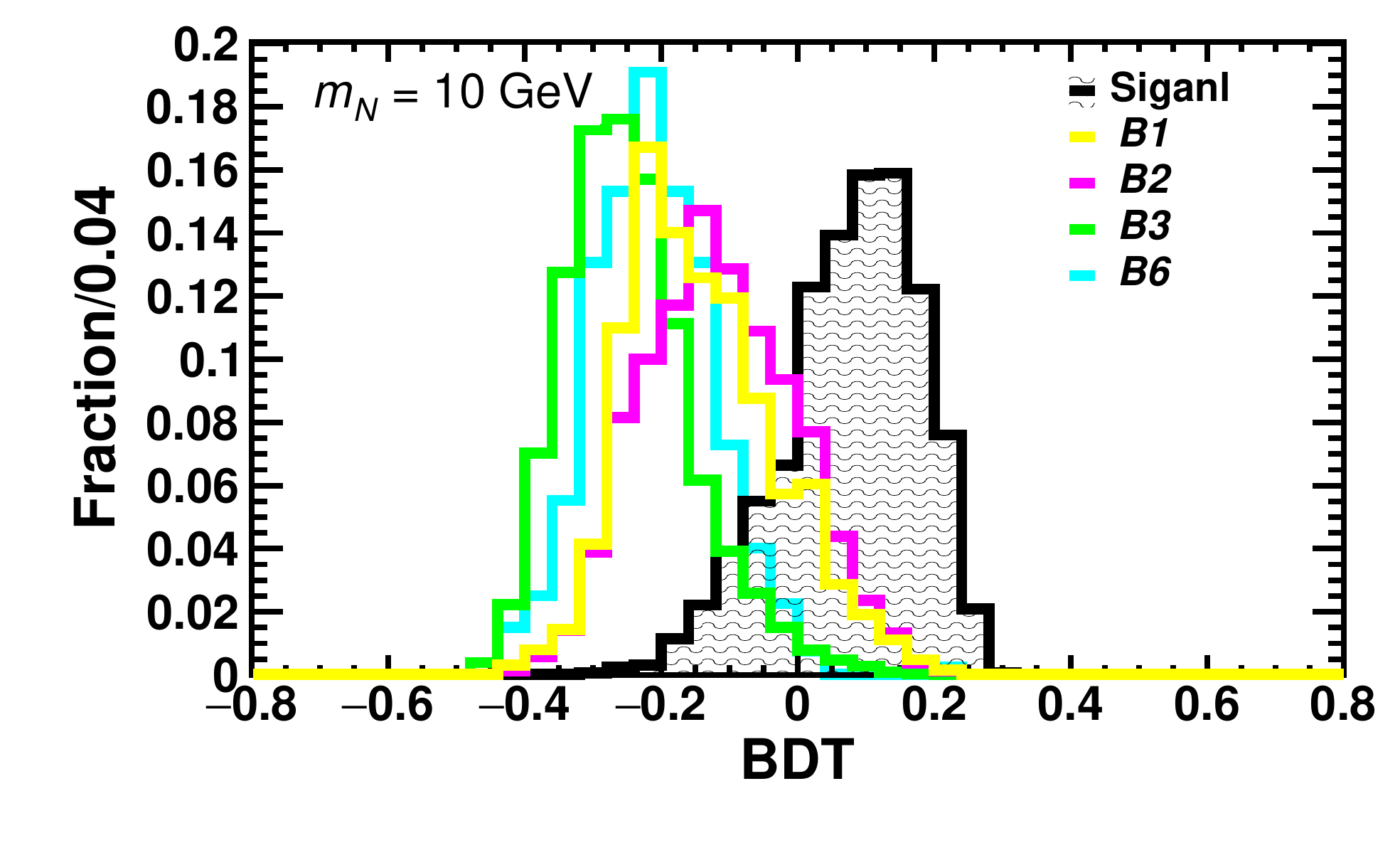}
		\includegraphics[width=4.5cm,height=3cm]{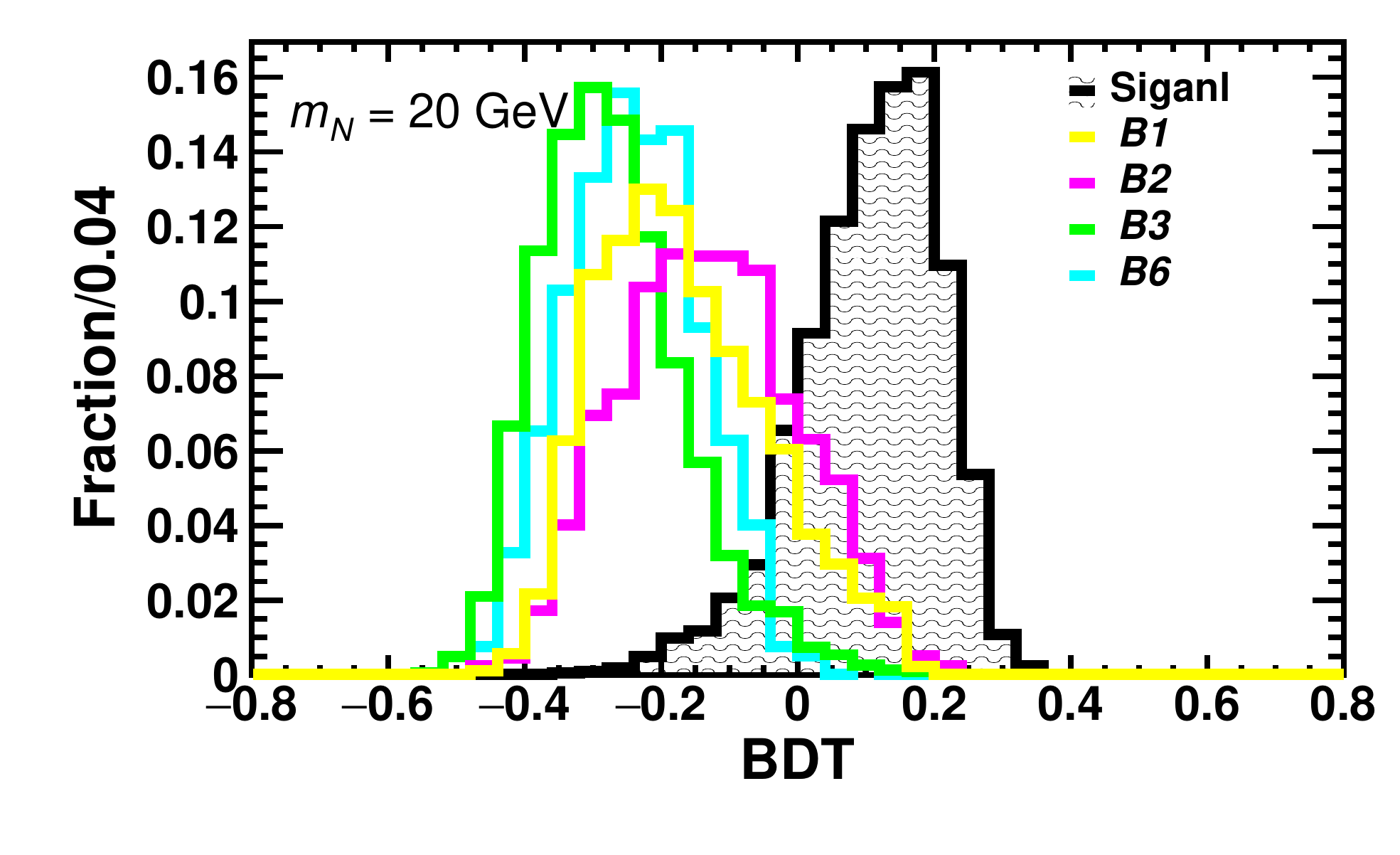}
	}
\end{figure}
\vspace{-1.0cm}
\begin{figure}[H] 
	\centering
	\addtocounter{figure}{1}
	\subfigure{
		\includegraphics[width=4.5cm,height=3cm]{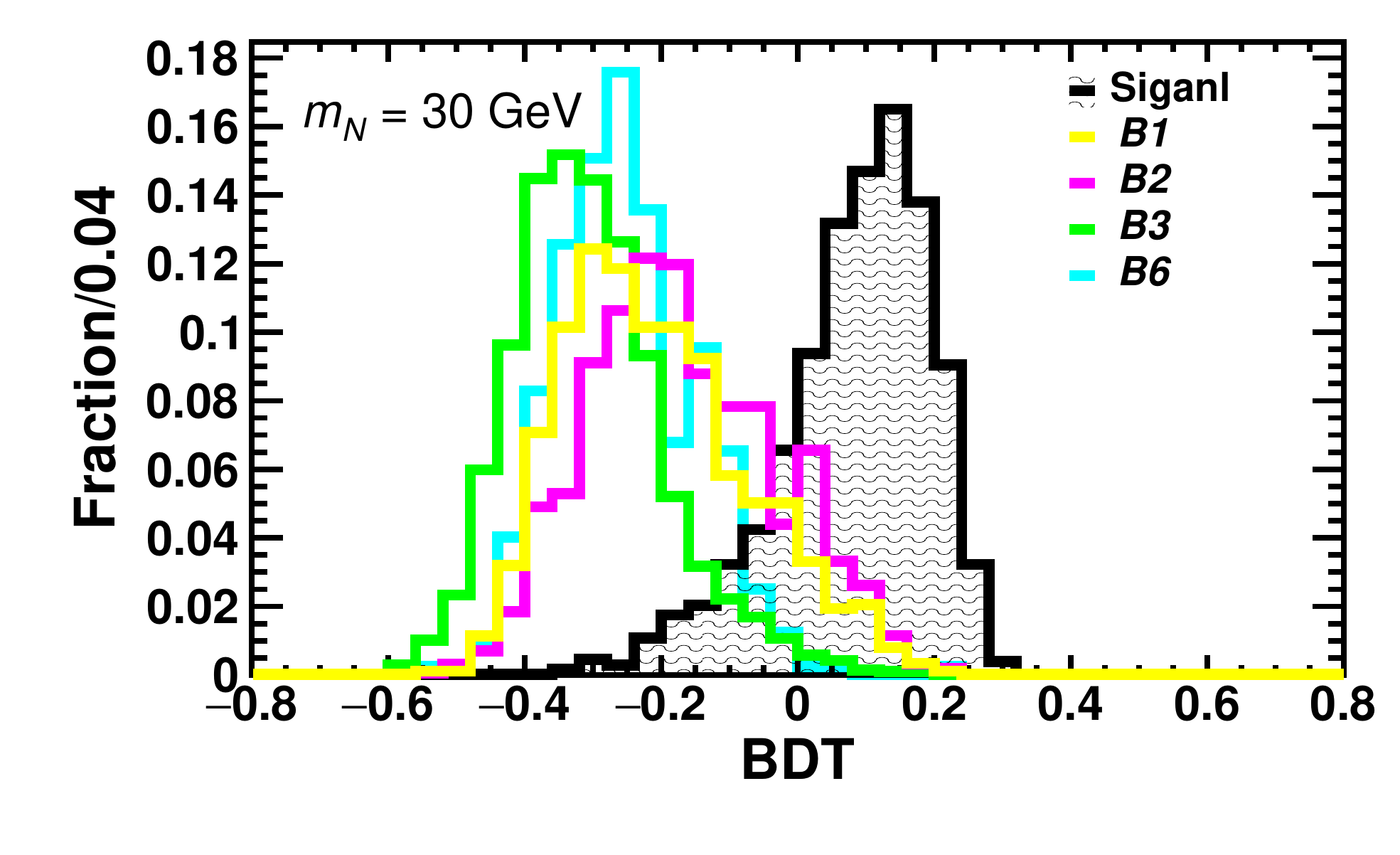}
		\includegraphics[width=4.5cm,height=3cm]{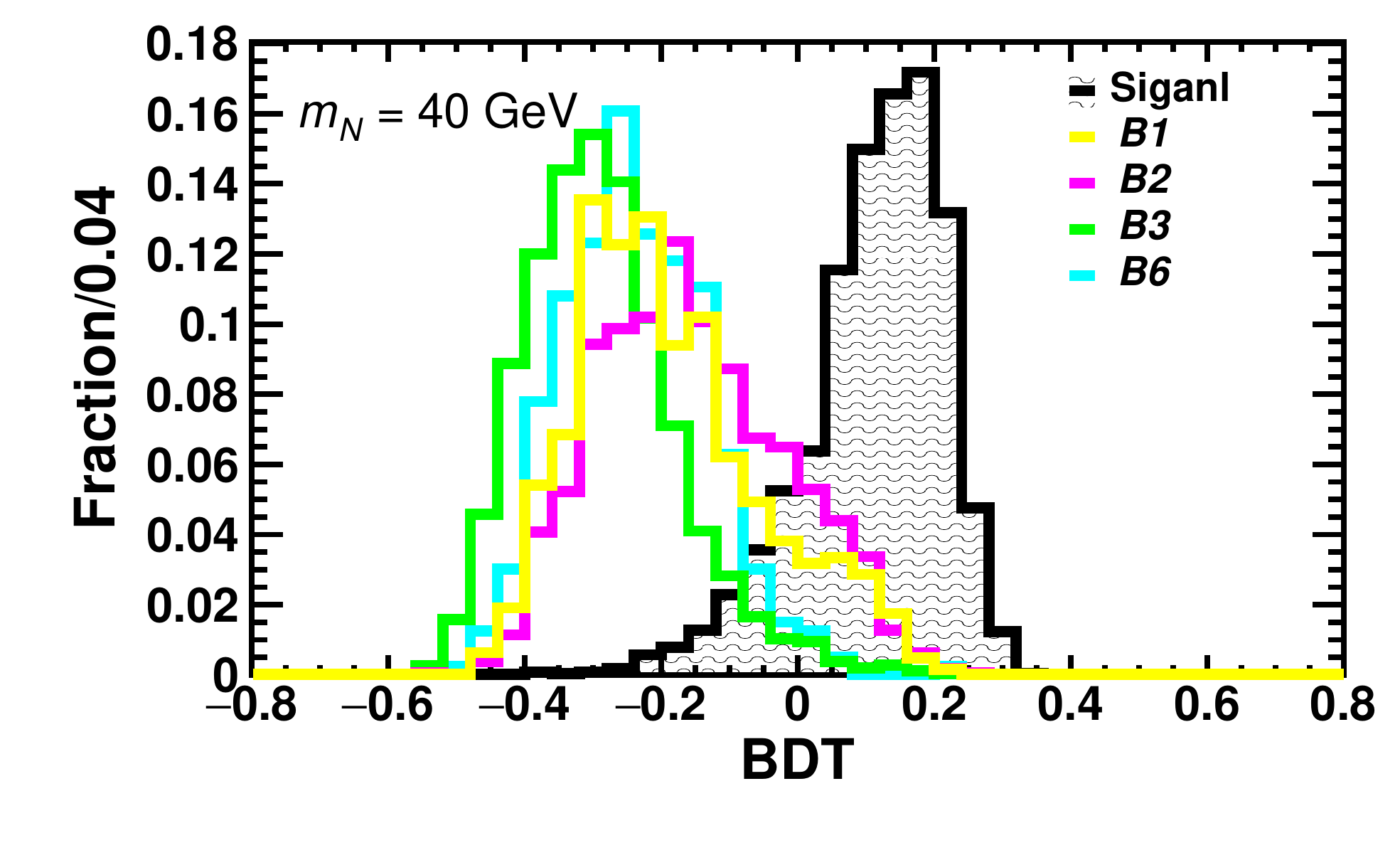}
	}
\end{figure}
\vspace{-1.0cm}
\begin{figure}[H] 
	\centering
	\addtocounter{figure}{-1}
	\subfigure{
		\includegraphics[width=4.5cm,height=3cm]{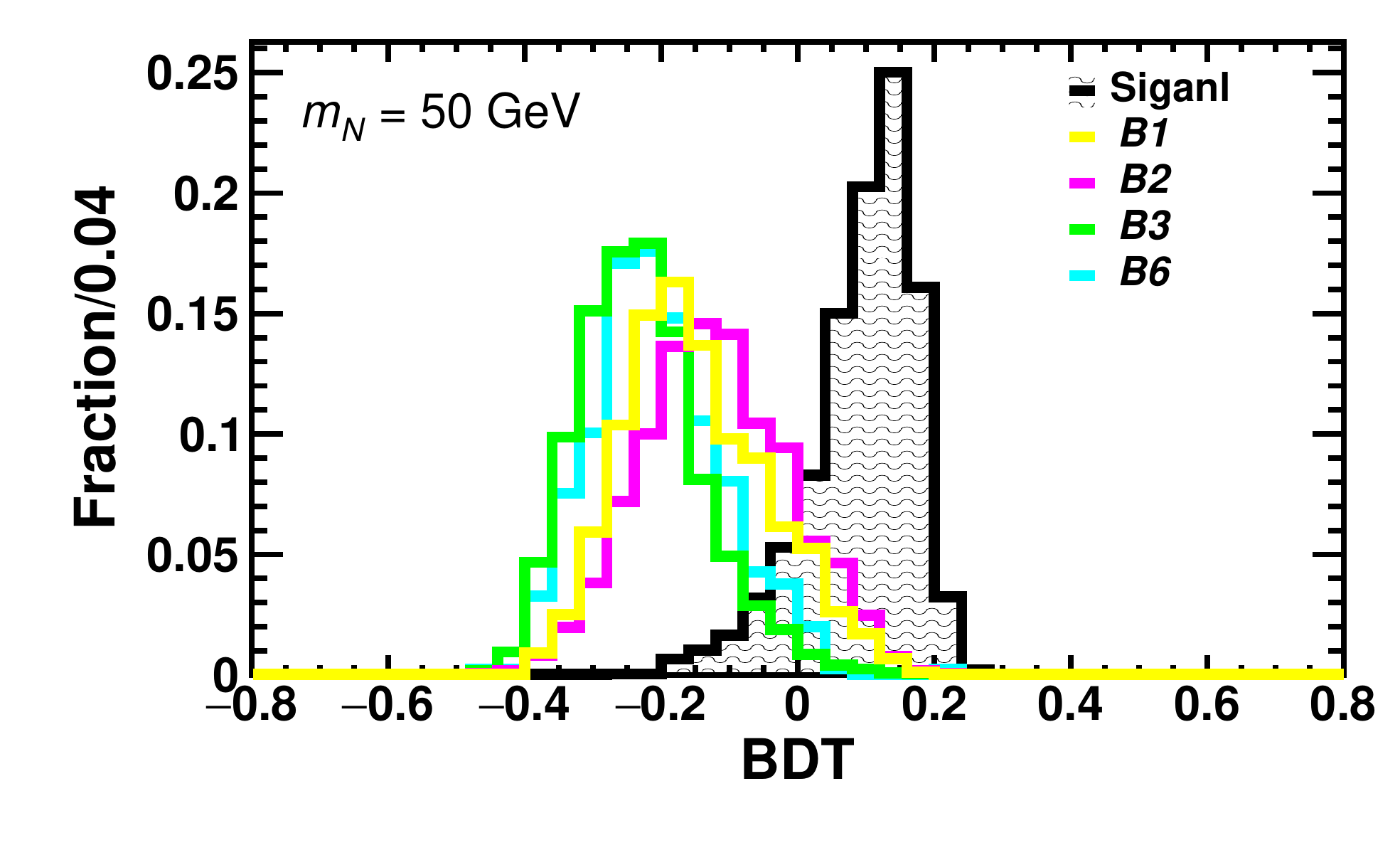}
		\includegraphics[width=4.5cm,height=3cm]{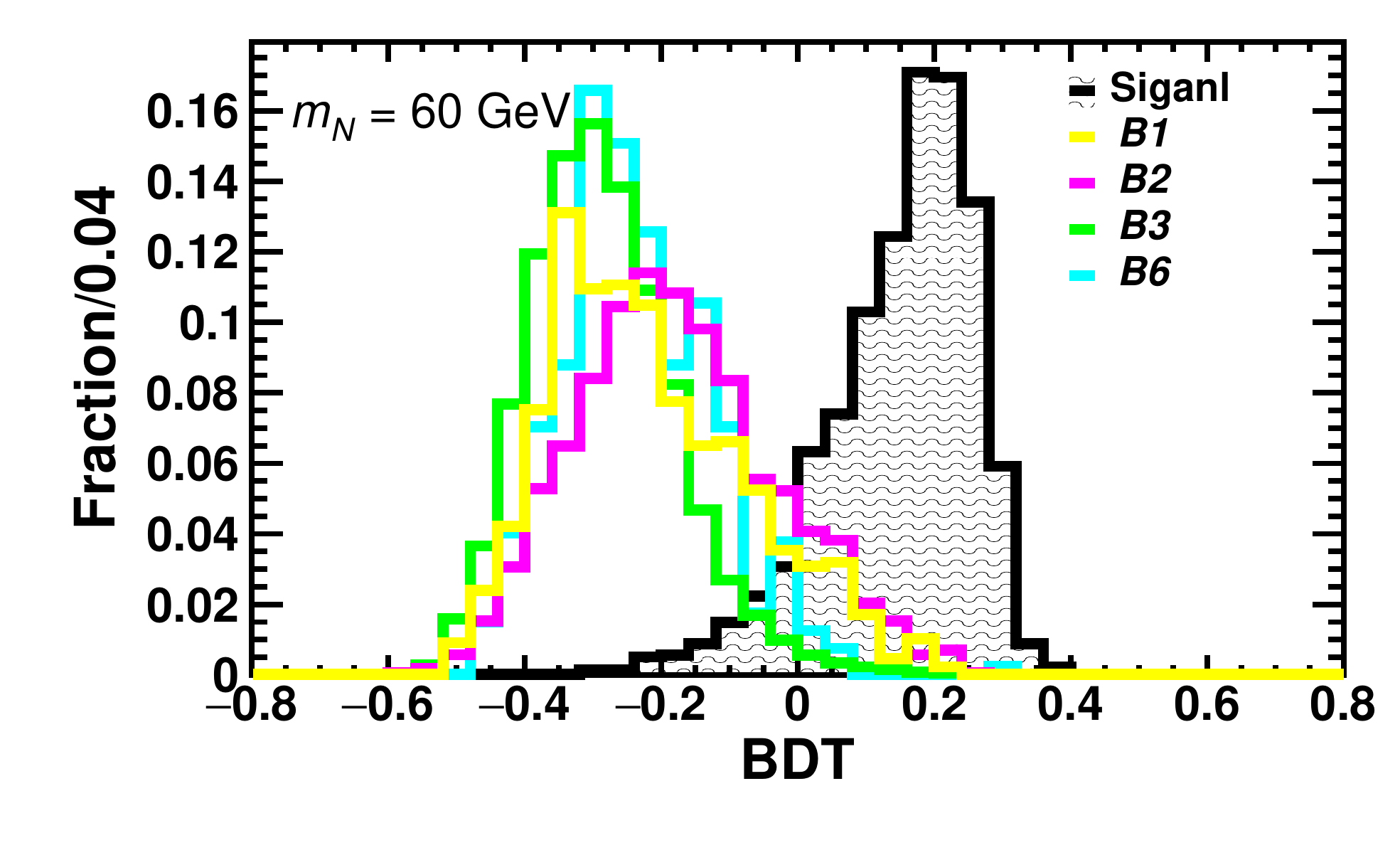}
	}
\end{figure}
\vspace{-1.0cm}
\begin{figure}[H] 
	\centering
	\addtocounter{figure}{1}
	\subfigure{
		\includegraphics[width=4.5cm,height=3cm]{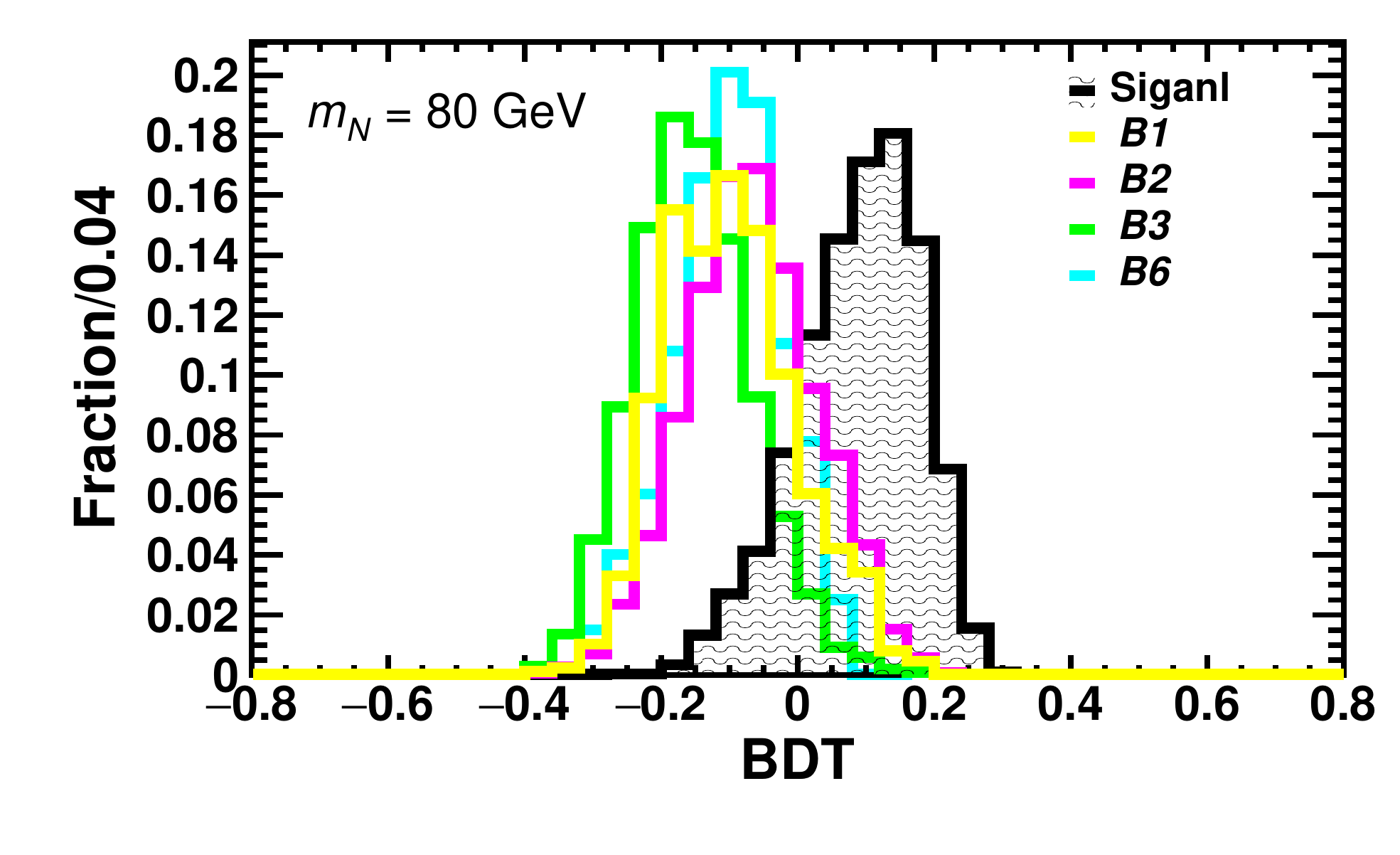}
		\includegraphics[width=4.5cm,height=3cm]{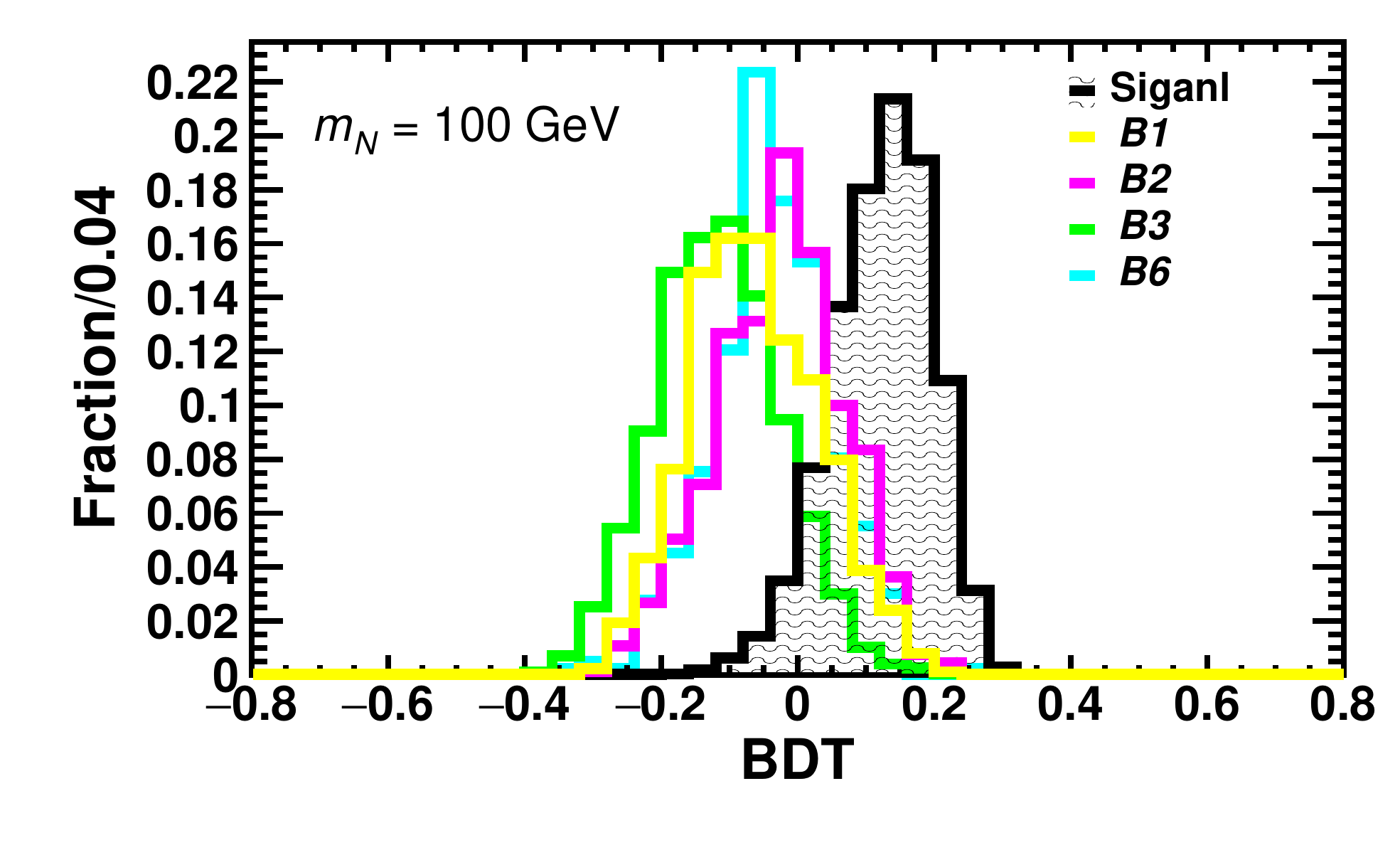}
	}
\end{figure}
\vspace{-1.0cm}
\begin{figure}[H] 
	\centering
	\addtocounter{figure}{-1}
	\subfigure{
		\includegraphics[width=4.5cm,height=3cm]{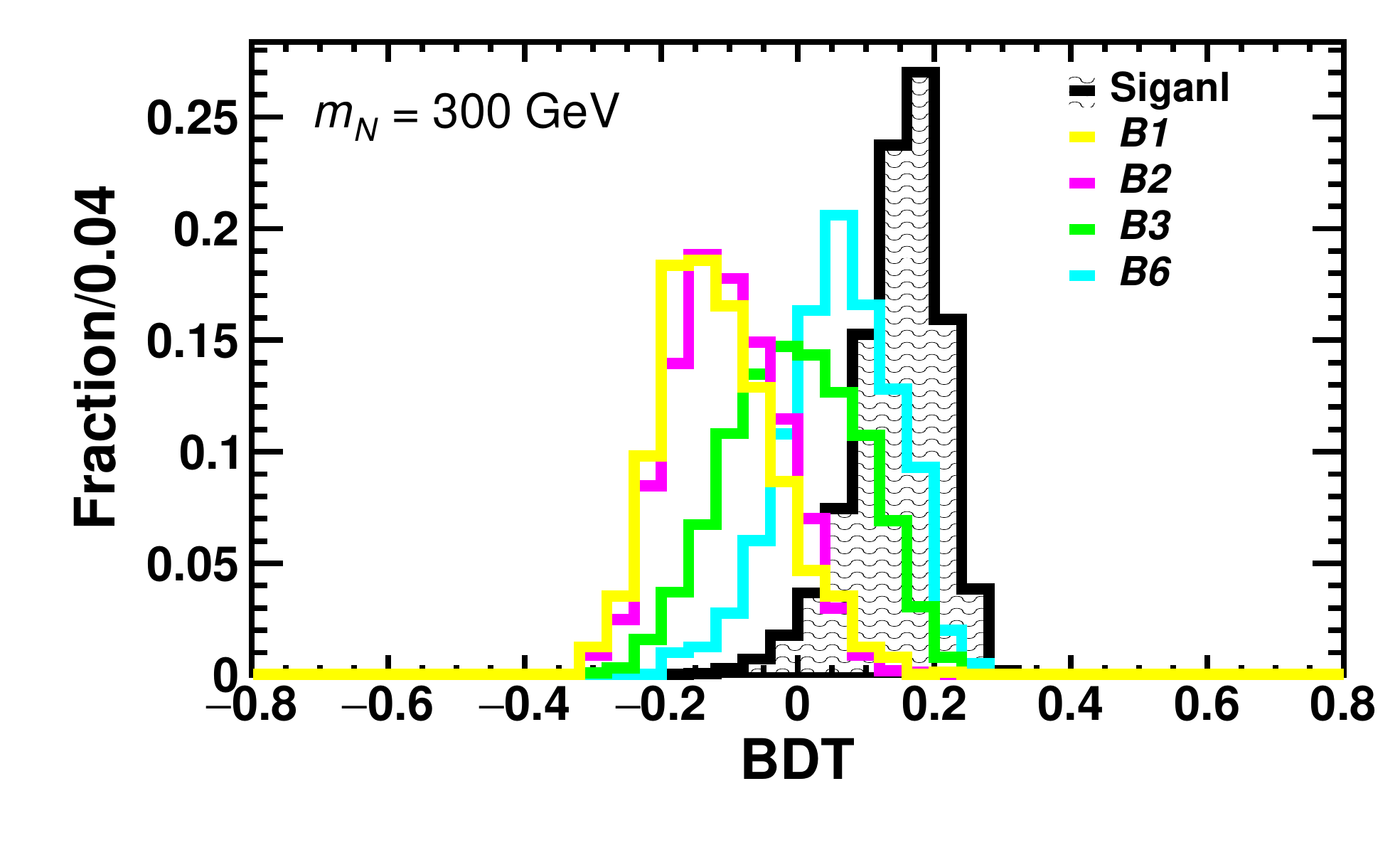}
		\includegraphics[width=4.5cm,height=3cm]{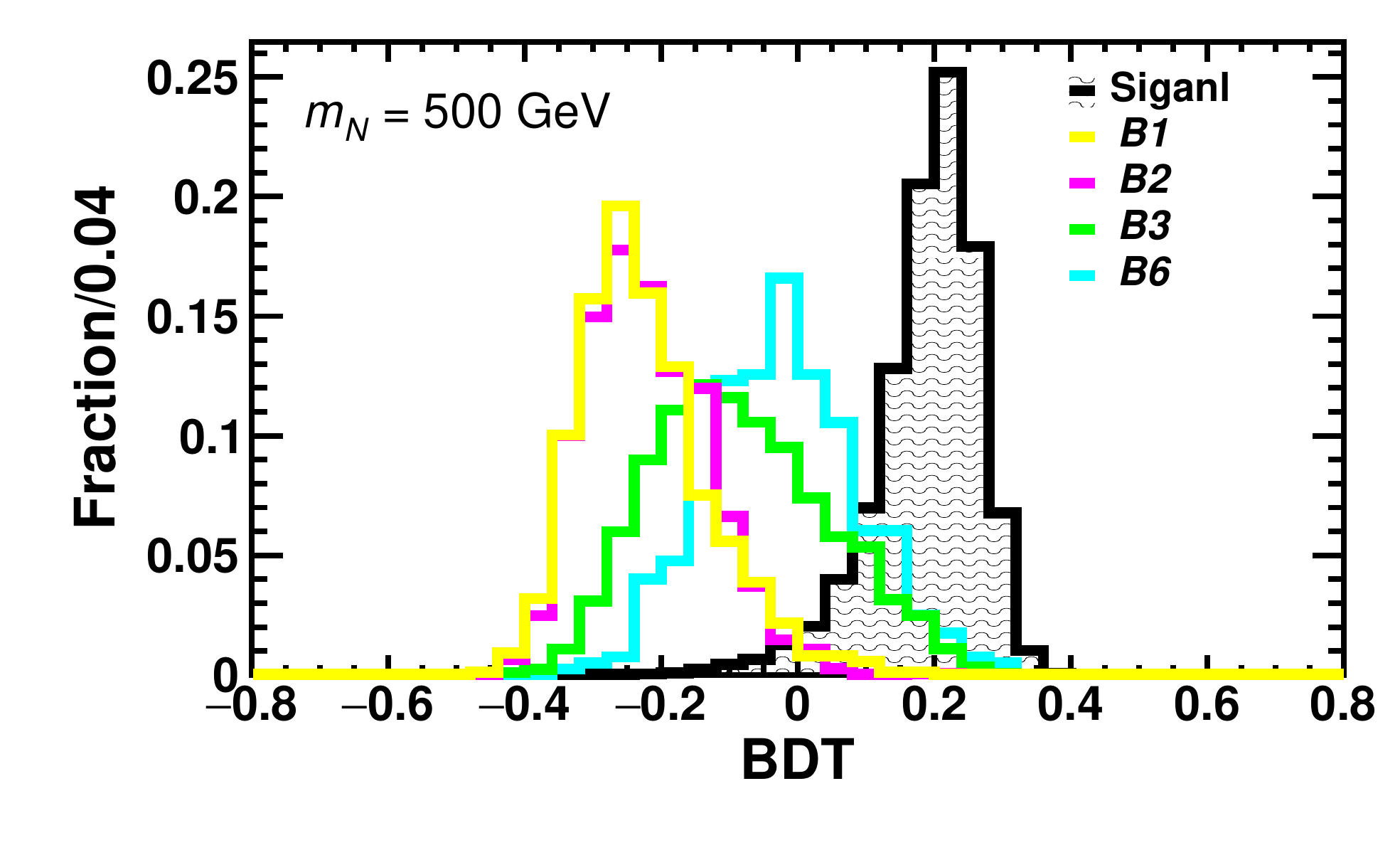}
	}
\caption{
Distributions of BDT responses for the signal (black, filled) and four dominant background processes at the HL-LHC in the scenarios with different $m_N$ assumptions.
}
\label{fig:BDT14TeV}
\end{figure}

\begin{figure}[H] 
	\centering
	\addtocounter{figure}{-1}
	\subfigure{
		\includegraphics[width=4.5cm,height=3cm]{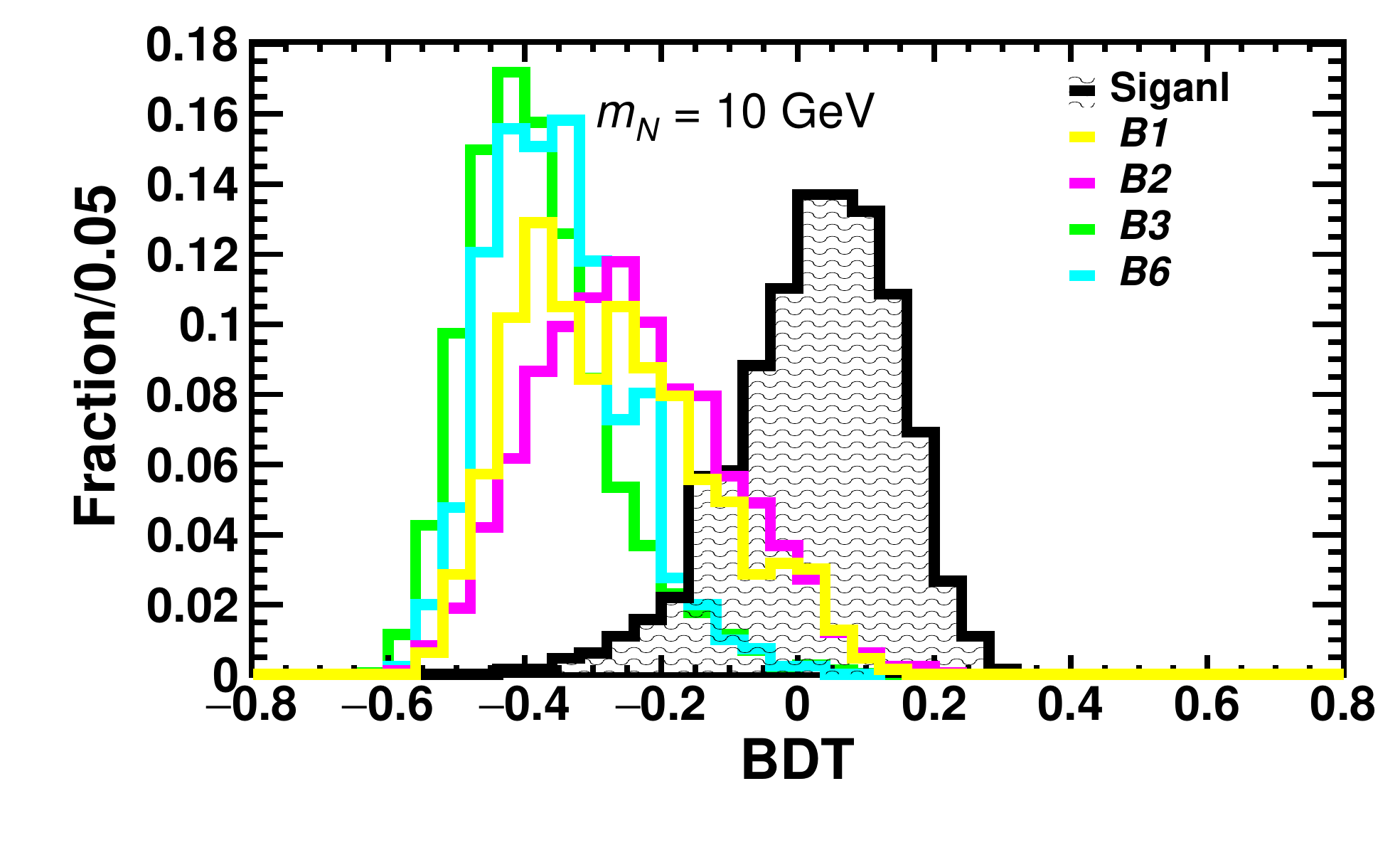}
		\includegraphics[width=4.5cm,height=3cm]{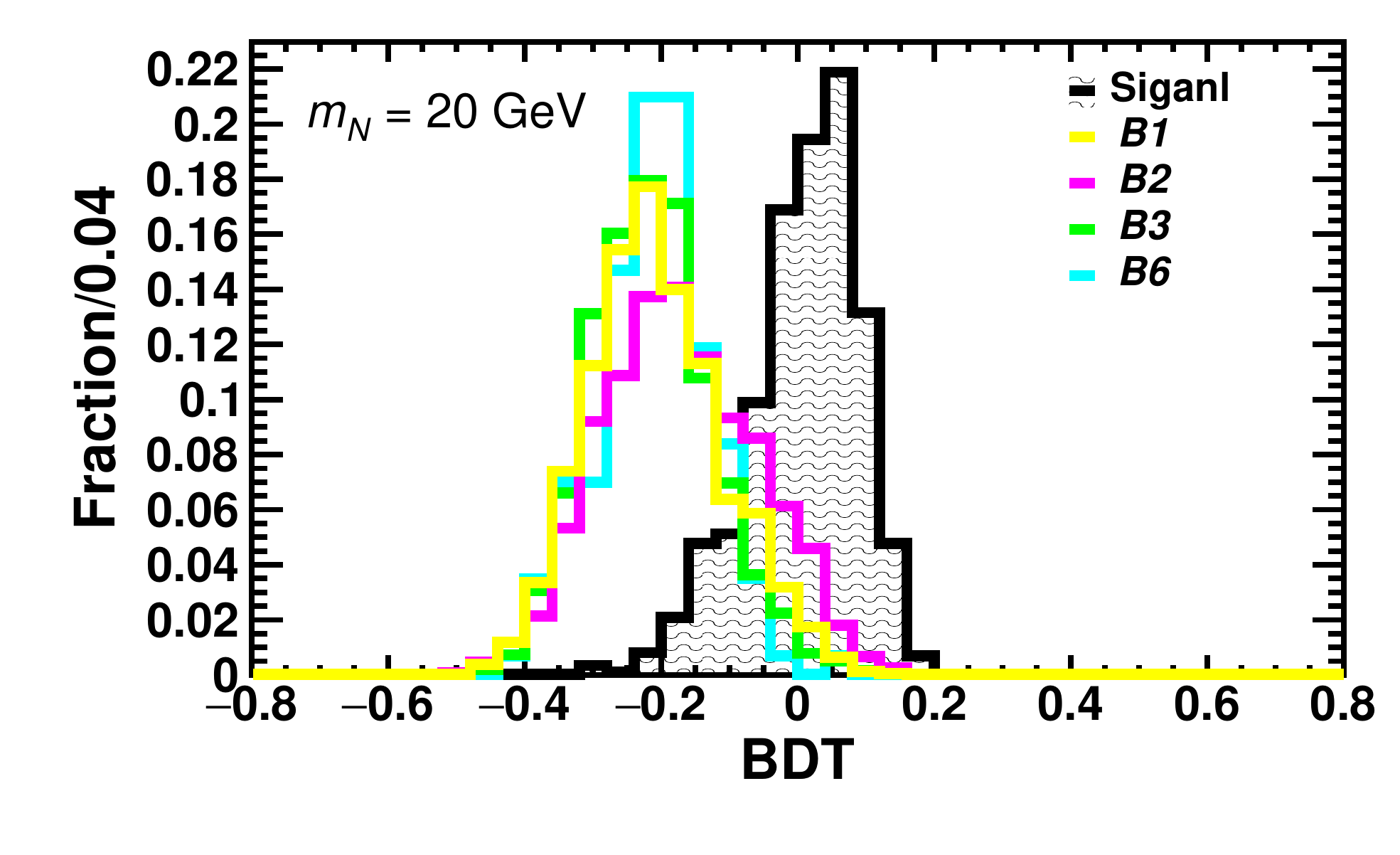}
	}
\end{figure}
\vspace{-1.0cm}
\begin{figure}[H] 
	\centering
	\addtocounter{figure}{1}
	\subfigure{
		\includegraphics[width=4.5cm,height=3cm]{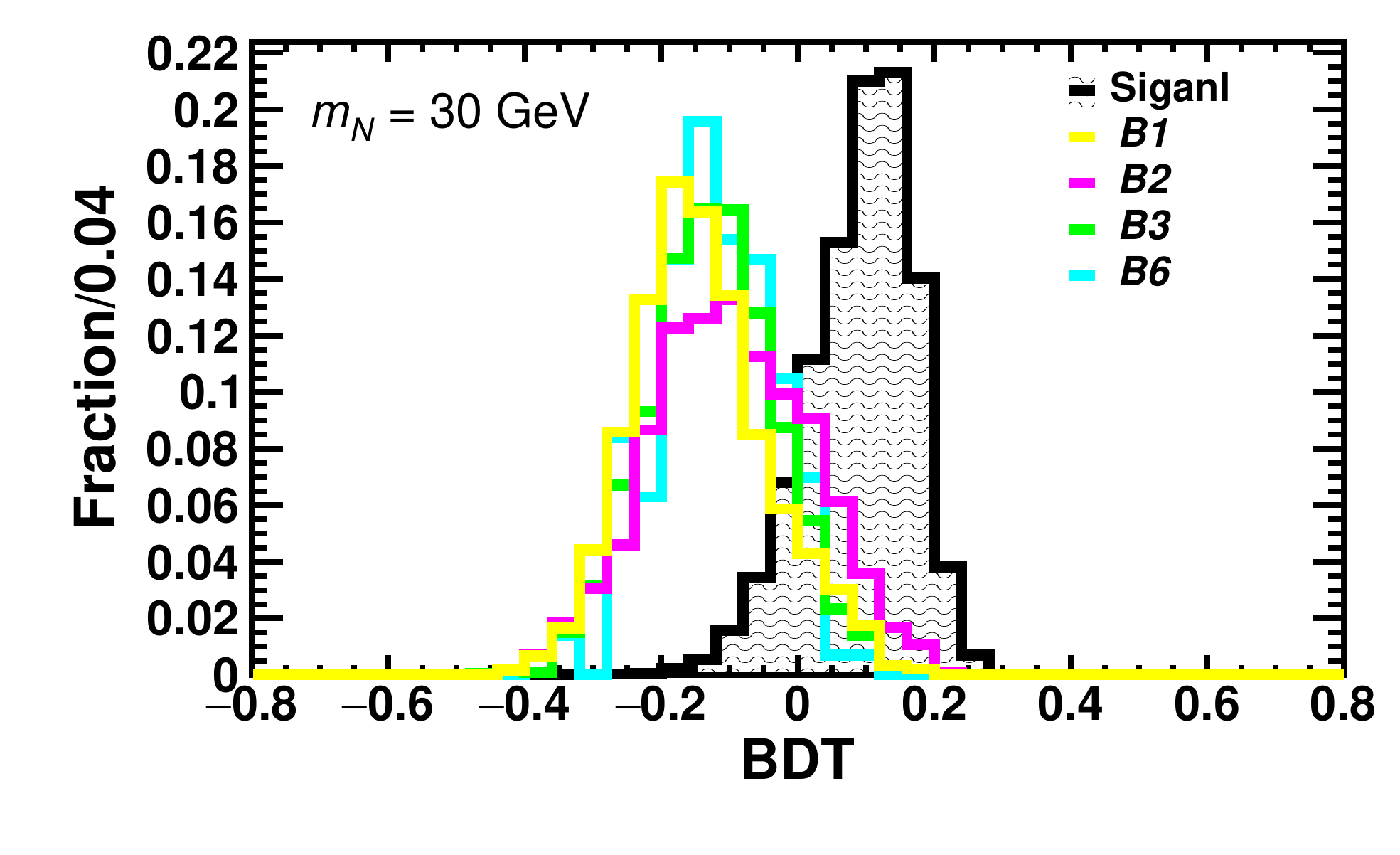}
		\includegraphics[width=4.5cm,height=3cm]{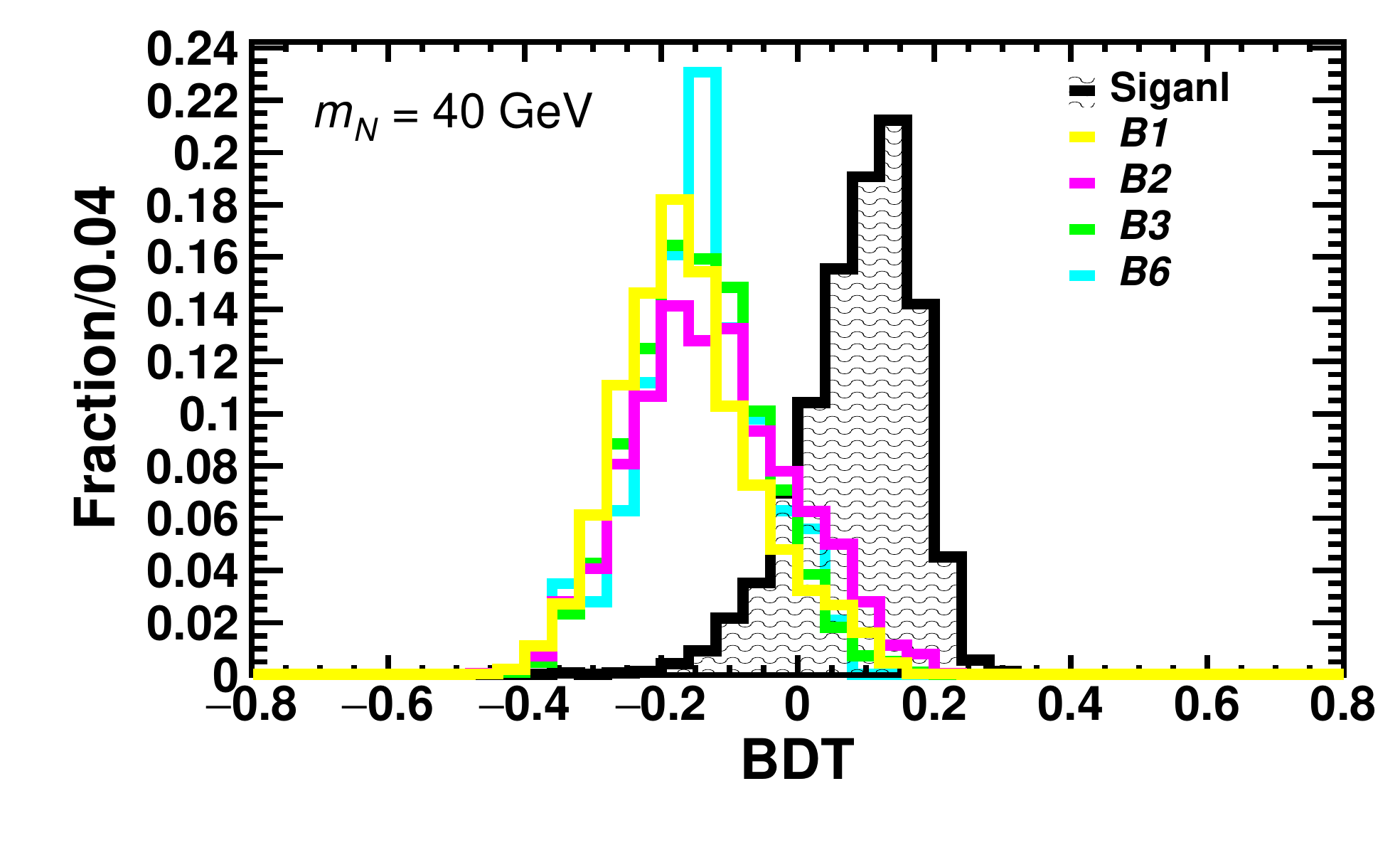}
	}
\end{figure}
\vspace{-1.0cm}
\begin{figure}[H] 
	\centering
	\addtocounter{figure}{-1}
	\subfigure{
		\includegraphics[width=4.5cm,height=3cm]{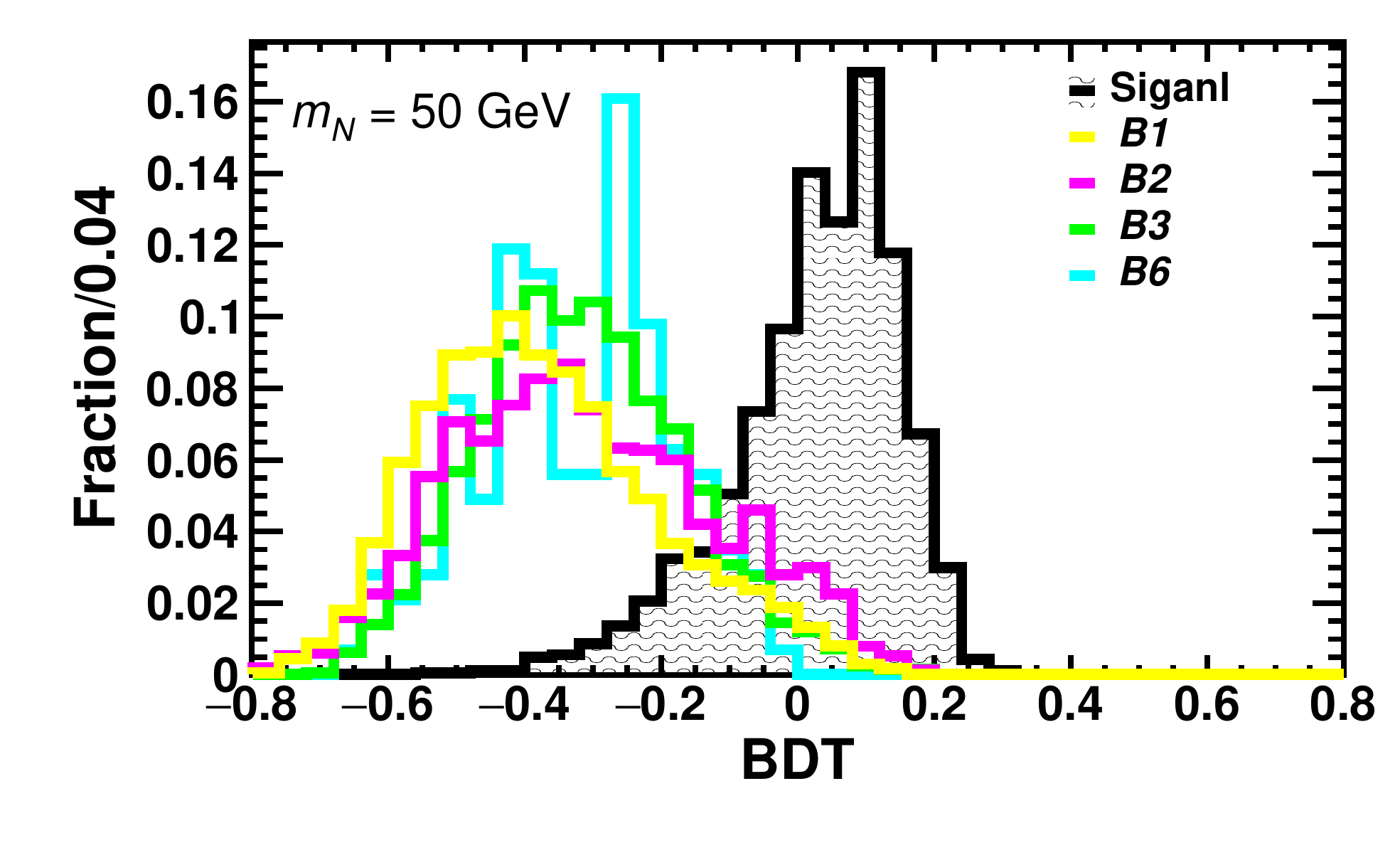}
		\includegraphics[width=4.5cm,height=3cm]{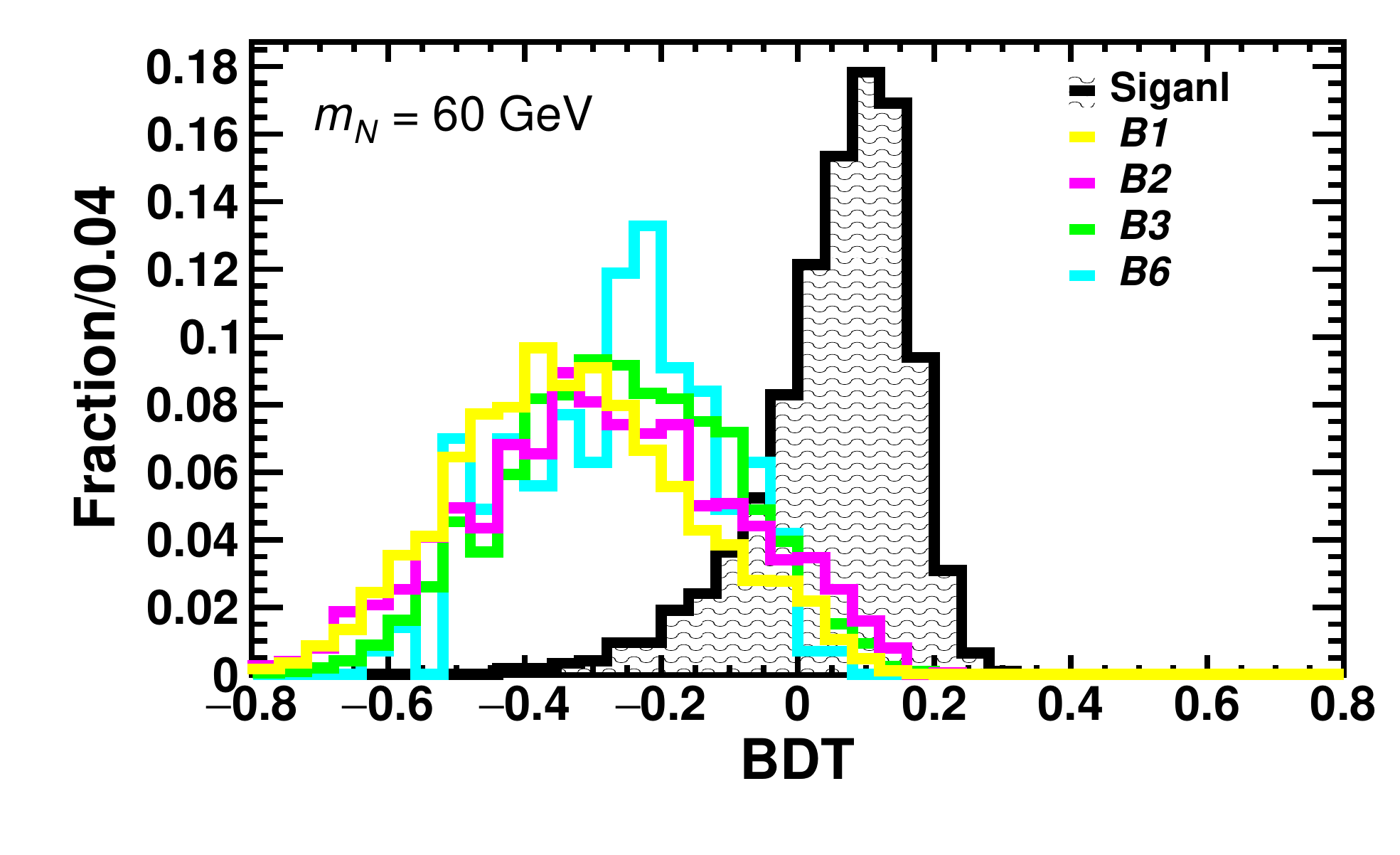}
	}
\end{figure}
\vspace{-1.0cm}
\begin{figure}[H] 
	\centering
	\addtocounter{figure}{1}
	\subfigure{
		\includegraphics[width=4.5cm,height=3cm]{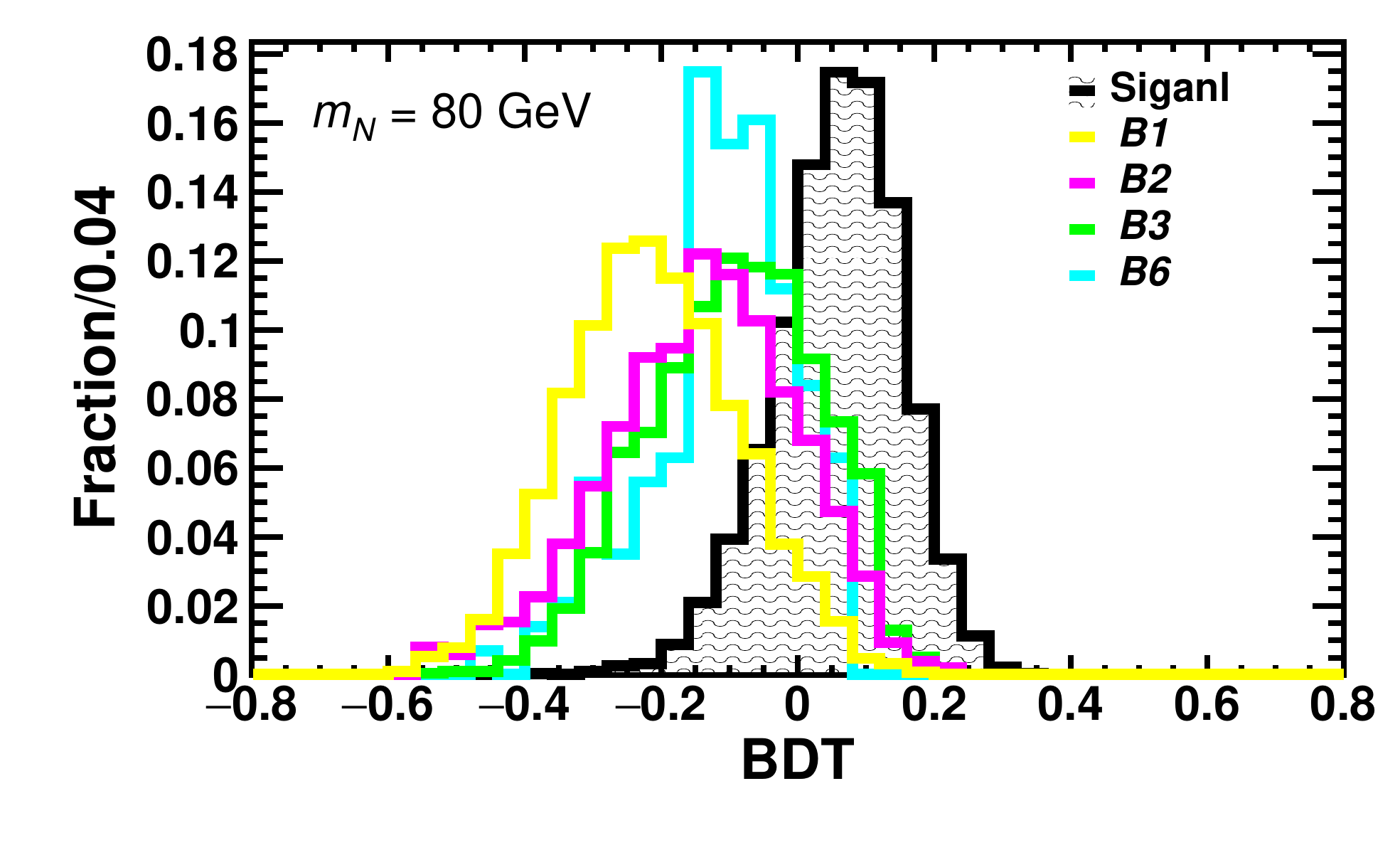}
		\includegraphics[width=4.5cm,height=3cm]{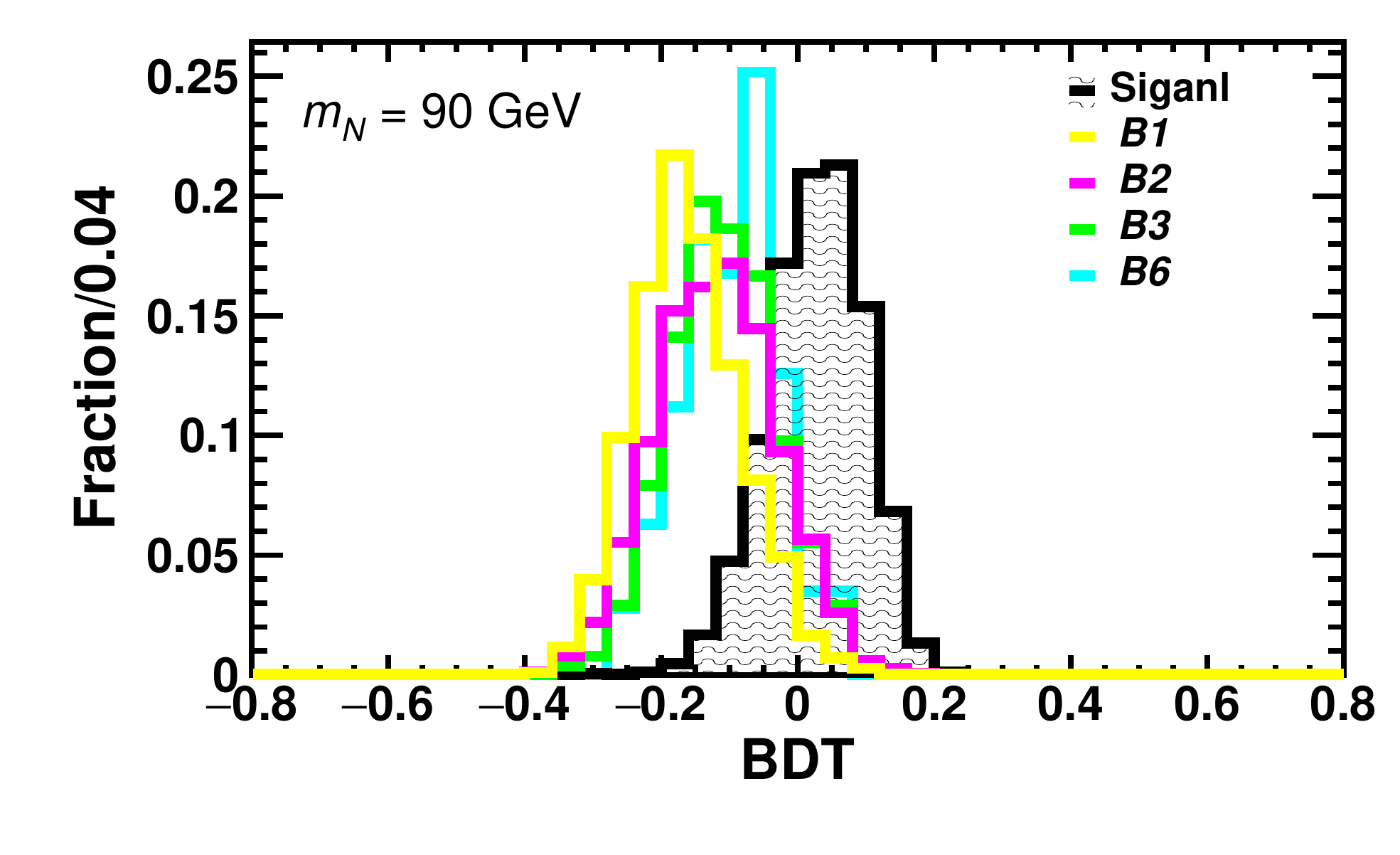}
	}
\end{figure}
\vspace{-1.0cm}
\begin{figure}[H] 
	\centering
	\addtocounter{figure}{-1}
	\subfigure{
		\includegraphics[width=4.5cm,height=3cm]{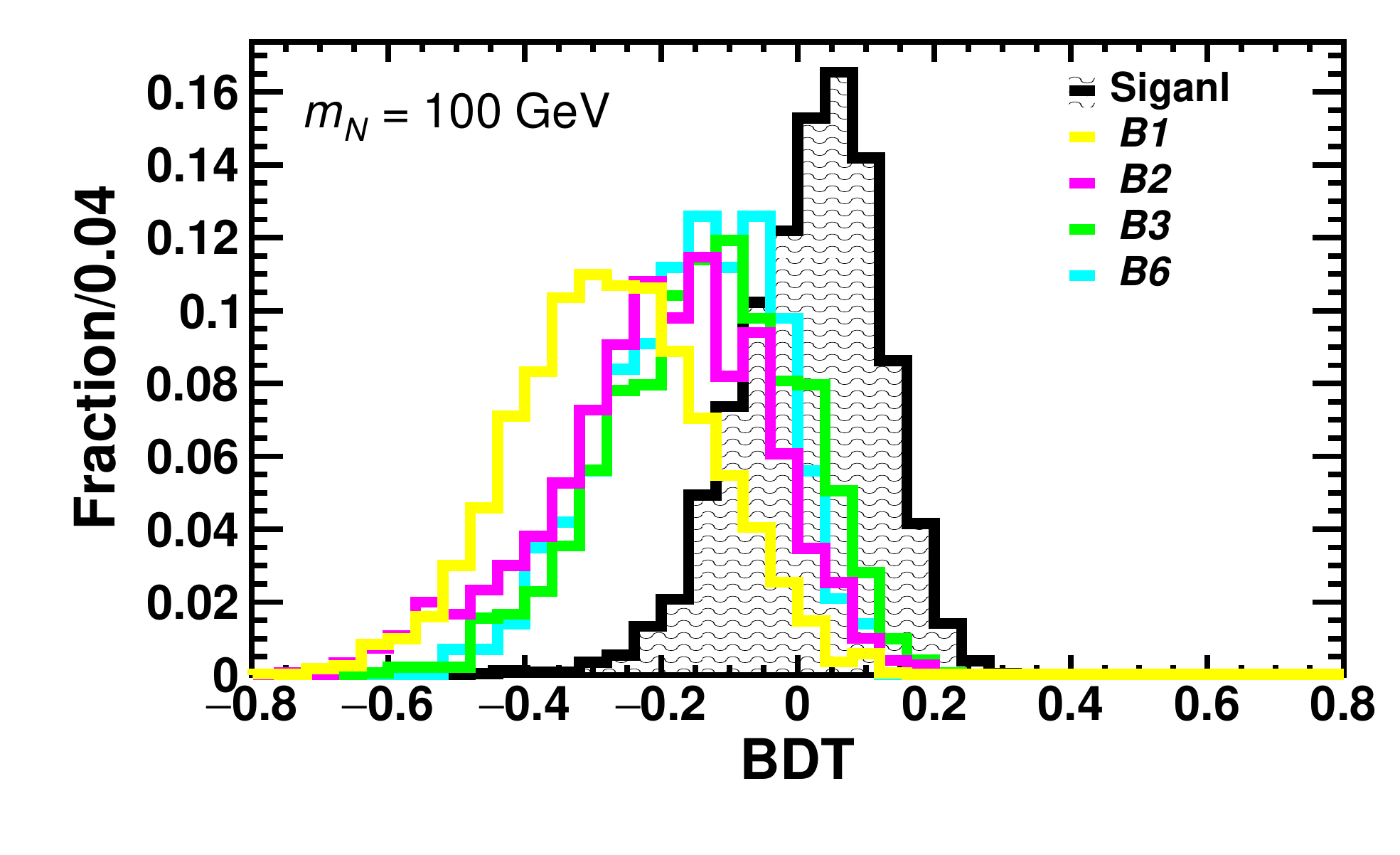}
		\includegraphics[width=4.5cm,height=3cm]{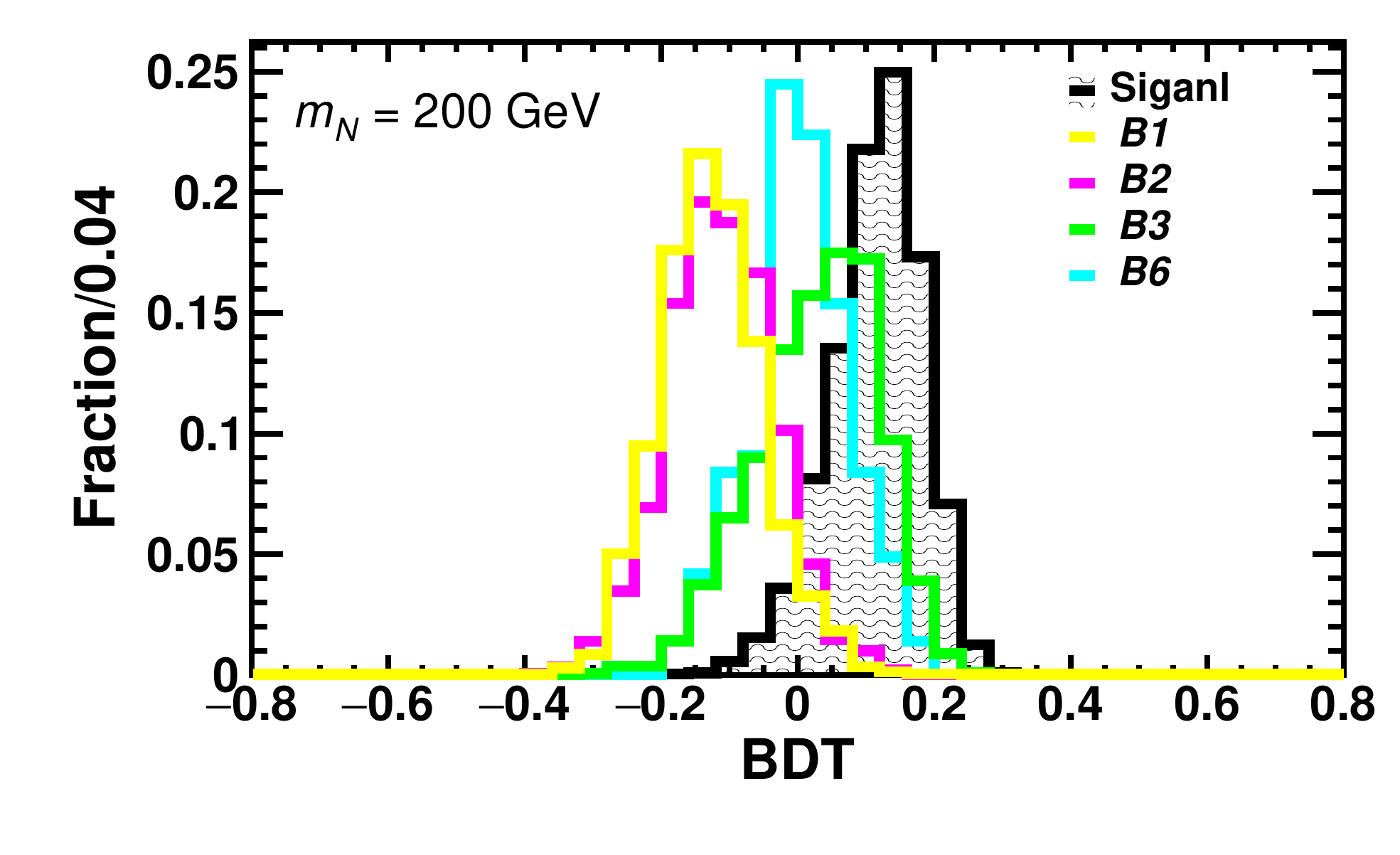}
	}
\end{figure}
\vspace{-1.0cm}
\begin{figure}[H] 
	\centering
	\addtocounter{figure}{1}
	\subfigure{
		\includegraphics[width=4.5cm,height=3cm]{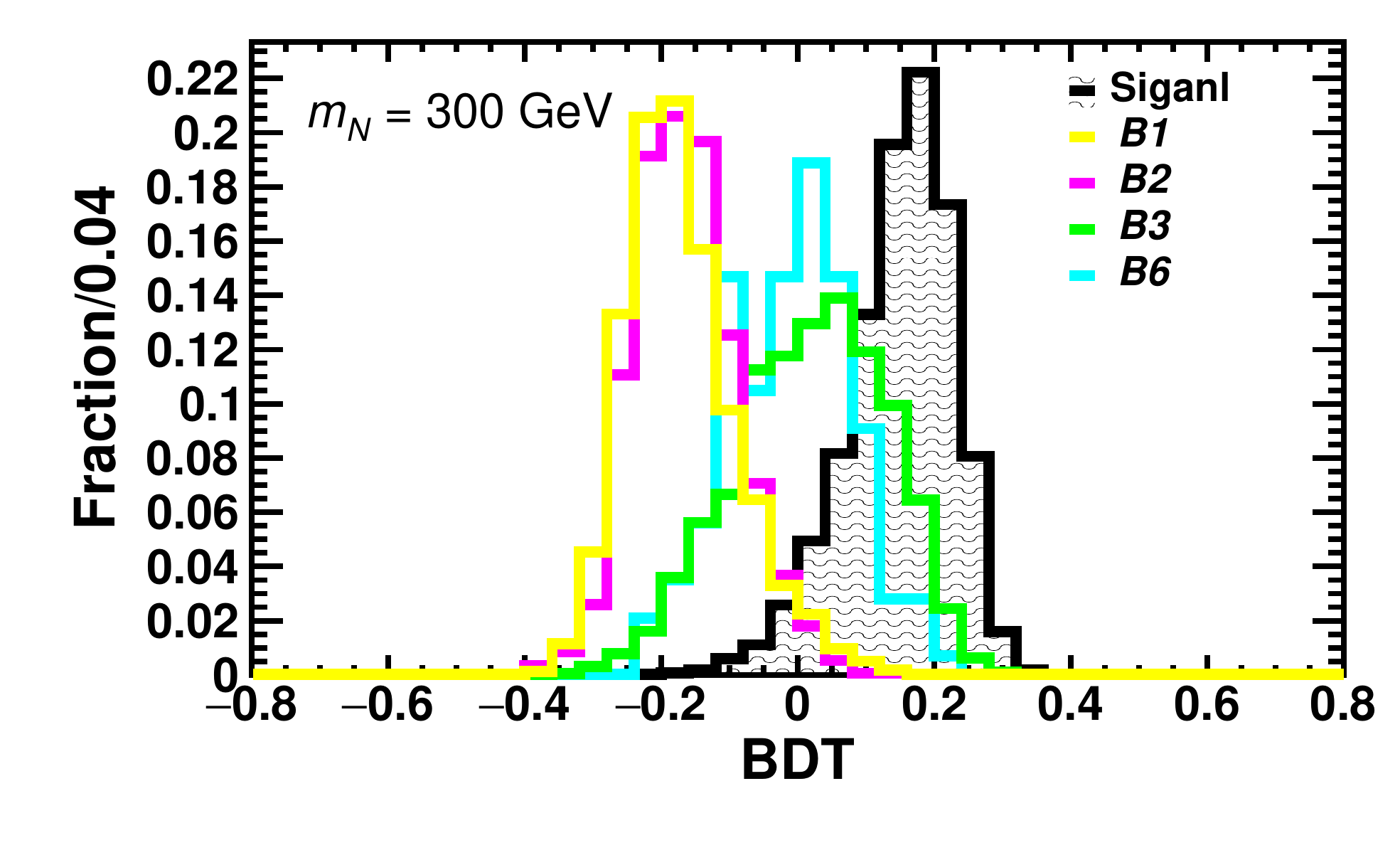}
		\includegraphics[width=4.5cm,height=3cm]{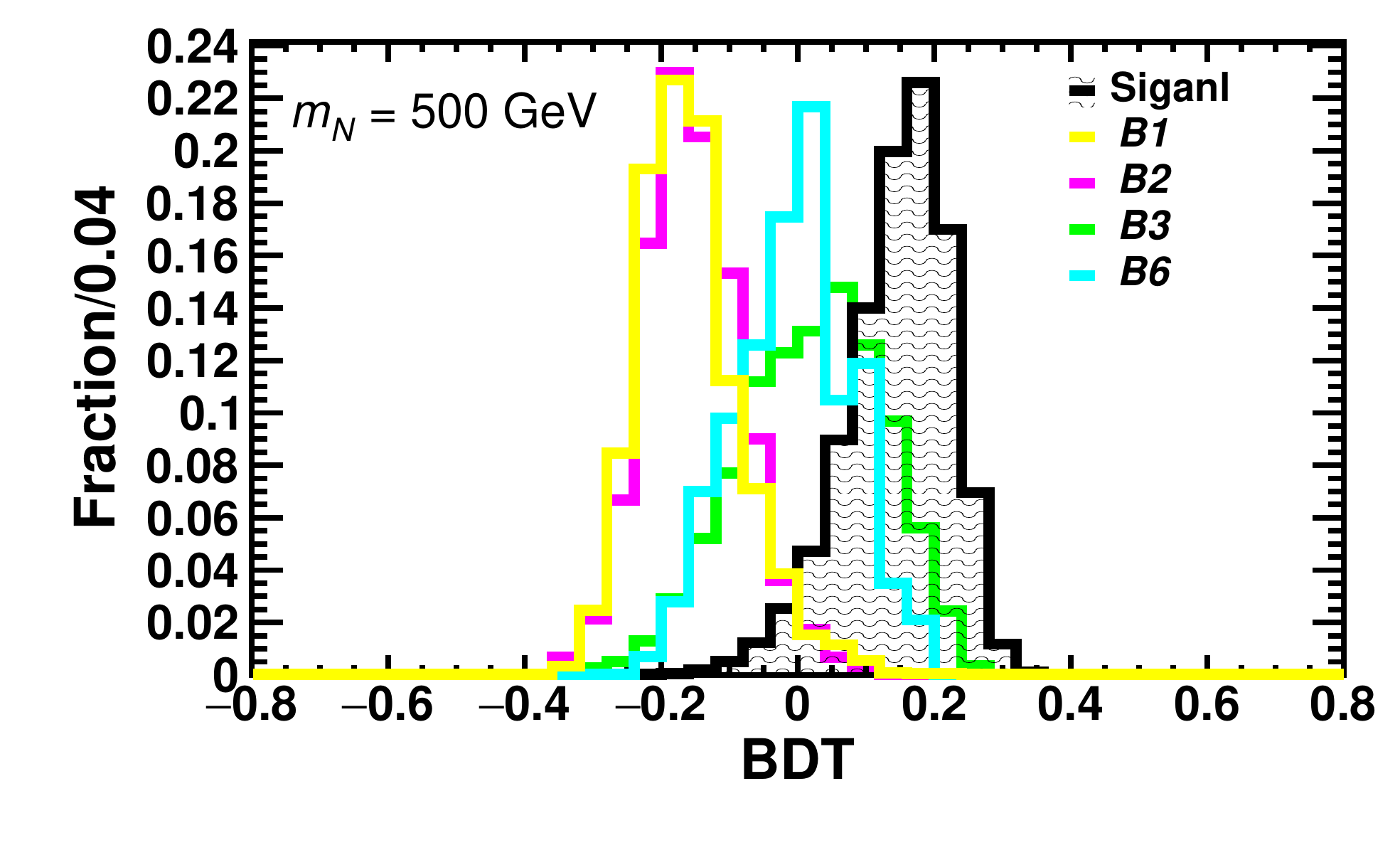}
	}
\end{figure}
\vspace{-1.0cm}
\begin{figure}[H] 
	\centering
	\addtocounter{figure}{-1}
	\subfigure{
		\includegraphics[width=4.5cm,height=3cm]{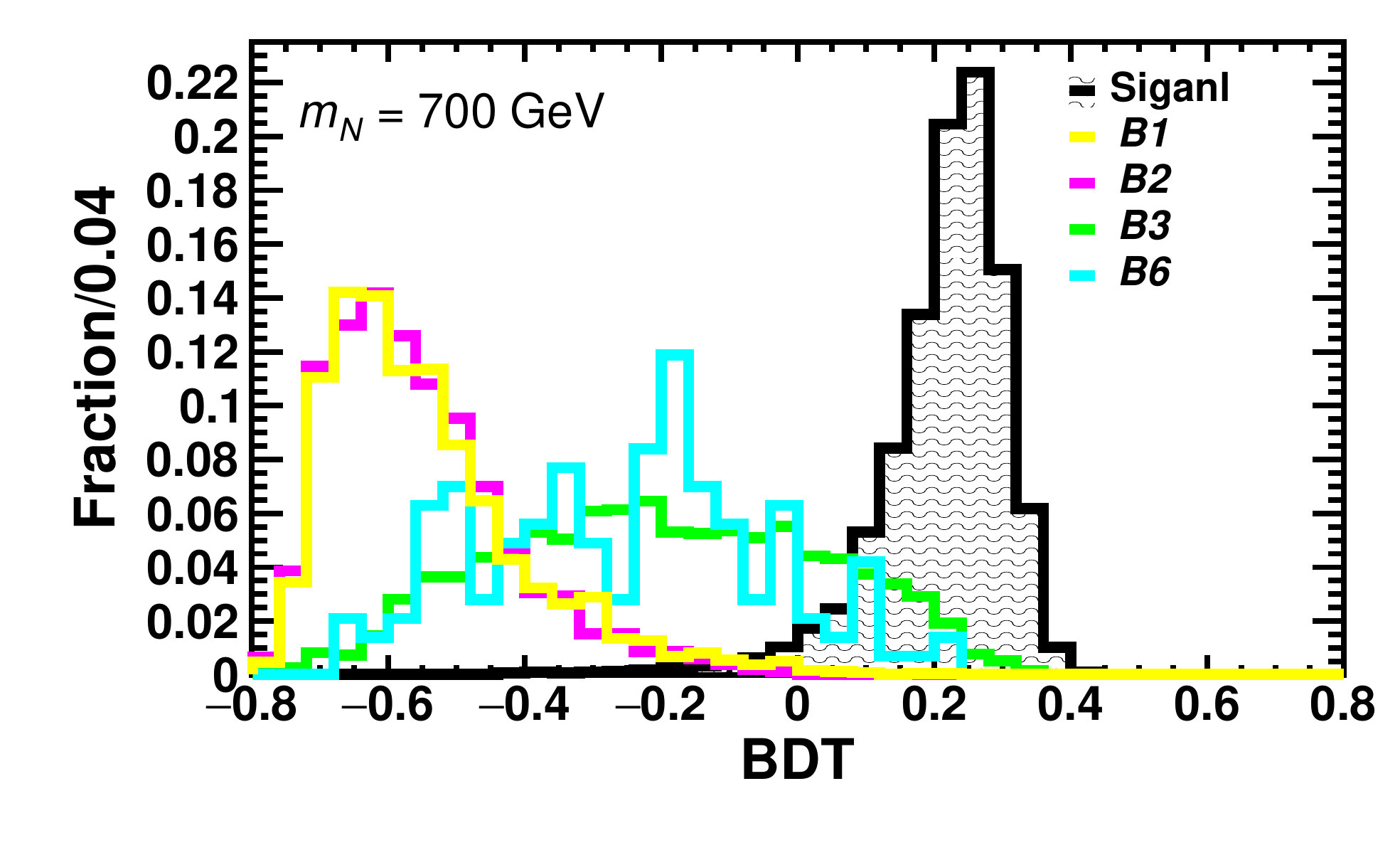}
		\includegraphics[width=4.5cm,height=3cm]{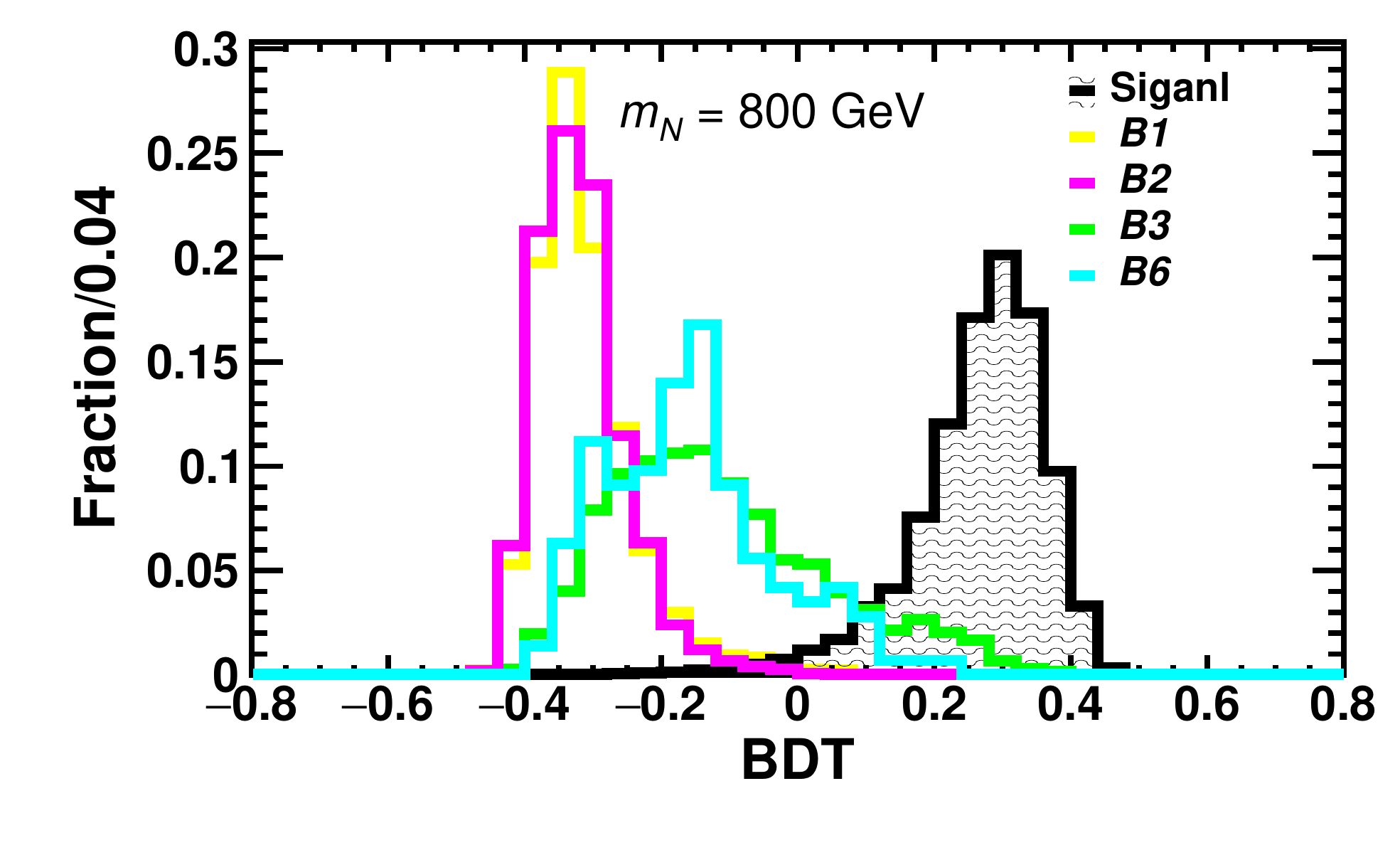}
	}
\end{figure}
\vspace{-1.0cm}
\begin{figure}[H] 
	\centering
	\addtocounter{figure}{1}
	\subfigure{
		\includegraphics[width=4.5cm,height=3cm]{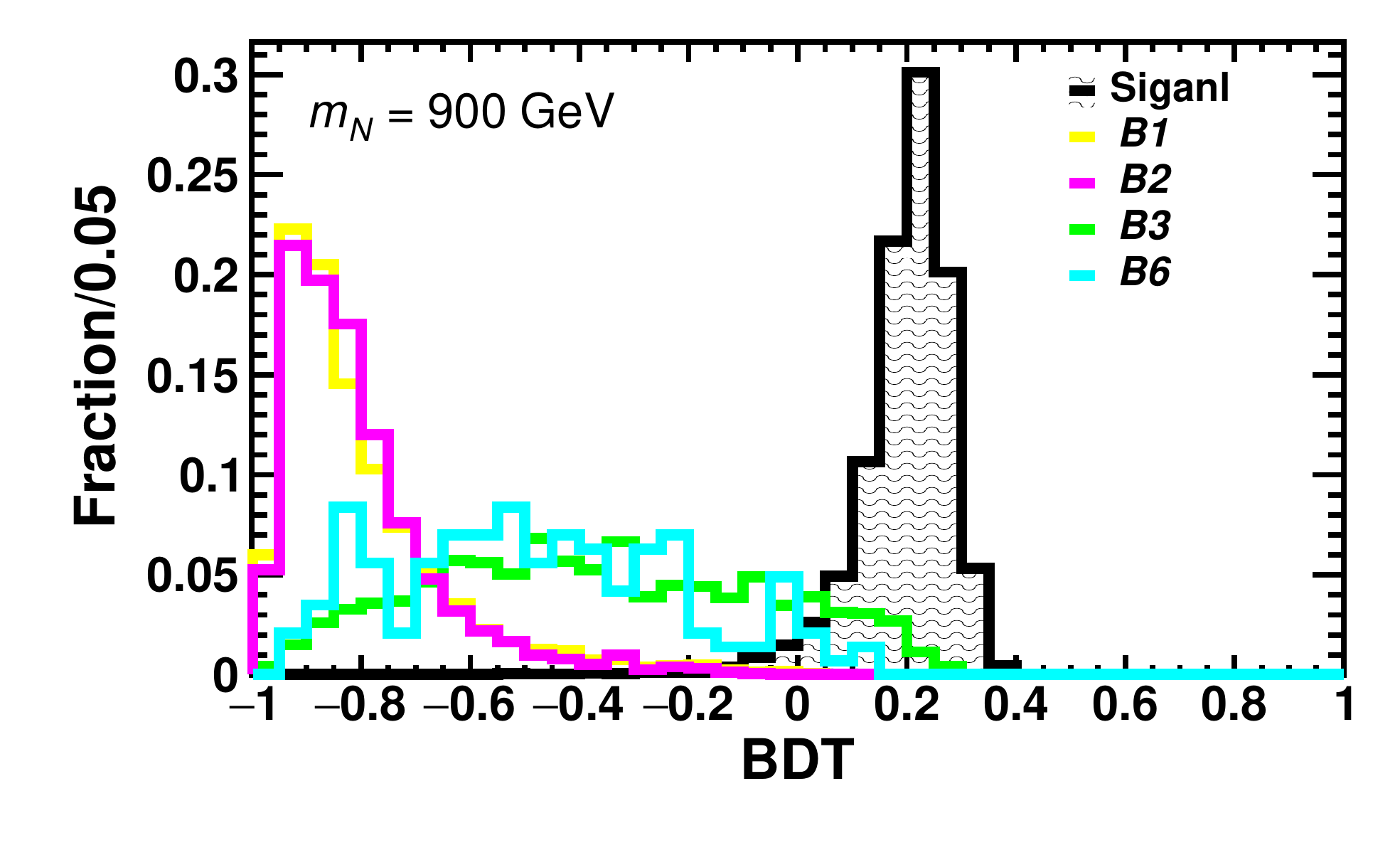}
		\includegraphics[width=4.5cm,height=3cm]{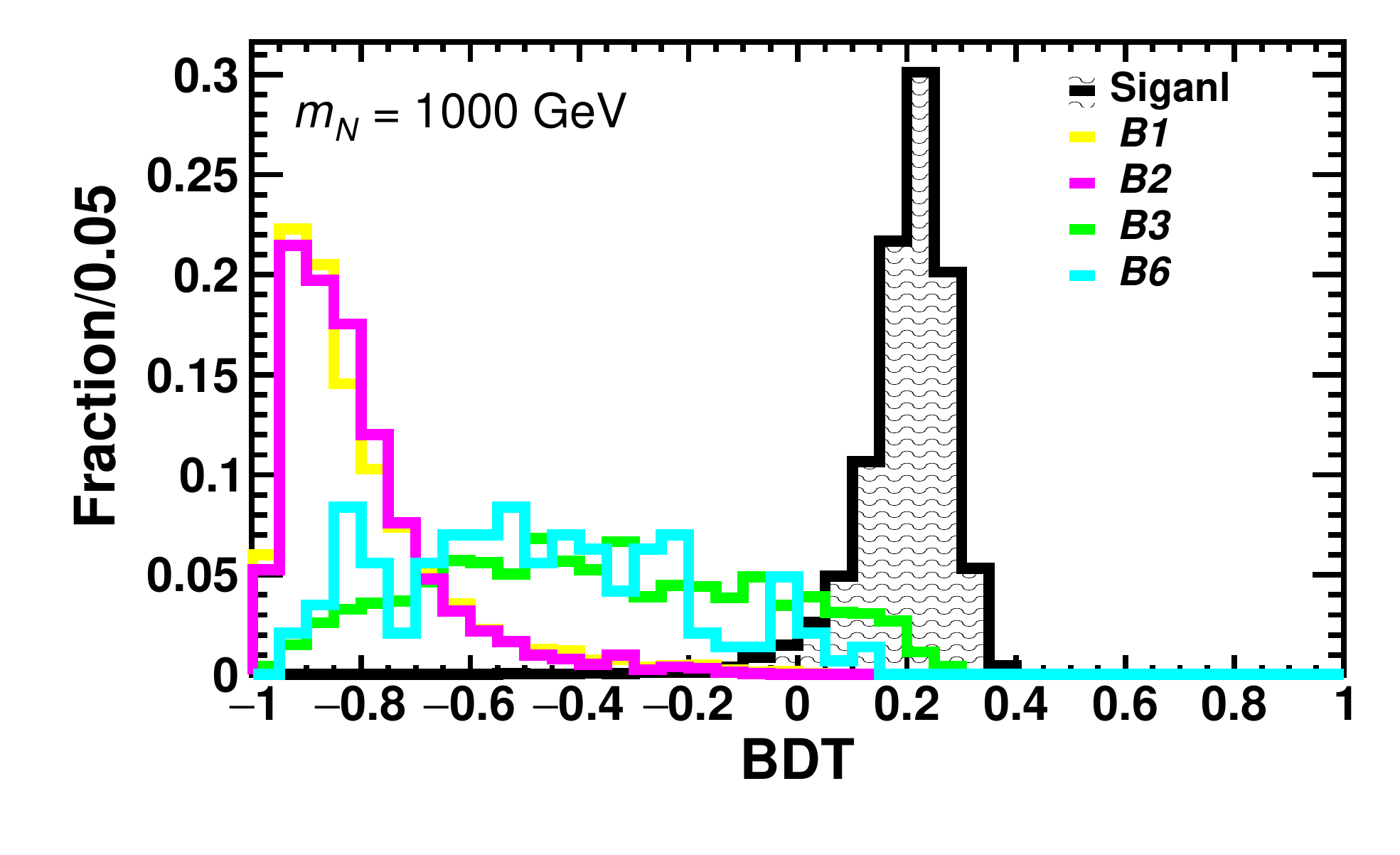}
	}
\caption{
Distributions of BDT responses for the signal (black, filled) and four dominant background processes at the SppC/FCC-hh in the scenarios with different $m_N$ assumptions.
}
\label{fig:BDT100TeV}
\end{figure}

%\newpage
%\vspace{2 cm}
%\newpage
\section{The selection efficiency table}
\label{appendix:efficiencies}

\begin{table*}[t]
	\centering
	\begin{ruledtabular}
		\begin{tabular}{ccccccccc}
$m_N$ & selection & signal & B1& B2 & B3 & B4 & B5& B6\\
\hline

   \multirow{2}{*}{20 GeV} 
	&preselection&$4.16\mltp10^{-4}$&$1.66\mltp10^{-5}$&$1.85\mltp10^{-4}$&$3.84\mltp10^{-4}$&$3.91\mltp10^{-5}$&$8.87\mltp 10^{-5}$&$2.26\mltp10^{-5}$\\
	&BDT$>$0.079&$6.30\mltp10^{-1}$&$4.32\mltp10^{-2}$&$5.41\mltp10^{-2}$&$5.07\mltp10^{-3}$&$7.33\mltp10^{-3}$&$8.49\mltp10^{-3}$&$5.01\mltp10^{-3}$\\
			\hline
   \multirow{2}{*}{30 GeV} 
	&preselection&$5.90\mltp10^{-4}$&$1.66\mltp10^{-5}$&$1.85\mltp10^{-4}$&$3.84\mltp10^{-4}$&$3.91\mltp10^{-5}$&$8.87\mltp 10^{-5}$&$2.26\mltp10^{-5}$\\
	&BDT$>$0.151&$3.07\mltp10^{-1}$&$4.67\mltp10^{-3}$&$6.36\mltp10^{-3}$&$1.14\mltp10^{-3}$&$\cdots$&$\cdots$&$2.51\mltp10^{-3}$\\
\hline

\multirow{2}{*}{40 GeV} 
	&preselection&$7.57\mltp10^{-4}$&$1.66\mltp10^{-5}$&$1.85\mltp10^{-4}$&$3.84\mltp10^{-4}$&$3.91\mltp10^{-5}$&$8.87\mltp 10^{-5}$&$2.26\mltp10^{-5}$\\
	&BDT$>$0.129&$4.93\mltp10^{-1}$&$2.23\mltp10^{-2}$&$1.72\mltp10^{-2}$&$3.43\mltp10^{-3}$&$7.33\mltp10^{-3}$&$4.24\mltp10^{-3}$&$2.51\mltp10^{-3}$\\
			\hline

   \multirow{2}{*}{50 GeV} 
	&preselection&$8.13\mltp10^{-4}$&$1.66\mltp10^{-5}$&$1.85\mltp10^{-4}$&$3.84\mltp10^{-4}$&$3.91\mltp10^{-5}$&$8.87\mltp 10^{-5}$&$2.26\mltp10^{-5}$\\
	&BDT$>$0.116&$4.67\mltp10^{-1}$&$8.17\mltp10^{-3}$&$1.15\mltp10^{-2}$&$1.47\mltp10^{-3}$&$3.66\mltp10^{-3}$&$\cdots$ &$2.51\mltp10^{-3}$\\
 \hline

\multirow{2}{*}{60 GeV} 
	&preselection&$7.16\mltp10^{-4}$&$1.66\mltp10^{-5}$&$1.85\mltp10^{-4}$&$3.84\mltp10^{-4}$&$3.91\mltp10^{-5}$&$8.87\mltp 10^{-5}$&$2.26\mltp10^{-5}$\\
	& BDT$>$0.136  &$6.22\mltp10^{-1}$&$1.63\mltp10^{-2}$&$1.97\mltp10^{-2}$&$3.11\mltp10^{-3}$&$3.66\mltp10^{-3}$&$4.24\mltp10^{-3}$&$2.51\mltp10^{-3}$\\                                  
			\hline
\multirow{2}{*}{80 GeV} 
	&preselection&$7.16\mltp10^{-4}$&$1.66\mltp10^{-5}$&$1.85\mltp10^{-4}$&$3.84\mltp10^{-4}$&$3.91\mltp10^{-5}$&$8.87\mltp 10^{-5}$&$2.26\mltp10^{-5}$\\
	& BDT$>$0.130  &$3.64\mltp10^{-1}$&$1.05\mltp10^{-1}$&$1.53\mltp10^{-2}$&$2.45\mltp10^{-3}$&$3.66\mltp10^{-3}$&$7.07\mltp10^{-3}$&$2.51\mltp10^{-3}$\\                        
			\hline
\multirow{2}{*}{100 GeV} 
	&preselection&$2.00\mltp10^{-3}$&$1.66\mltp10^{-5}$&$1.85\mltp10^{-4}$&$3.84\mltp10^{-4}$&$3.91\mltp10^{-5}$&$8.87\mltp 10^{-5}$&$2.26\mltp10^{-5}$\\
	& BDT$>$0.142  &$4.33\mltp10^{-1}$&$1.87\mltp10^{-1}$&$2.29\mltp10^{-2}$&$1.69\mltp10^{-3}$&$1.83\mltp10^{-3}$&$5.66\mltp10^{-3}$&$5.01\mltp10^{-3}$\\
			\hline
\multirow{2}{*}{300 GeV} 
	&preselection&$1.07\mltp10^{-2}$&$1.66\mltp10^{-5}$&$1.85\mltp10^{-4}$&$3.84\mltp10^{-4}$&$3.91\mltp10^{-5}$&$8.87\mltp 10^{-5}$&$2.26\mltp10^{-5}$\\
	& BDT$>$0.155  &$5.02\mltp10^{-1}$&$2.33\mltp10^{-3}$&$1.27\mltp10^{-3}$&$4.33\mltp10^{-2}$&$3.30\mltp10^{-2}$&$1.41\mltp10^{-2}$&$1.33\mltp10^{-1}$\\
			\hline
\multirow{2}{*}{500 GeV} 
	&preselection&$1.28\mltp10^{-2}$&$1.66\mltp10^{-5}$&$1.85\mltp10^{-4}$&$3.84\mltp10^{-4}$&$3.91\mltp10^{-5}$&$8.87\mltp 10^{-5}$&$2.26\mltp10^{-5}$\\
	& BDT$>$0.210  &$4.73\mltp10^{-1}$& $\cdots$ &$6.36\mltp10^{-4}$&$1.32\mltp10^{-2}$&$1.10\mltp10^{-2}$& $\cdots$ &$2.01\mltp10^{-2}$\\
		\end{tabular}
	\end{ruledtabular}
\caption{
Selection efficiencies of preselection and BDT cuts for both signal and background processes at the HL-LHC in the scenarios with different $m_N$ assumptions,
where ``$\cdots$" means the number of events can be reduced to be negligible.
}
\label{tab:CutEffiHLLHC}
\end{table*}

\vspace{-30.0cm}

\begin{table*}[t]
	\centering
	\begin{ruledtabular}
		\begin{tabular}{ccccccccc}
			$m_N$ & selection & signal & B1& B2 & B3 & B4 & B5& B6\\
			\hline

   			\multirow{2}{*}{20 GeV} 
	&preselection&$4.30\mltp10^{-4}$&$1.37\mltp10^{-4}$&$5.27\mltp10^{-5}$&$2.47\mltp10^{-4}$&$2.23\mltp10^{-4}$&$3.12\mltp10^{-5}$&$5.30\mltp10^{-5}$\\
	&BDT$>$0.071&$9.81\mltp10^{-2}$&$2.98\mltp10^{-3}$&$1.13\mltp10^{-2}$&$2.61\mltp10^{-3}$&$1.09\mltp10^{-2}$&$1.30\mltp10^{-2}$ &$\cdots$\\
			\hline
   			\multirow{2}{*}{30 GeV} 
	&preselection&$5.29\mltp10^{-4}$&$1.37\mltp10^{-4}$&$5.27\mltp10^{-5}$&$2.47\mltp10^{-4}$&$2.23\mltp10^{-4}$&$3.12\mltp10^{-5}$&$5.30\mltp10^{-5}$\\
	&BDT$>$0.126&$3.71\mltp10^{-1}$&$5.11\mltp10^{-3}$&$2.33\mltp10^{-2}$&$4.18\mltp10^{-3}$&$1.45\mltp10^{-2}$&$4.55\mltp10^{-2}$ &$\cdots$\\
			\hline
   
			\multirow{2}{*}{40 GeV} 
	&preselection&$6.12\mltp10^{-4}$&$1.37\mltp10^{-4}$&$5.27\mltp10^{-5}$&$2.47\mltp10^{-4}$&$2.23\mltp10^{-4}$&$3.12\mltp10^{-5}$&$5.30\mltp10^{-5}$\\
	&BDT$>$0.144&$2.76\mltp10^{-1}$&$2.98\mltp10^{-3}$&$1.13\mltp10^{-2}$&$1.57\mltp10^{-3}$&$3.62\mltp10^{-3}$&$1.95\mltp10^{-2}$ &$\cdots$\\
			\hline

   			\multirow{2}{*}{50 GeV} 
	&preselection&$6.13\mltp10^{-4}$&$1.37\mltp10^{-4}$&$5.27\mltp10^{-5}$&$2.47\mltp10^{-4}$&$2.23\mltp10^{-4}$&$3.12\mltp10^{-5}$&$5.30\mltp10^{-5}$\\
	&BDT$>$0.098 & $3.09\mltp10^{-1}$&$2.37\mltp10^{-4}$&$3.53\mltp10^{-4}$&$1.08\mltp10^{-4}$&$4.65\mltp10^{-5}$&$1.06\mltp10^{-3}$ &$\cdots$\\
			\hline
   
			\multirow{2}{*}{60 GeV} 
	&preselection&$5.69\mltp10^{-4}$&$1.37\mltp10^{-4}$&$5.27\mltp10^{-5}$&$2.47\mltp10^{-4}$&$2.23\mltp10^{-4}$&$3.12\mltp10^{-5}$&$5.30\mltp10^{-5}$\\
	& BDT$>$0.128 &$1.18\mltp10^{-1}$&$7.71\mltp10^{-4}$&$4.66\mltp10^{-3}$&$2.09\mltp10^{-3}$&$3.62\mltp10^{-3}$&$1.95\mltp10^{-2}$&$\cdots$\\
			\hline
			\multirow{2}{*}{80 GeV} 
	&preselection&$5.69\mltp10^{-4}$&$1.37\mltp10^{-4}$&$5.27\mltp10^{-5}$&$2.47\mltp10^{-4}$&$2.23\mltp10^{-4}$&$3.12\mltp10^{-5}$&$5.30\mltp10^{-5}$\\
	& BDT$>$0.076 &$4.52\mltp10^{-1}$&$9.77\mltp10^{-3}$&$4.73\mltp10^{-2}$&$8.62\mltp10^{-2}$&$6.16\mltp10^{-2}$&$1.49\mltp10^{-1}$&$1.39\mltp10^{-2}$\\    
			\hline
			\multirow{2}{*}{100 GeV} 
	&preselection&$2.55\mltp10^{-3}$&$1.37\mltp10^{-4}$&$5.27\mltp10^{-5}$&$2.47\mltp10^{-4}$&$2.23\mltp10^{-4}$&$3.12\mltp10^{-5}$&$5.30\mltp10^{-5}$\\
	& BDT$>$0.048 &$4.19\mltp10^{-1}$&$9.25\mltp10^{-3}$&$3.60\mltp10^{-2}$&$8.41\mltp10^{-2}$&$6.88\mltp10^{-2}$&$1.23\mltp10^{-1}$&$3.49\mltp10^{-2}$\\
			\hline
			\multirow{2}{*}{300 GeV} 
	&preselection&$1.10\mltp10^{-2}$&$1.37\mltp10^{-4}$&$5.27\mltp10^{-5}$&$2.47\mltp10^{-4}$&$2.23\mltp10^{-4}$&$3.12\mltp10^{-5}$&$5.30\mltp10^{-5}$\\
	& BDT$>$0.147 &$5.49\mltp10^{-1}$&$8.52\mltp10^{-4}$& $\cdots$ &$3.33\mltp10^{-3}$&$6.04\mltp10^{-4}$&$1.59\mltp10^{-3}$&$1.29\mltp10^{-4}$\\
			\hline
			\multirow{2}{*}{500 GeV} 
	&preselection&$1.22\mltp10^{-2}$&$1.37\mltp10^{-4}$&$5.27\mltp10^{-5}$&$2.47\mltp10^{-4}$&$2.23\mltp10^{-4}$&$3.12\mltp10^{-5}$&$5.30\mltp10^{-5}$\\
	& BDT$>$0.240 &$3.93\mltp10^{-1}$& $\cdots$ &$\cdots$&$9.25\mltp10^{-5}$&$4.65\mltp10^{-5}$& $\cdots$ &$\cdots$\\
				\hline
				\multirow{2}{*}{700 GeV} 
				&preselection&$1.20\mltp10^{-2}$&$1.37\mltp10^{-4}$&$5.27\mltp10^{-5}$&$2.47\mltp10^{-4}$&$2.23\mltp10^{-4}$&$3.12\mltp10^{-5}$&$5.30\mltp10^{-5}$\\
				& BDT$>$0.195 &$6.69\mltp10^{-1}$&$1.29\mltp10^{-3}$& $\cdots$ &$8.09\mltp10^{-2}$&$7.25\mltp10^{-2}$&$3.90\mltp10^{-2}$&$1.39\mltp10^{-2}$\\
				\hline
			\multirow{2}{*}{900 GeV} 
			&preselection&$1.19\mltp10^{-2}$&$1.37\mltp10^{-4}$&$5.27\mltp10^{-5}$&$2.47\mltp10^{-4}$&$2.23\mltp10^{-4}$&$3.12\mltp10^{-5}$&$5.30\mltp10^{-5}$\\
			& BDT$>$0.178 &$6.74\mltp10^{-1}$&$2.57\mltp10^{-4}$& $\cdots$ &$2.87\mltp10^{-2}$&$1.09\mltp10^{-2}$&$1.30\mltp10^{-2}$& $\cdots$\\
\end{tabular}
\end{ruledtabular}
\caption{
Selection efficiencies of preselection and BDT cuts for both signal and background processes at the SppC/FCC-hh in the scenarios with different $m_N$ assumptions,
where ``$\cdots$" means the number of events can be reduced to be negligible.
}
\label{tab:CutEffi100}
\end{table*}

\section{Effects of varying assumptions of mixing parameters }
\label{appendix:Effmixings}

To estimate the effects of varying assumptions of mixing parameters quantitatively, we compare the results for three assumptions of mixing parameters: (i) $|V_{\tau N}|^2 \neq 0 $ and $|V_{e N}|^2 = |V_{\mu N}|^2 = 0$; (ii) $|V_{\mu N}|^2 \neq 0 $ and $|V_{e N}|^2 = |V_{\tau N}|^2 = 0$; (iii) $|V_{\tau N}|^2 = |V_{\mu N}|^2 \neq 0 $ and $|V_{e N}|^2 = 0$.
We make case (i) as the theoretical hypothesis, and test the results of cases (ii) and (iii).
Specifically, we firstly apply the same preselection cuts as described in the article for all three cases. 
Then, the BTD training and test are performed by inputting the signal of case (i) and SM background.
This trained program is finally applied to all three cases, and the same BDT cut is adopted.
In this way, the effects on results can be estimated as changing kinematics for different cases.

\begin{figure}[H]
	\centering
	\subfigure{
		\includegraphics[width=4.5cm,height=3cm]{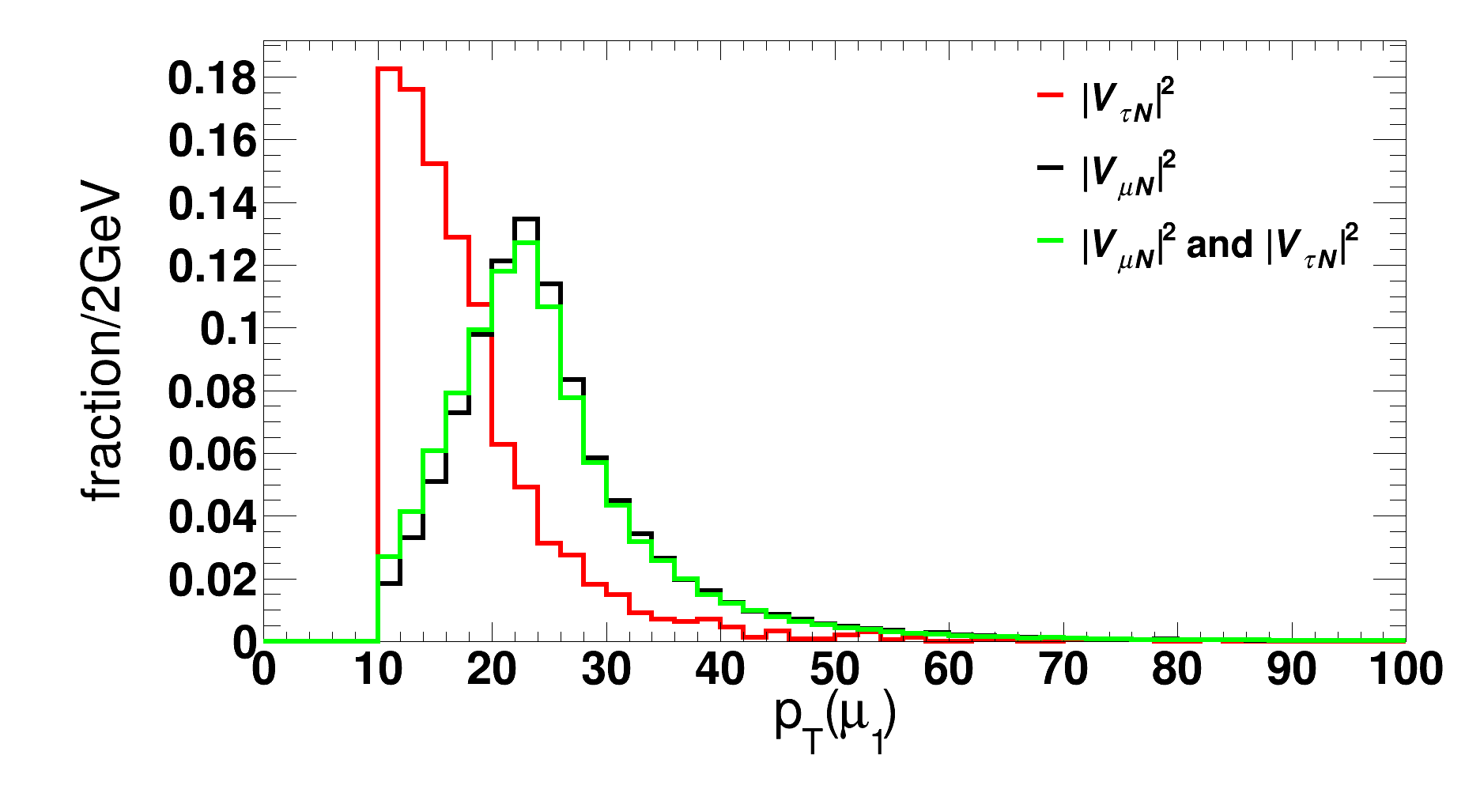}
		\includegraphics[width=4.5cm,height=3cm]{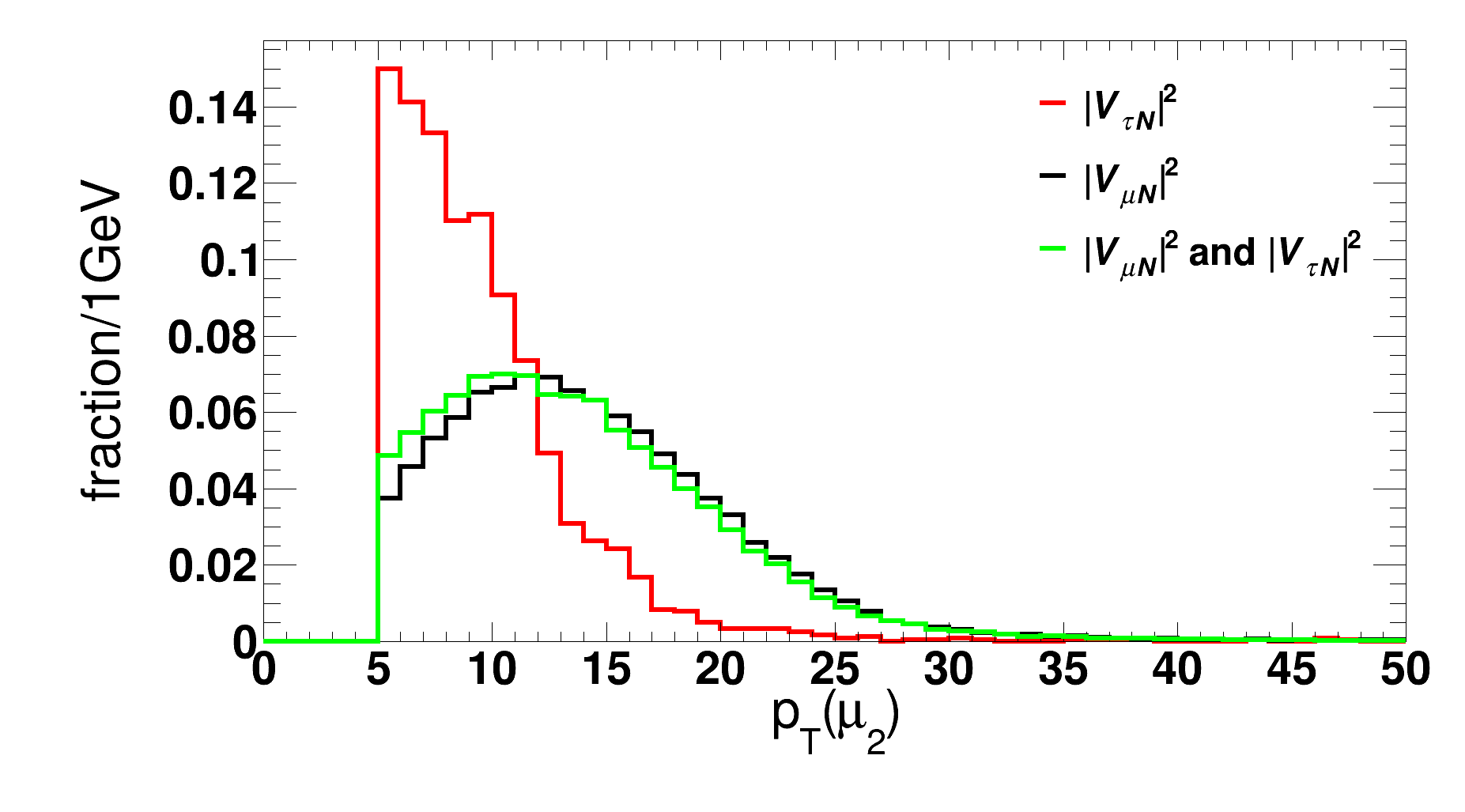}
	}
\end{figure}
\vspace{-1.0cm}
\begin{figure}[H] 
	\centering
	\addtocounter{figure}{0}
	\subfigure{
		\includegraphics[width=4.5cm,height=3cm]{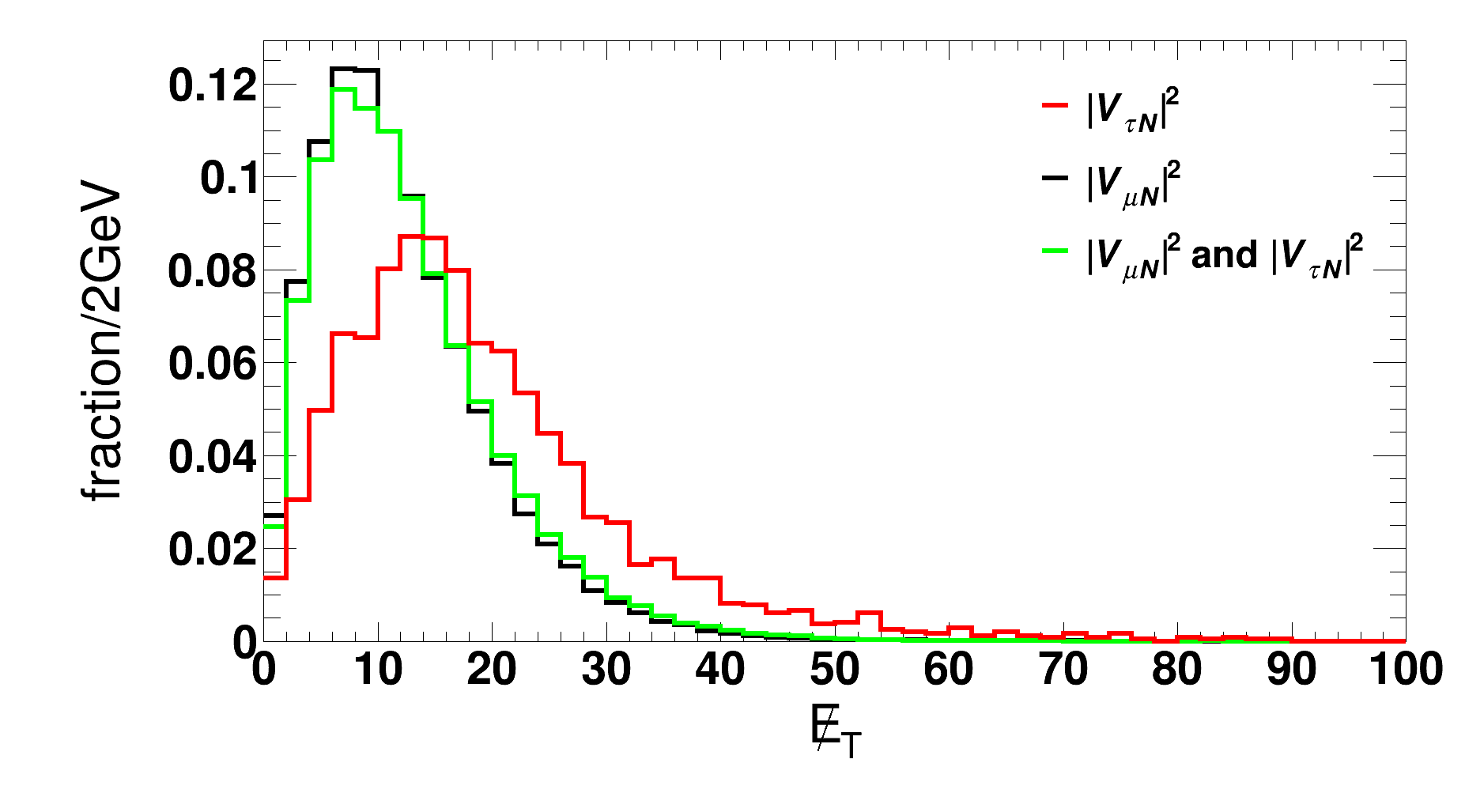}
		\includegraphics[width=4.5cm,height=3cm]{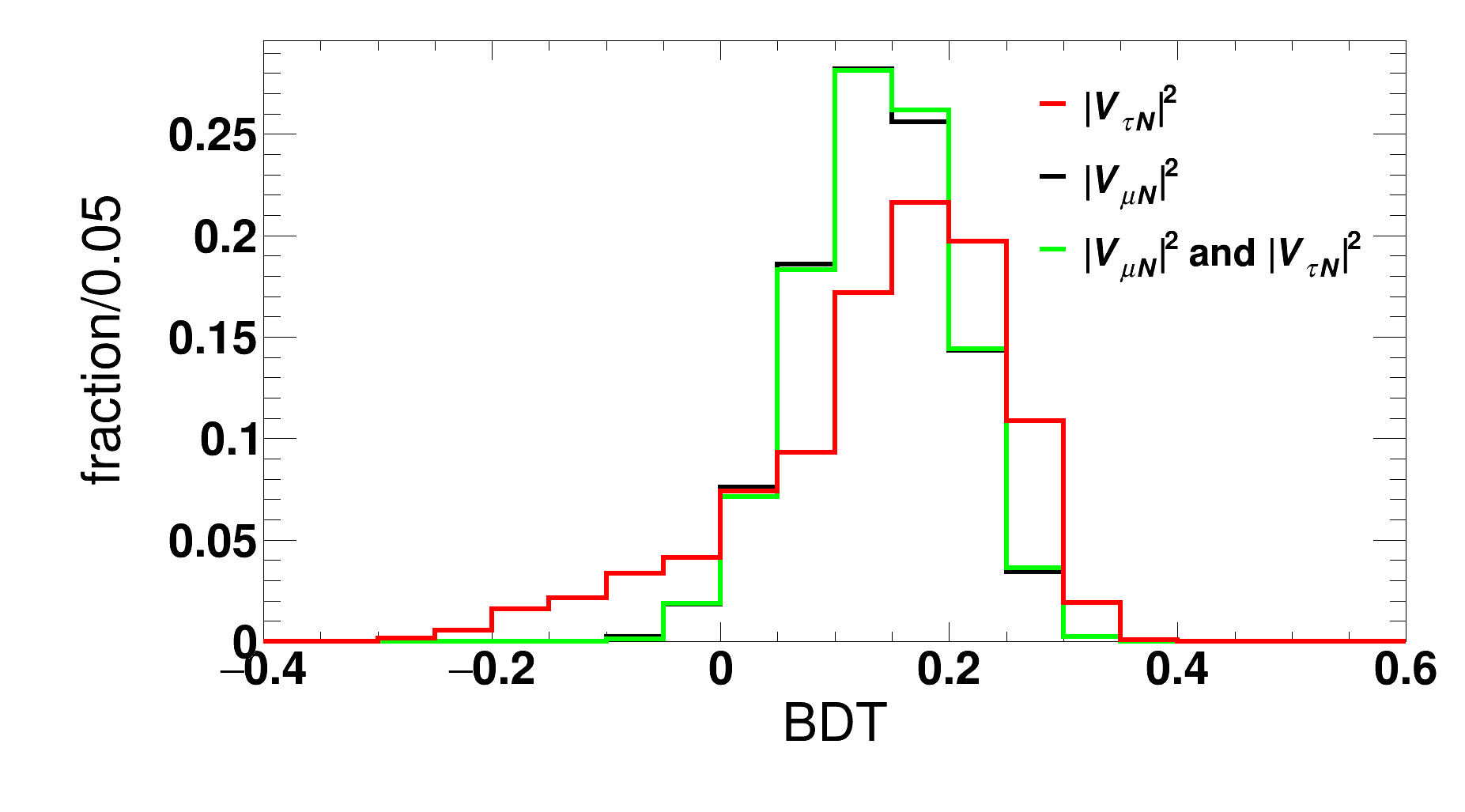}
	}
\caption{
The distributions of kinematic observables and BDT response after applying the preselection cuts when $m_N = 50$ GeV, for three assumptions of mixing parameters. 
}
\label{fig:kin}
\end{figure}

In Fig.~\ref{fig:kin}, 
we show distributions of kinematic observables and BDT response after applying the preselection cuts when $m_N = 50$ GeV for three cases.
One observes that compared with cases (ii) and (iii), the transverse momenta of final state muons are indeed softer, while the missing energy is larger in case (i).
We note that the case (ii) and case (iii) have very similar distributions. 
This is because due to small branching ratio of $\tau^- \to \mu^- \bar{\nu}_\mu \nu_\tau$, most signal events in case (iii) are still consisting of muons directly from $N$ vertex, rather than the tau decays.

\begin{table}[H]
\centering
\begin{tabular}{ccccc}
\hline
\hline
$m_N$ & selection & case (i) & case (ii) & case (iii) \\
\hline
\multirow{3}{*}{50 GeV} 
	&preselection&$8.13\mltp10^{-4}$&$1.26\mltp10^{-1}$&$3.78\mltp10^{-2}$ \\
 &BDT$>$0.116&$4.67\mltp10^{-1}$&$6.29\mltp10^{-1}$&$6.42\mltp10^{-1}$ \\
\cline{2-5}
               &total &$3.80\mltp10^{-4}$&$7.95\mltp10^{-2}$&$2.42\mltp10^{-2}$ \\
\hline
\hline
\end{tabular}
\caption{
Selection efficiencies of preselection and BDT cuts for signals in three cases at the HL-LHC when $m_N =$ 50 GeV. The last line indicates the total cut efficiencies.
}
\label{tab:effs}
\end{table}

Table~\ref{tab:effs}
shows the select efficiencies of preselection and BDT cuts for all three signal cases.
One observes that compared with case (i), the total efficiency increases by a factor of 209 and 64 in the cases (ii) and (iii), respectively.
This is mainly because the small branching ratio of $\tau^- \to \mu^- \bar{\nu}_\mu \nu_\tau$ and softer muons render signal events in case (i) passing preselection cuts less efficiently.
This is also the main reason why limits on the mixing parameters in case (i) are usually weaker than those in cases (ii) and (iii).
However, we emphasize again that cases (ii) and (iii) require the existence of sizeable $|V_{\mu N}|^2$, while case (i) depends on  $|V_{\tau N}|^2$ only.
In this sense, it is inappropriate to compare the limits for different cases, and this study focusing on case (i) still has its unique significance.
In future experiment, once the signal is observed, three cases can be distinguishable based on kinematics.

%%%%%%%%%%%%%%%%%%%%%%%%%%%%%%%%%%%%%%%%%
%%%%%%%%%%%%%%%%%%%%%%%%%%%%%%%%%%%%%%%%%
\begin{acknowledgments}
\noindent
We thank Boping Chen, Haiyong Gu, Filmon Andom Ghebretinsae, Hao Sun, Minglun Tian and Zeren Simon Wang for helpful discussions. 
L.B. and K.W. are supported by the National Natural Science Foundation of China under grant no.~11905162, 
the Excellent Young Talents Program of the Wuhan University of Technology under grant no.~40122102, and the research program of the Wuhan University of Technology under grant no.~2020IB024.
Y.N.M. is supported by  by the National Natural Science Foundation of China under grant no.~12205227 and the Fundamental Research Funds for the Central Universities (WUT:~2022IVA052).
The simulation and analysis work of this paper was completed with the computational cluster provided by the Theoretical Physics Group at the Department of Physics, School of Sciences, Wuhan University of Technology.
\end{acknowledgments}

%%%%%%%%%%%%%%%%%%%%%%%%%%%%%%%%%%%%%%%%%
%%%%%%%%%%%%%%%%%%%%%%%%%%%%%%%%%%%%%%%%%
	
\bibliography{Refs}

\bibliographystyle{h-physrev5}
	
\end{document}